\documentclass[a4paper,11pt]{article}
\pdfoutput=1 

\usepackage{jheppub}

\usepackage{comment}
\usepackage{xcolor}
\usepackage[utf8]{inputenc}
\usepackage{amsmath}
\usepackage{float}
\usepackage{amssymb}
\usepackage{graphicx}
\usepackage{multirow}
\usepackage[normalem]{ulem}

\usepackage{slashed}
\usepackage{mathtools}
\usepackage{pifont}

\usepackage{tikz-feynman}
\tikzfeynmanset{compat=1.1.0}
\usetikzlibrary{arrows,automata,positioning}
\tikzset{
	vertex/.style={circle,draw, minimum size=1.5em},
	edge/.style={->,> = latex'}
}

\definecolor{darkgreen}{rgb}{0.0, 0.5, 0.0}

\newdimen\arrowsize
\newdimen\mylw
\pgfkeys{/my arrows/chemeq/.style={draw,thick,double distance=2pt,onearc-onearc}}
\pgfkeys{/my arrows/size/.code={\pgfsetarrowoptions{onearc}{#1}}}
\def\myalw{.4pt}
\pgfarrowsdeclare{onearc}{onearc}{%
	\mylw=\myalw
	\pgfarrowsleftextend{-\pgfgetarrowoptions{onearc}-.5\mylw}
	\pgfarrowsrightextend{0pt}
}{%
	\pgfsetdash{}{0pt}
	\mylw=\pgflinewidth
	\pgfsetlinewidth{\myalw}
	\advance\arrowsize by.5\pgflinewidth
	\pgfpathmoveto{\pgfpoint{-\pgfgetarrowoptions{onearc}}{-\pgfgetarrowoptions{onearc}-.5\mylw}}
	\pgfpatharc{180}{90}{\pgfgetarrowoptions{onearc}}
	\pgfusepathqstroke
}

\title{\boldmath 
	Dark Matter in a Three-Brane Randall-Sundrum Scenario
    out of the Evanescent Limit}

\author{Andrea Donini,\,}
\author{Miguel G. Folgado,\,}
\author{Alejandro Muñoz-Ovalle}
\affiliation{Departamento de F\'isica Te\'orica, Universidad de Valencia and IFIC, Universidad de Valencia-CSIC,
C/ Catedr\'atico Jos\'e Beltr\'an, 2 | E-46980 Paterna, Spain}

\emailAdd{donini@ific.uv.es}
\emailAdd{migarfol@ific.uv.es}
\emailAdd{almuo@ific.uv.es}

\abstract{
The Nature of Dark Matter (DM), that constitutes  approximately 25\% of the energy density in the Universe, is still eluding us. 
An intriguing possibility is
that DM does indeed interacts with SM particles only gravitationally (the only mean by which we have detected
it so far), albeit in an extra-dimensional scenario
yet it has not been possible to detect it by some non-gravitational means. In a three-brane Randall-Sundrum
setup, with DM located on a Deep Infra-Red 
GeV-TeV brane, and the SM on an Infra-Red TeV-PeV one, it was shown to be possible to recover the observed DM relic abundance and somewhat relax the hierarchy problem, whilst avoiding LHC stringent bounds on DM and KK graviton masses that constrain severely similar two-brane setups.
The phenomenological results, however, have been obtained under
the assumption that the bulk curvatures on the left ($k_1$) and the right ($k_2$) of the intermediate IR-brane are identical, 
$k_2 \to k_1$. Since the brane tension $\sigma_{\rm IR}$ of the intermediate brane is proportional to $k_2 - k_1$ and, therefore, vanishes,
it is clear that this limit (for which the IR-brane becomes {\em evanescent}) is unphysical. We could say that this is {\em no brane}: 
if the brane tension of a brane vanishes, 
there is no brane in the bulk (pun intended).
In this paper, therefore, we study in detail the theoretical framework needed to explore this interesting phenomenological possibility, 
{\em out of the evanescent brane limit}. We show that most of the 
formul\ae $\,$used in the {\em evanescent limit} are still valid
for ${\cal O}(1)$ differences between $k_1$ and $k_2$ (thus, introducing no new unjustified hierarchy in the bulk).
Once the relevant couplings of radions and KK gravitons are computed, we study the (enlarged) parameter space of the model looking
for the region in which the relic DM abundance is recovered.
}

\begin{document}
	\maketitle
	\flushbottom

\section{Introduction} 
\label{sec:intro}

The existence of Dark Matter (DM) in the Universe is
an experimental evidence, stemming from many different sets of astrophysical and cosmological data, where its impact
on gravitational fields at very different length scales. 
The analysis
of these data sets seems to infer that some non-relativistic matter must be added to the energy content of the Universe.
The most simple explanation for this is that a huge amount
of massive particles with very small interaction with Standard
Model (SM) matter is distributed around visible astrophysical 
structures ({\em e.g.}, galaxies and clusters). However, 
we have not been able to detect any new massive
particle in collider experiments on Earth so far.

Several papers in the literature have explored the possibility
that DM is indeed made of massive particles that do interact
only gravitationally, albeit in an extra-dimensional 
setup, thus enhancing the strength of gravitational interactions. Remind that extra-dimensional BSM scenarios
have been proposed in order to solve the Hierarchy
Problem, to offer a first step to a theory of Quantum Gravity,
and maybe linking it with an underlying String Theory. It could thus be interesting to use them to solve one of the other open problem of the SM.
Is it possible that enhanced gravitational interactions
offer a way to explain the observed DM relic abundance
through a thermal freeze-out mechanism? In order to 
explore this proposal, using a Randall-Sundrum (RS) setup 
is a popular option (see Ref.~\cite{RandallLisa1999LMHf} for RS1 and \cite{RandallLisa1999AAtC} for RS2 models, respectively). 
The parameter space of a RS1 model with DM and the SM 
located at the IR-brane was studied in Refs.~\cite{Rizzo:2018joy,Folgado:2019sgz,Folgado:2020vjb,deGiorgi:2021xvm,chivukula2024limitskaluzakleinportaldark, Chivukula:2025pmk} and \cite{Lee:2013bua,Carrillo-Monteverde:2018phy,Lee:2024wes} (with a little more emphasis on the relation between RS models and the holographic principle 
\cite{tHooft:1993dmi,Susskind:1994vu,Maldacena:1997re} in the latter). The 
outcome of these studies is that the observed DM relic abundance can be reproduced mainly through ${\rm DM \, DM} \to {\rm SM \, SM}$ processes in a tiny portion of the parameter space, 
where most of it is excluded by LHC resonance searches.
The mathematical details of this setup, that are non-trivial due to the delicate cancellations of unitarity-violating terms
between triple graviton vertices and diagrams that involve the radion ({\em i.e.}, the scalar component of the 5D metric, $h_{55}$) have been worked out in Refs.~\cite{de_Giorgi_2021,2311.01507,Chivukula_2019,Chivukula_2020,Chivukula_2020b,Chivukula:2023qrt}.
In Ref.~\cite{Folgado:2019gie}, the same hypothesis
was studied, albeit in a different geometry, the so-called
{\em Clockwork/Linear Dilaton} model \cite{Antoniadis:2001sw,Cox:2012ee,Giudice:2017fmj}, with similar
results. On the other hand, several papers studied
the possibility to implement the same setup, whilst
adopting the {\em freeze-in} scenario \cite{Bernal:2020fvw,Bernal:2020yqg,deGiorgi:2022yha}.

\begin{figure}[htbp]
	\centering
\includegraphics[width=0.5\textwidth]{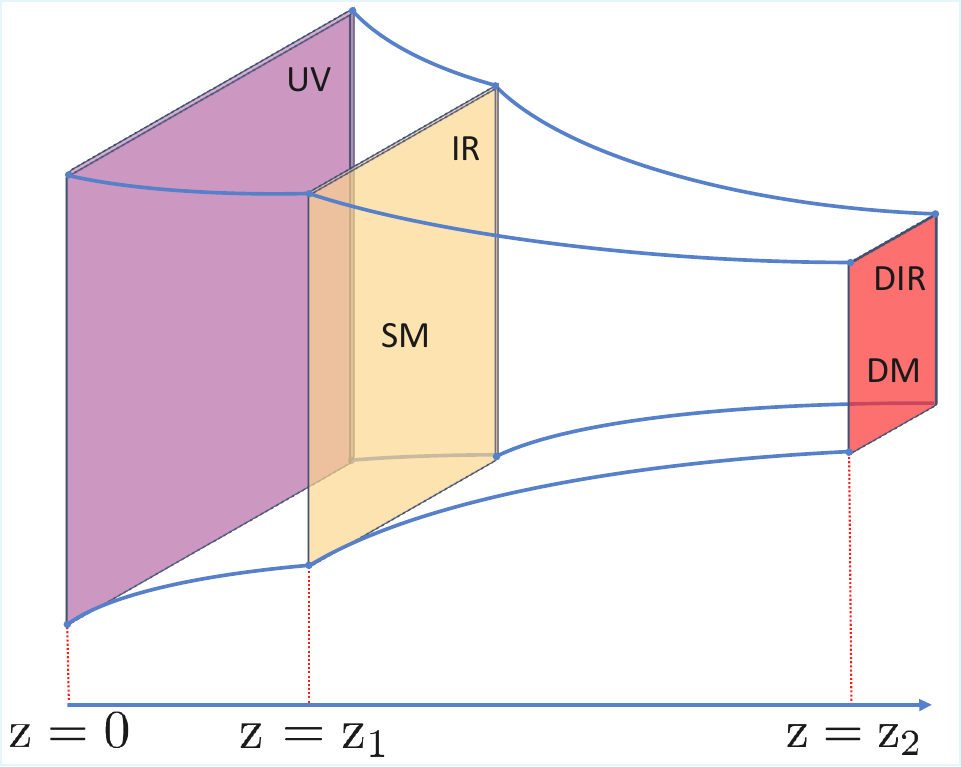}
 	\caption{Schematic representation of the three-brane setup used in this work in conformal coordinates. }\label{fig:3branes}
\end{figure}

In this paper, we address the same problem ({\em i.e.}, explaining the observed relic DM abundance with brane-localized DM in an extra-dimensional RS setup), albeit introducing
three branes in a RS1 geometry: the UV Planck-scale brane, the IR intermediate brane (where the SM lives), and the
Deep IR (DIR) brane (where the DM resides). Whereas the UV-
and the DIR-branes are located at the fixed point of a $Z_2$-orbifold, as in a standard RS1 setup, the intermediate IR-brane
is located at some tunable point $y_1$ within the segment
$y \in [0, y_2]$ (where $y_2 = \pi r_c$, being $r_c$ the orbifold compactification radius). In order to maximize the
phenomenological impact, in conformal coordinates 
the IR-brane is located at $z_1$, with $1/z_1\sim {\rm TeV-PeV}$ scale, whereas the DIR-brane is at $z_2$, with $1/z_2\sim {\rm  GeV-TeV}$ scale. In Fig.~\ref{fig:3branes} we illustrate the three-brane framework considered in this work.
 
Using this particular choice of $y_1$ and $y_2$,
interactions of KK gravitons and radions with DM at the DIR-brane
are fully enhanced due to the extra-dimension, whereas their interactions with SM at the IR-brane, or SM-DM interactions
mediated by KK gravitons and radions are only partially enhanced, 
thus being effectively suppressed with respect to the former.
A similar three-brane setup was first proposed in Ref.~\cite{Kogan:1999wc,Kogan:2000cv,Kogan:2000xc}
and recently revisited in Refs.~\cite{Seung1,Seung2}. 
The phenomenology of this setup, together with a comprehensive
study of the experimental bounds from collider and Direct/Indirect 
Dark Matter searches, was studied for the case of scalar DM in Ref.~\cite{Donini:2025cpl}.
The outcome is that the region of the parameter
space where it is possible to reproduce the observed DM relic
abundance is much larger than in the two-brane setup, mainly
due to the fact that the LHC bounds are largely ineffective, 
as they are suppressed by powers of the IR scale, 
that can be much larger than the DIR one.
In particular, it was found that DM primarily undergoes annihilations into gravitons/radions, with radions decaying eventually into SM well before the onset of Big Bang Nucleosynthesis. This strongly differs
from two-brane setups, where DM dominantly annihilates into SM particles (through a graviton/radion exchange) and the DM relic abundance is reproduced in a different (and tiny) region of the parameter space. 

These results have
been obtained under a significant simplification, though: in a setup 
with three branes, the bulk curvature to the left and to the 
right of the intermediate brane ($k_1$ and $k_2$, respectively) 
will generally differ. However, computations are greatly
simplified in the limit $k_2 \to k_1$, an approximation 
that has been used in Ref.~\cite{Donini:2025cpl} and
also in Refs.~\cite{Cacciapaglia,Lee,Koutroulis:2024wjl}, where
similar results have been obtained, albeit
with different phenomenological assumptions\footnote{The first
reference study the case of a freeze-in scenario; 
the second reference is set
in the context of conformal field theory, where the dilaton plays the role of the radion, and to reconcile the relic abundance of GeV-scale DM and abide by constraints from indirect detection, they considered forbidden DM; in the third reference, DM is made out of Dirac fermions.}
It should be stressed that, even though computations 
are greatly simplified by this approximation, the geometrical setup 
in the limit $k_2 \to k_1$ is rather unphysical. 
This is because the brane
tensions of the three-branes are tuned to some specific
values in order to get a stable Anti-de Sitter background
geometry on the orbifold. For this reason, the brane tension
of the intermediate brane is proportional to the difference
$k_2 - k_1$. In the limit $k_2 \to k_1$, then, the brane
tension of the intermediate brane vanishes, thus making
the brane {\em evanescent}. If we assume that, physically
speaking, a brane is a topological defect with lower-dimensionality than the bulk, characterized
by a certain 4D energy density (the brane tension), in the
{\em evanescent limit} the intermediate brane effectively
becomes only a mathematical concept. Any mechanism that
could localize SM fields relies on the fact that the brane
has some physical property, related to its tension
(see, {\em e.g.}, Refs.~\cite{Bajc:1999mh,DeRujula:2000he,Dimopoulos:2000ej,Dubovsky:2001pe,Shifman:2003uh,Dvali:2007nm,Guerrero:2009ac}). It is therefore
reasonable to compute how the departure from the evanescent limit affects the results of Ref.~\cite{Donini:2025cpl}.
This is what we do in this paper: in Sects.~\ref{sec:twobranes}
and \ref{sec:GWtwobranes} we remind the details
of the two-brane RS1 setup and its stabilization through
the Goldberger-Wise mechanism \cite{GW1,GW2,GW3}; in Sect.~\ref{sect:threebranes} we extend the treatment
to the case of three branes and work out the wave-function
of the KK gravitons and their couplings to DM and SM fields; 
in Sect.~\ref{sect:threebranesradions} we apply the
Goldberger-Wise mechanism to the three-brane setup, compute
the masses of the scalar modes that stabilize the segments
$y \in [0,y_1]$ and $y \in [y_1, y_2]$ of the orbifold as a function
of the parameters of the model and, eventually, derive
their couplings with DM and SM fields; in Sect.~\ref{sec:pheno} we perform
a first study of the enlarged parameter space of the 
three-brane setup out of the evanescent limit to look
for regions in agreement with present experimental constraints and that reproduce the observed relic DM abundance; in 
Sect.~\ref{sec:conc} we eventually draw our conclusions.

\section{Warming up: a two-brane Warped Extra-Dimensions reminder}
\label{sec:twobranes}	
	
The popular Randall-Sundrum (RS1) scenario~\cite{RandallLisa1999AAtC} considers a non-factorizable 5-dimensional metric in the form:
\begin{equation}
ds^2 = \bar G^{(5)}_{MN} \,  dx^M dx^N  = e^{- 2 A(y)} \, \eta_{\mu\nu} \, dx^\mu \, dx^\nu - dy^2,
\end{equation}
where $\bar G^{(5)}_{MN}$ is the background metric, $A(y) = k |y|$, with $k$ the curvature along the 5-th dimension, and the signature of the metric is $(+,-,-,-,-)$. 
The extra spatial dimension, denoted by
$y$, is compactified on a circle with radius $r_c$, subject to a $Z_2$ orbifold symmetry. This identification renders the points $y=0$ and $y=\pi r_c$ as fixed, singular loci. Two branes are situated at these orbifold fixed points: the so-called ultraviolet (UV) brane at $y=0$, and the infrared (IR) brane at $y=\pi r_c$. This configuration corresponds to the conventional RS1 setup, wherein the Standard Model fields are typically localized on the IR brane. In this framework, all fundamental scales — such as the curvature scale $k$ and the 5D Planck scale $M_5$ — are assumed to be of the order of the 4D Planck mass.  The action in 5D is:
\begin{equation}
    S_\text{RS}= - M_5^3\,\int d^{4}x \int_0^{\pi r_c} dy\, \sqrt{\bar G^{(5)}} \, \left[ R^{\left(5\right)} + 2 \, \Lambda_{5} \right], \label{eq:RSaction}
\end{equation}
where $R^{\left(5\right)}$ is the 5D Ricci scalar, 
$\bar G ^{(5)}$ is the determinant of the 5D
metric $\bar G^{(5)}_{MN}$ given above and $\Lambda_5$ is the 5D cosmological constant.
By solving Einstein's equations of General Relativity, the following relation between the 5D cosmological constant $\Lambda_{5}$ and the curvature $k$ may be obtained \cite{RandallLisa1999LMHf}:
\begin{equation} \label{k curvature}
    k=\sqrt{-\frac{\Lambda_{5}}{6}} \, .
\end{equation}
This setup exhibits a 5D bulk geometry characterized by anti-de Sitter spacetime, associated with a negative 5D cosmological constant $\Lambda_5$. Within this framework, the 4D Planck mass is related to the fundamental 5D scale $M_5$ through the following relation:
\begin{equation}
    \label{eq:MplanckRS}
    \bar M_{\rm P}^2 = \frac{M_5^3}{k} \, \left (1 -e^{-2 \pi k r_c} \right ) = \frac{M_5^3}{k} \, \left (1 - \bar \omega^2 
    \right ) \, ,
\end{equation}
where $\bar M_{\rm P}$ is the reduced Planck mass, $\bar M_{\rm P} = M_{\rm P}/\sqrt{8 \pi}$ and $\bar \omega = \exp{(- \pi \,  k \, r_c)}$. For a different choice of the location of IR- and UV-branes see, {\em e.g.}, App.~C of Ref.~\cite{Giudice:2016yja}.

To ensure the stability of the anti-de Sitter background geometry along the compactified segment $y \in [0,\pi r_c]$, it is necessary to incorporate additional localized terms in the action. These terms must be introduced at the orbifold fixed points, specifically at $y = 0$ and $y = \pi r_c$.
\begin{equation}
    \label{eq:RSbraneterms}
    S_{\rm brane} = \sum_{i = \rm UV, \rm IR} 
    \int d^4 x \int_0^{\pi r_c} dy \sqrt{- \bar g_i^{(4)}} \delta (y - y_i) 
    \left \{ - \sigma_i + \dots \right \} \, , 
\end{equation}
where $\sigma_i$ are the brane tensions and $\dots$ refer to the Lagrangian density of fields that can be localized either on the UV or the IR brane.
The determinant of the induced metric, $\bar g_i{(4)} = \bar G^{(5)}/ \bar G^{(5)}_{55}$, is just $\bar G^{(5)}(x,y)$ computed at the brane locations, since $\bar G^{(5)}_{55} = -1$.

As previously emphasized, in the two-brane RS1 model all fundamental mass parameters are naturally of order $\bar{M}_{\rm P}$. However, energy scales associated with fields confined to the IR brane are exponentially suppressed relative to the Planck scale due to the warped geometry. In particular, a bare Higgs mass $m_0\sim M_{\rm P}$ becomes exponentially redshifted at the position of the IR brane. This results in a physical Higgs mass of the order of the electroweak scale, arising from the warped background metric evaluated at $y = \pi r_c$. After canonical normalization of the Higgs kinetic term, the physical mass is given by
\begin{equation}
    m_H =  \exp{\left ( - \pi \, k \, r_c \right )} \times m_0 = \bar \omega \, m_0 \, ,
\end{equation}
where $m_H$ denotes the observed Higgs mass. By appropriately choosing the product $k \, r_c$, it is possible to naturally generate $m_H\sim 1$ TeV, thereby offering a possible 
solution of the SM hierarchy problem.
 
The values of $\sigma_i$ must be chosen properly to glue the metric in the intervals $y \in ]-\pi r_c, 0[$ and $]0, \pi r_c[$, while at the same time 
enforcing reflectivity at the $y = 0$ and $y = \pi r_c$: 
\begin{equation}
\sigma_{\rm UV}= - \sigma_{\rm IR} =  6 \,M_5^3 \,k \, .
\label{eq:branetensionfinetuningRS}
\end{equation}
In the case of a weak gravitational field, we can expand the metric over the background metric $\bar g^{(5)}_{MN}$ such that:
\begin{equation}
\label{eq:gravitonexp1}
G^{(5)}_{MN} = \bar G^{(5)}_{MN} + \frac{1}{M_5^{3/2}} \, h^{(5)}_{MN} (x,y) + \dots \, .
\end{equation}
The 5D graviton field $h^{(5)}_{MN}(x,y)$ can be then decomposed in a Kaluza-Klein tower of 4D fields as follows: 
    \begin{equation}
    	h^{(5)}_{MN}(x,y) = \sum_n h_{MN}^n(x) \chi^{(n)}(y) \, .
    \end{equation}
From this, we infer that the wave-function $\chi^{(n)}(y)$ has dimension $[\chi^{(n)}(y)] = 1/2$. In the 5D theory, the graviton decomposes into three distinct Kaluza–Klein (KK) towers of four-dimensional fields: a tower of massive spin-2 modes $h_{\mu\nu}^{n} (x)$, commonly referred to as {\em KK gravitons}; a tower of spin-1 modes $h_{\mu 5}^n (x)$, known as {\em KK graviphotons}; and a tower of scalar modes $h_{55}^n$, identified as {\em KK graviscalars}.

In the so-called {\em unitary gauge}, in which only the physical degrees of freedom are manifest, it can be shown (in the case of a flat extra dimension; see Ref.~\cite{Giudice:1998ck}) that the KK graviscalars are absorbed by the KK graviphotons, thereby endowing the latter with mass. Subsequently, the KK graviphotons themselves are ``eaten" by the spin-2 KK modes, providing the necessary longitudinal components for the massive KK gravitons. As a result, the physical spectrum of the model consists only 
of the massless modes 
$h_{\mu\nu}^0$, $h_{\mu 5}^0$ and $h_{55}^0$, together with a tower of massive KK gravitons, each carrying 5 d.o.f.'s. It is also established that the massless graviphoton does not couple to SM matter~\cite{Giudice:1998ck}, and it is therefore typically neglected in phenomenological analyses. 

Notice that eq.~(\ref{eq:gravitonexp1}) is not the only way to introduce gravitons ({\em i.e.} fluctuations over the metric), though. Traditionally, when a warped extra-dimension is considered, the scalar fluctuation $h_{55}(x,y)$ (the {\em graviscalar}) is introduced in a different parametrization \cite{Csaki:2000zn}: 
\begin{equation}
\label{eq:perturbedmetric}
ds^2 = e^{-2 A } \left [ \left ( 1 - 2 F \right )\, \eta_{\mu\nu} + E_{, \mu \nu} \right ] dx^2 - ( 1 + G )^2 dy^2 = \left ( \bar G_{MN} + \delta G_{MN} \right ) \, dx^M dx^N \, ,
\end{equation}
where
\begin{equation}
\delta G_{MN} = \left (
\begin{array}{cc}
e^{-2 A} \, \left [ \left ( 1 - 2 F \right )\, \eta_{\mu\nu} + E_{, \mu \nu} \right ] & 0 \\
& \\
0 & - ( 1 + G )^2 \, .
\end{array}
\right )
\end{equation}
In this expression, $F,E$ and $G$ are adimensional scalar fluctuations. In a two-brane setup, it can be shown that $E$ can be gauged away by choosing $E = 0$, whereas $F$ and $G$
are proportional due to the Einstein equation for $R_{\mu 5}$. Therefore, we are left with one dynamical scalar degree of freedom, conventionally chosen to be $F(x,y)$. Also in this
case it can be shown that, as it was the case for a flat extra-dimension, KK excitations of $F$ can be removed from the physical spectrum and, eventually, only one massless 4D
scalar field is present. The two parametrizations 
of the perturbed metric in eqs.~(\ref{eq:gravitonexp1})
and (\ref{eq:perturbedmetric}) may look different, but they
are standard choices once we adopt a given perspective
of what General Relativity is: the former parametrization 
is a straightforward way to introduce quantum degree of freedoms, the gravitons, treating GR as a non-renormalizable effective low-energy model of some underlying renormalizable Quantum Theory of Gravity; on the other hand, the latter parametrization is the standard way to introduce perturbations of the background metric treating GR as the fundamental classical theory of gravity. In the first case we will 
use gravitons in quantum field theory computations
over the background metric (see, {\em e.g.}, Ref.~\cite{Birrell:1982ix}), whereas
in the second case we will solve the classical Einstein's
equations to assess the impact of perturbations on $\bar G_{MN}$.

We will first study the eigenfunctions in the extra-dimension of the KK graviton modes. They can be obtained solving the equation of motion:
\begin{equation}
\left \{ \partial_{y}^2 - 4 k \partial_{y} + m_{n}^2 e^{2 k y} \right \} \chi^{(n)} (y) = 0
\end{equation} 
in the interval $y \in [0, \pi r_c]$. 
Introducing a coordinate $z$ appropriately chosen \cite{Csaki:2004ay}, it can be shown that the metric can be written as 
$ds^2 = 1/g^2(z) \left ( \eta_{\mu\nu} dx^\mu dx^\nu - dz^2\right)$, 
where:
\begin{equation}
    z = \frac{1}{k} \left ( e^{k y} - 1 \right ) \; \qquad {\rm for\,\,} \;  y \in [0, L = \pi r_c] \, ,
\end{equation}
and the ``conformal weight" is:
\begin{equation}
\label{eq:gztwobranes}
    g(z) = k z + 1 \, .
\end{equation}
In terms of the conformal coordinate $z$, the KK graviton wave-function is:
\begin{equation}
\label{eq:eigenfunctions2branes}
\left \{
\begin{array}{l}
 \hat \chi^{(0)} (z) = \frac{A_0}{\left [ g(z) \right ]^{3/2}} \, , \qquad m_0 = 0 \, , \\
 \\
    \hat \chi^{(n)} (z) =  \sqrt{\frac{g(z)}{k}}\, \left [ 
    A_{1n} \, Y_2 \left ( \frac{m_n}{k} \, g(z)\right ) + 
    A_{2n} \, J_2 \left ( \frac{m_n}{k} \, g(z)\right ) 
    \right ] \, ,
    \end{array}
    \right .
\end{equation}
where $\chi^{(n)} (y)  = \hat \chi^{(n)}(z) / \sqrt{g(z)} $, and $J_2$ and $Y_2$ are the Bessel function of the first and second kind, respectively. 

The non-zero mass spectrum of KK gravitons, denoted $m_n$, along with the corresponding normalization coefficients $A_{in}$, can be determined by imposing appropriate boundary conditions arising from the orbifold compactification. Specifically, the wave-functions must satisfy periodicity under translations around the compact dimension, $\chi^{(n)} (y) = \chi^{(n)} (y + 2 \pi r_c)$, as well as invariance under the $Z_2$ reflection symmetry, $\chi^{(n)}(y) = \chi^{(n)} (-y)$. Moreover, continuity of the wave-function across the brane locations must be enforced, i.e., $\chi^{(n)}(y \to 0^-, \pi r_c) = \chi^{(n)}(y \to 0^+, -\pi r_c)$, along with the appropriate jump conditions for its first derivative, which account for the localized energy densities at the branes. These derivative discontinuities are required to consistently match the metric across each brane and are derived from the Einstein equations with brane sources~\cite{Shiromizu:1999wj}. Applying these conditions, one obtains:
\begin{equation}
\label{eq:BCtwobranes}
\left \{
\begin{array}{l}
Y_1 \left ( \frac{m_n}{k}\right ) + \frac{ A_{2n} }{ A_{1n} } \,  J_1 \left ( \frac{m_n}{k} \right ) = 0 \, ,\\
\\
Y_1 \left ( \frac{m_n}{k} \, e^{\pi k r_c} \right ) +  \frac{A_{2n}}{A_{1n}} \, J_1 \left ( \frac{m_n}{k} \, e^{\pi k r_c }\right ) = 0 \, .
\end{array}
\right .
\end{equation}
Notice that the same result is obtained in two steps when using
the parametrization of the perturbed metric in eq.~(\ref{eq:perturbedmetric}): first, we implement the boundary conditions (BC) relative to the discontinuity of the first derivative of the background metric due to the brane tension terms (at least) localized at $y = 0, \pi r_c$. Then, we solve the Einstein equation for the tensor fluctuation $h_{\mu\nu}$ of the metric, imposing continuity of the wave function of the KK gravitons and of its first derivative at the brane locations \cite{Csaki:2000zn}.
Introducing the warping factor $\bar \omega = \exp(- \pi k r_c)$ and the mass $m_n$ as $m_n = k \, \left (x_n + \delta x_n \right ) \, \bar \omega $, where $x_n$ are the zeroes of
the Bessel function of the first kind $J_1 (x)$, we see that $m_n \ll k$. Assuming that the shifts $\delta x_n$ are small quantities, from the first equation in (\ref{eq:BCtwobranes}) 
we get at leading order in $\bar \omega$: 
\begin{equation}
\label{eq:twobranesJ2coeff}
A_{2n} = \frac{4}{\pi} \, \frac{1}{x_n^2 \, \bar \omega^2} \, A_{1n} + \dots \, ,
\end{equation}
from which we immediately see that $A_{2n} \gg A_{1n} $, and substituting in the second equation we find
\begin{equation}
\label{eq:twobranesmassshift}
\delta x_n = \frac{\pi}{4} \, x_n^2 \, \frac{Y_1(x_n)}{J_2(x_n)} \, \bar \omega^2 + \dots \, ,
\end{equation}
that is indeed small.
This means that the mass spectrum is \cite{Davoudiasl:1999jd}: 
\begin{equation}
\label{eq:twobranesmaspectrum}
m_n = k \, x_n \, \bar \omega + {\cal O}(\bar \omega^3) \, , 
\end{equation}
{\em i.e.} at leading order they are proportional to the $J_1$ Bessel function zeroes and are ${\cal O} (\bar \omega M_{\rm P})$. 
The normalisation factors $A_0$ and $A_{1n}$ can be obtained using the orthonormalisation condition: 
\begin{equation}
\label{eq:twobranesnormalizationcond}
    2 \int_0^{\bar z}  dz  \hat \chi^{(m)} (z) \, \hat \chi^{(m)} (z)  = 1 \qquad m = 0, 1, \dots \, .
\end{equation}
For the zero mode we get: 
\begin{equation}
\label{eq:twobranesnormalizationfactor0}
A_0 = \sqrt{\frac{k}{1- \bar \omega^2}} \, ,
\end{equation}
whereas for the $n$-th mode: 
\begin{eqnarray}
\label{eq:twobranenormalization}
1 &=& 2 \int_0^{z(\pi r_c)} \, dz \, \frac{g(z)}{k} \,
\left [ A_{1n} \, Y_2 \left ( \frac{m_n}{k} \, g(z) \right ) + A_{2n} \, J_2 \left ( \frac{m_n}{k} \, g(z) \right ) \right ]^2 
    \nonumber \\
    &\simeq& \frac{32}{\pi^2} \, \frac{1}{x_n^6 \,
    \bar \omega^6} \, \frac{A_{2n}^2}{k^2} \, \int_{x_n \, 
    \bar \omega}^{x_n} \, du \, u \, J^2_2(u) \, ,
\end{eqnarray} 
where $u = x_n \, \bar\omega \, g(z)$. Using standard Bessel function integration tables \cite{BesselIntegrals}, eventually:
\begin{equation}
\label{eq:twobranescoefficientA2n}
A_{2n} = k \, \frac{1}{J_2(x_n)} \, \bar \omega + \dots \, .
\end{equation}
from which:
\begin{equation}
\label{eq:twobranesnormalizationfactorn}
A_{1n} = \frac{\pi}{4} \, k \, \frac{x_n^2}{J_2(x_n)} \, \bar \omega^3 + \dots \, ,
\end{equation}

Therefore, at leading order in $\bar \omega$, the wave-function for the $n$-th KK graviton in terms of the conformal coordinate $z$ is:
\begin{equation}
\label{eq:twobranefinaleigenfunction}
    \hat \chi^{(n)}(z) = \sqrt{k \, g(z)} \, 
    \left [ \frac{\pi}{4} \, \frac{x_n^2}{J_2 (x_n)} \, \bar \omega^3 \,
    Y_2 \left (\frac{m_n}{k} g(z) \right ) -
     \frac{1}{J_2 (x_n)} \, \bar \omega \, 
     J_2 \left (\frac{m_n}{k} g(z) \right )\right ] \, .
\end{equation}
For low-lying modes ($n$ small), the KK gravitons are well separated in mass, with a non-uniform spacing determined by the zeroes of the Bessel function $J_1$. These modes are typically treated as individual resonances in collider searches, such as those conducted at the LHC. In contrast, for large $n$, the spacing between adjacent KK modes becomes approximately constant, leading to a quasi-continuous spectrum.

It can also be shown that the couplings of KK gravitons to fields localized on the IR brane —such as the SM fields in the conventional RS1 framework— are universal and characterized by a single effective scale, denoted $\Lambda_{\rm IR, 2b}$:
\begin{equation}
{\cal L} = -\frac{1}{\bar M_{\rm P}} \, T^{\mu \nu}(x) h^{0}_{\mu \nu}(x) -\frac{1}{\Lambda_{\rm IR, 2b}} \, \sum_{n=1} T^{\mu \nu}(x) h^{n}_{\mu \nu}(x) \, ,
\end{equation}
where: 
\begin{equation}
\Lambda_{\rm IR, 2b} = \bar \omega \, \bar M_{\rm P}
\end{equation}
(with the label ``2b" reminding that this scale refers to the two-branes setup).
The universality of the coupling does not hold for fields localized on the UV brane. In this case, the coupling between KK gravitons and UV localized fields depends on the KK mode number $n$. Specifically, within the two-brane RS1 setup, the coupling of the $n^{\rm th}$ KK graviton to a 4D field confined to the UV brane (at $y=0$) is suppressed and scales with a $n$-dependent factor:
\begin{equation}
\label{eq:twobranesKKcouplingUV}
\Lambda^n_{\rm UV, 2b} = \frac{c_n}{\bar \omega^3} \, \bar M_{\rm P} \, ; \qquad \qquad c_n = 8 \frac{J_2(x_n)}{x_n^2} \, .
\end{equation}

\section{Stabilization of the two-brane setup: excerpts}
\label{sec:GWtwobranesShort}

We present here a short reminder of how to stabilize the length 
of the extra-dimension within the {\em Goldberger-Wise mechanism}
for two branes in an anti-de Sitter 5D metric. The interested reader is referred to the related App.~\ref{sec:GWtwobranes}, where full details are given in a (hopefully) clear way.

If the size of the extra-dimension is unstabilized, the graviscalar zero-mode (either $h_{55}^{(0)}$ or $F$, depending on the specific
parametrization of the metric) is a massless Goldstone boson: 
since the same RS1 background metric can be obtained for any choice of $r_c$, choosing any specific value for it implies a spontaneous breaking dilatation invariance. In order to give it
a mass, we must introduce an explicit breaking of the symmetry. 
Adding a scalar bulk field $\varphi$ with both a bulk potential $U(\varphi)$ and two localized potentials $V_i(\varphi)$ chosen appropriately, it can be shown that an effective potential for
the graviscalar field is generated, and minimizing it we can 
fix the length of the extra-dimension and, at the same time, 
induce a mass term for the would-be Goldstone. This mechanism
was first suggested in Refs.~\cite{GW1,GW2} and it is widely
known as {\em Goldberger-Wise mechanism}. 

Consider the following action for the scalar bulk field: 
\begin{eqnarray}
{\cal S} = {\cal S_{\rm grav}} &+& \int d^4 x \int_0^{r_c} dy \, \sqrt{\bar G^{(5)}} \, \left ( \frac{1}{2} \, \partial_M \varphi \partial^M \varphi - U(\varphi) \right ) \nonumber \\
&-& \sum_{j={\rm UV, IR}} 
\int d^4x \int_0^{\pi r_c} dy \, \sqrt{- \bar g_j} \, \delta (y - y_j) \, V_j (\varphi) \, ,
\end{eqnarray}
where $y_{\rm UV} = 0$ and $y_{\rm IR} = \pi r_c$. The brane terms $V_j$ are:
\begin{equation}
\left \{
\begin{array}{l}
V_{\rm UV} (\varphi) = \mu_{\rm UV} \left ( \varphi^2 - \frac{v_{\rm UV}^2}{\pi r_c} \right )^2 \, , \\
\\
V_{\rm IR} (\varphi) = \mu_{\rm IR} \left ( \varphi^2 - \frac{v_{\rm IR}^2}{\pi r_c} \right )^2 \, ,
\end{array}
\right .
\end{equation}
with $v_{\rm UV} > v_{\rm IR}$. A factor $\pi r_c $ has
been introduced so as to normalize properly the VEV's $v_i$.
The parameters $\mu_j$ have the dimension of an inverse mass squared, whereas the VEV's $v_j$ have the dimension of a mass.

The bulk field can be decomposed as follows: 
\begin{equation}
    \varphi (x,y) = \varphi^{(0)} (x, y) + \delta \varphi  (x,y)\, ,
\end{equation}
where:
\begin{equation}
\label{eq:zero modeRS2b}
    \varphi^{(0)} (x,y) =  \left [ \frac{1}{\pi r_c} + \phi^{(0)}_1 (x) \right ] \, \varphi^{(0)} (y) \, , 
\end{equation}
being $\varphi^{(0)}(y)$ the zero mode wave-function\footnote{Notice that the zero mode of the bulk field $\varphi$ is the only field that may develop a ($y$-dependent) VEV.}, and the 
KK modes can be expanded on 4D fields as usual: 
\begin{equation}
\delta \varphi (x,y) = \sum_{n= 1}^\infty \phi^{(n)} (x) \, \varphi^{(n)} (y) \, .
\end{equation}

After solving the equation of motion in the extra-dimension (see App.~\ref{sec:GWtwobranes}), we get for the eigenfunctions of the KK modes of the scalar bulk field: 
\begin{equation}
\hat \varphi^{(n)} (z) =   2 \sqrt{2} \frac{k}{x_{2 n} \, J_0 (x_{2 n})} \, J_\nu \left [ x_{2 n} \, 
\bar \omega g(z) \, \right ] + {\cal O} \left ( \bar \omega^{2 \nu} \right ) \, ,
\end{equation}
where $\nu = 2 \sqrt{1 + \epsilon}$ (with $\epsilon = m^2/4 k^2$), 
$x_{2n}$ is the $n$-th zero of the Bessel function $J_2(x)$
and $g(z)$ has been defined in eq.~(\ref{eq:gztwobranes}). After normalization, the action of the bulk scalar KK modes with $n \geq 1$ is: 
\begin{equation}
{\cal S} [ \phi^{(n)}] = \frac{1}{2} \, \sum_{n = 1}^\infty \, \int d^4 x \, \left ( \partial_\mu \phi^{(n)} \partial^\mu \phi^{(n)} - m_n^2 \phi^{(n)} \, \phi^{(n)} \right )  \, .
\end{equation}
In the absence of back-reaction, these fields do not interact with brane fields, as their wave-function vanishes at $y = 0, \pi r_c$ \cite{Csaki:2000zn}. The masses of the KK modes are: 
\begin{equation}
m_n = k \, (x_{\nu n} + \delta x_{\nu n} ) \, \bar \omega \, ,
\end{equation}
with $x_{\nu n}$ an ${\cal O}(\epsilon)$ perturbation of $x_{2n}$ and
$\delta x_{\nu n}$ a coefficient ${\cal O}(\bar \omega^{2 \nu})$. 

In order to derive the action for the zero-mode, we can integrate
over the extra-dimension to get: 
\begin{eqnarray}
{\cal S} [ \phi^{(0)} ] = 
\frac{1}{2} \, \int d^4 x \left \{ K (\omega) \partial_\mu \phi^{(0)}(x) \partial^\mu \phi^{(0)} (x)  - V (\omega) \, \left [ \frac{1}{\pi r_c} + \phi^{(0)}(x) \right ] 
\left [ \frac{1}{\pi r_c} + \phi^{(0)} (x) \right ] \right \} 
\nonumber \\
\end{eqnarray}
where
\begin{equation}
\left \{
\begin{array}{l}
K (\omega) = \int_0^{\pi r_c} dy e^{- 2 k y} \varphi^{(0)} (y) \varphi^{(0)} (y) \, , \\
\\
V (\omega) = \int_0^{\pi r_c} dy e^{- 4 k y} \left [m^2  \varphi^{(0)} (y) \varphi^{(0)} (y) + \partial_5  \varphi^{(0)} (y) \partial_5 \varphi^{(0)}(y) \right ] = 
(\pi k r_c) \, v_{\rm UV}^2  \, F_\nu (\omega) \, ,
\end{array}
\right .
\end{equation}
with
\begin{equation}
F_\nu (\omega) = \frac{1}{\left ( 1 - \omega^{2\nu} \right ) } \, 
 \left [ 
\left ( \nu + 2  \right ) \omega^{2\nu} \, \left ( 1 - \omega^{2 - \nu} \, R \right )^2 +
\left ( \nu - 2 \right )  \, \left ( 1 - \omega^{2 + \nu} \, R \right )^2
\right ]
\end{equation}
and $R$ is the ratio of the VEV's on the two branes, $R = v_{\rm IR}/v_{\rm UV}$. Minimizing\footnote{The whole point is that
the effective potential is a function of $\omega$, whereas $\bar \omega$ is the value for which $V(\omega)$ is minimized.}  $V(\omega)$  with respect to  $\omega$, we get at leading order in $\epsilon$ the following relation: 
\begin{equation}
\label{eq:minimumRS1twobranesTEXT}
\bar \omega^{- \epsilon} \, R = 1 \qquad \longrightarrow \qquad \pi k r_c = \frac{1}{\epsilon} \, \ln \left ( \frac{v_{\rm UV}}{v_{\rm IR}}\right ) + {\cal O} \left (\epsilon \right ) + \dots \, ,
\end{equation}
where $\dots$ stand for terms of higher order in $\bar \omega$.

How does the graviscalar get a mass within this approach? 
We can define the following metric: 
\begin{eqnarray}
ds^2 &=& \tilde G_{MN} \, dx^M \, dx^N =  e^{-2 \, k \,  \theta \, T(x)} \, g(x) \, dx^2 - T^2(x) \, d \theta^2  \nonumber \\
&=& e^{-2 k r_c \left [1 + \frac{\delta T(x)}{r_c} \right ] \theta} \, g(x) \, dx^2 
- r_c^2  \left [1 + \frac{\delta T(x)}{r_c} \right ]^2 d \theta^2 \, ,
\end{eqnarray}
where $T(x)$ has the dimension of a coordinate (and whose VEV is $ \langle T(x) \rangle = r_c$), whilst $\theta \in [0,\pi]$. The metric $g_{\mu \nu} (x) = \eta_{\mu\nu} + \dots$ includes the standard 4D graviton. Computing the gravitational action in 5D for this metric, 
after integration over the extra-dimension we get: 
\begin{equation}
{\cal S}_{5D} = {\cal S}_{4D} 
+ \, 3 \, \frac{M_5^3}{k} \, \int d^4 x \, \sqrt{-g} \, \left [ \partial_\mu \left ( e^{-\pi k T} \right ) \right ]^2 + \dots \, ,
\end{equation}
where $\dots$ stand for other terms (full details are given in App.~\ref{sec:GWtwobranes}). The second term represents the kinetic term for a massless scalar field. If we normalize the field so as to get a canonical kinetic term: 
\begin{equation}
r(x) = \sqrt{\frac{6 M_5^3}{k} } \, e^{- \pi k T(x)}
= \bar r \, e^{\delta r (x)/ \bar r} \sim \bar r + \delta r (x) \, ,
\end{equation}
where, eventually, we get 
\begin{equation}
{\cal S}[r] = \frac{1}{2} \int d^4 x \sqrt{-g} \, 
\partial_\mu \delta r \, \partial^\mu \delta r \, .
\end{equation}
for the quantum fluctuation $\delta r (x)$ at the minimum of the 
effective potential, $\bar r$. This field is what is commonly known
as {\em radion}. The potential $V(\omega)$ is, indeed, an effective potential for $r(x)$ that, in turn, is related to the original scalar perturbation of the metric ($T$, under the metric ansatz used above). Minimizing $V(\omega)$ corresponds to giving a VEV to $T$, such that $\langle T \rangle = r_c$. The action for the two scalar would-be 
massless fields (the graviscalar and the bulk zero-mode) is, then: 
\begin{equation}
{\cal S} [r, \phi^{(0)}] = \frac{1}{2} \int d^4 x \sqrt{-g} \, \left \{ \partial_\mu \delta r \partial^\mu \delta r + K(r) \, \partial_\mu \phi^{(0)} \partial^\mu \phi^{(0)} - V(r) \left [\frac{1}{\pi r_c} + \phi^{(0)} \right ]^2 
\right \} \, .
\end{equation}
If we now expand the potential over the minimum, and we keep
terms of the lagrangian up to second order in the quantum fluctuations, we may recognize a mass term for the radion field: 
\begin{equation}
{\cal S} [r, \phi^{(0)}] = \frac{1}{2} \int d^4 x \sqrt{-g} \, \left \{ \partial_\mu \delta r \partial^\mu \delta r + K(\bar r) \, \partial_\mu \phi^{(0)} \partial^\mu \phi^{(0)} 
- m_r^2 \, \delta r^2 - V(\bar r) \, \phi^{(0)} \phi^{(0)}
\right \} \, ,
\end{equation}
where the radion mass is: 
\begin{equation}
\label{eq:radionmass2branes}
m_r^2 \sim 
\epsilon^2 \, 
\left ( \frac{k^2}{M_{\rm P}^2 }\right ) \, 
\left ( \frac{v_{\rm IR}^2}{M^2_{\rm P} }\right ) \, 
\left [ \frac{\Lambda^2}{\ln \left ( v_{\rm UV} / v_{\rm IR}\right ) }\right ] + \dots \, ,
\end{equation}
where we have used eq.~(\ref{eq:minimumRS1twobranesTEXT}).
Under the standard assumption that scales in the RS1 setup are ${\cal O} (M_{\rm P})$, we see that the radion mass is warped down by the exponential factor and, therefore, 
it should be ${\cal O} (\Lambda)$. However, due to the coefficient $\epsilon^2$, it can actually be much lighter than 
$\Lambda$ and, therefore, give a phenomenology different
from that of the KK gravitons~\cite{GW2}.

After rescaling the field $\phi^{(0)}$ in order to have a canonically normalized kinetic term, we have for its mass term:
\begin{equation}
m_0^2 =\frac{V(\bar r)}{K(\bar r)} = 2 k^2 \, \left ( \frac{m^2}{k^2} \right ) + \dots \, ,
\end{equation}
where $\dots$ stand for terms of higher-order in $\bar \omega$ and $\epsilon$. We can see that, contrary to the case of the radion mass, the leading-order mass of the bulk field zero mode $\phi^{(0)}$ is  ${\cal O}(m)$. Since in the RS paradigm all scales are assumed to be ${\cal O} (M_{\rm P})$, the bulk field zero mode is much heavier than the KK modes and than the radion, and it decouples from the low-energy spectrum of the model. 

In Ref.~\cite{Csaki:2000zn} a different method was used to derive the radion mass. In that paper, the (classical) coupled equations of motion of a scalar bulk field and the perturbation of the metric in
a RS background are studied. When the back-reaction of the bulk 
field over the metric (parametrized by a back-reaction parameter $\ell^2$) is taken into account, it was found that the mass of the graviscalar zero-mode is non-vanishing and proportional to $\ell^2$. Since the back-reaction should be small in order not to modify significantly the background metric, the radion mass could be lighter than the typical mass of the KK excitations of the graviton. 
We should clarify that the two approaches are intrinsecally different from a physical point of view: in the Goldberger-Wise paper \cite{GW2} the effective potential to be minimised represents the effect of the quantum fluctuations of the bulk field $\varphi$ when integrated in an unperturbed anti-de Sitter 5D metric $\bar G_{MN}$. On the other hand, in Ref.~\cite{Csaki:2000zn}, the classical equations of motion of the metric and a scalar field are integrated, to derive a perturbed metric $G_{MN}$ whose scalar degree of freedom $F$ is massive due to the classical interaction with the bulk scalar field (that, in general relativity, is represented by the curvature induced by the field on the space-time). It is not surprising, then, that the dependence of the radion mass on the parameter of the model differs between the two approaches: in Ref.~\cite{GW2} we find $m_r^2 \propto \epsilon^2 \, \Lambda^2$, and in Ref.~\cite{Csaki:2000zn} we have $m_r^2 \propto \ell^2$. In principle, both terms should be included, and the
effect of the quantum fluctuations of $\varphi$ should be computed
over the perturbed metric $G_{MN}$, so that the radion mass should
depend on both the bulk field mass $m$ and the back-reaction $\ell$. This, however, is something that is never done in the literature. 
The formal rationale is that the Goldberger-Wise mechanism is 
said to be studied within the so-called {\em probe approximation}: 
as long as the VEV of the bulk field at the location of the two 
branes is $v_{\rm UV, IR} \ll M_5$ and $\epsilon \ll 1$ (something that we have already used), then we can safely neglect the back-reaction
of the bulk field over the metric. Working in any one of the two 
approaches is not a problem, though: the significant message to be taken home is that it exists in the spectrum of the theory a light scalar mode whose mass $m_r$ can be significantly smaller than 
the KK gravitons mass, that are ${\cal O}(\Lambda)$. This mass
should be treated as an additional free parameter of the model, $m_r$, 
and its relation with other parameters of the model is indeed irrelevant.

\section{A ``$+ + -$" three-brane Randall-Sundrum scenario}
\label{sect:threebranes}

In order to lessen the impact of LHC data on the allowed parameter space of a model with DM in the extra-dimension,  we consider the possibility to split the brane that may solve the hierarchy problem (the IR brane), from a second one where DM particles live (the Deep Infra-Red, or DIR brane). In this way, we gain substantial freedom in order to address the two problems separately. A three branes model was first presented long ago, in Ref.~\cite{Lykken:1999nb}, with the goal of merging the virtues of the RS1 model discussed above
\cite{RandallLisa1999LMHf} with the so-called RS2 model \cite{RandallLisa1999AAtC} (that differs from the first in that the second brane is moved to infinity). The phenomenology
of the two two-brane models is quite different: whereas the first model is designed to address the SM hierarchy problem by locating SM fields at the $y = \pi r_c$ fixed point
of the orbifold (thus achieving that energy scales as seen as from the IR-brane point of view are warped down to the electro-weak scale, even though they are ${\cal O}(M_{\rm P})$
at the fundamental level), the second model aims at showing that 4D gravity can be recovered in a 5D space-time if the curvature in the extra dimension is ``large enough" to prevent
low-energy excitations of the graviton to enter it. Once three branes are considered several options arise, though. In order to reproduce a background metric valid in all the space-time, different configurations are possible depending on the sign of the brane tension terms $\sigma_i$. Two options are typically considered: the ``$+ - +$" option~\cite{Kogan:1999wc} 
(in which first and third branes have a positive tension, whereas the tension of the middle one is negative) and the ``$+ + -$" option~\cite{Lykken:1999nb}. 
In the former case,  it was shown that the first KK mode of the graviton is extremely light. This mode, together with the graviton zero mode, gives rise to an effective 4D bi-gravity theory \cite{Kogan:2000cv}. At the same time, it accomplishes the goal of extending the RS2 model so as to address the hierarchy problem (taking advantage of the intermediate brane).
We are instead interested here in the latter three-brane scenario,  the ``$+ + -$" configuration. This model is better suited to phenomenology, as it allows us to play with the location of different branes and, thus, achieving different warpings of the energy scales we are interested in.  A $Z_2$ orbifold symmetry with compactification radius $r_c$ is also considered, as in the RS1 setup. Two branes are still located at the orbifold fixed points, $y = 0$ and $y =  L_2 = \pi r_c $, 
whereas the third brane is located at an arbitrary point in between, $y = L_1$. In the two bulk subregions,  $y \in ]0, L_1[$ and $y \in ]L_1, L_2[$, two different\footnote{There is no physical motivation for the two cosmological constants to be equal.} 5D cosmological 
constants are considered, $\Lambda_1$ and $\Lambda_2$. In order to get a stable background metric, the three-brane tensions
must be related to the two cosmological constants and to their difference.

The action of the model is given by:
\begin{eqnarray}
\label{eq:threebranesaction}
{\cal S} &=&  {\cal S}_{\rm grav} + {\cal S}_{\rm branes} = - M_5^3 \, \int d^4 x \int_0^{L_1} dy \sqrt{\bar G^{(5)}} \, 
\left \{ R^{(5)} + 2 \Lambda_1 \right \} 
\nonumber \\
&-& M_5^3 \, \int d^4 x \int_{L_1}^{L_2} dy \sqrt{\bar G^{(5)}} \, 
\left \{ R^{(5)} + 2 \Lambda_2 \right \} 
\nonumber \\
&+& \sum_{i = {\rm UV, IR, DIR}} 
\int d^4 x \int_0^{L_2} \sqrt{-g_i} \, \delta (y - y_i) \left [ -\sigma_i + \dots \right ]
\end{eqnarray}
where $y_{\rm UV} = 0, y_{\rm IR} = L_1$ and $y_{\rm DIR} = L_2$ are the branes location, 
$\Lambda_1$ and $\Lambda_2$ are the cosmological constants in the two bulk subregions; 
$\bar G_{MN}^{(5)}$ is the 5D background metric, with determinant $\bar G^{(5)}$, whereas $\bar g_i$ are the determinant of the induced metric on the three branes, $\bar g_i (x) = \bar G^{(5)}(x,y_i)/ \bar G_{55}^{(5)}$; eventually, $M_5$ is the fundamental 5D gravitational scale. The two curvatures $k_1, k_2$ relate to the two cosmological constants as follows: 
\begin{equation}
k_1 = \sqrt{\frac{-\Lambda_1}{6 M_5^3}} \, ; \qquad k_2 = \sqrt{\frac{- \Lambda_2}{6 M_5^3}} \, .
\end{equation}
The brane tensions $\sigma_i$ must be chosen appropriately in order to glue the background metric piecewise:
\begin{equation}
    \sigma_{\rm UV} = 6 M_5^3 k_1 \, ; \qquad 
    \sigma_{\rm IR} = 3 M_5^3 (k_2 - k_1)\, ; 
    \qquad \sigma_{\rm DIR} = - 6 M_5^3 k_2 \, ,
\end{equation}
where $k_2 > k_1$ in order to enforce the `+ + -" brane configuration.  
The dots in the brane actions stand for the corresponding 4D fields content of the model. In our case, the DM Lagrangian will 
be located at the DIR brane, whereas the SM is on the IR brane.
Notice that the two-brane RS1 setup is recovered in three
limits: when the intermediate brane goes to its left, $L_1 \to 0 \;   (\Delta L \to L_2)$; when it goes to its right, $L_1 \to L_2 \; (\Delta L \to 0)$; and, when $k_2 \to k_1$. In this latter case, the brane tension of the intermediate
brane vanishes and the brane becomes evanescent.

\subsection{``Nested" two-brane setup}

In Ref.~\cite{Kogan:2000xc} it was shown that the reduced Planck mass of the ``$+ + -$" model is related to the curvatures in the bulk subregions, $k_1$ and $k_2$, and to the lengths of the two segments, $L_1$ and $L_2 = L_1 + \Delta L$, as follows:
\begin{equation}
\label{eq:Mplanck3branes}
    \bar M_{\rm P}^2 = M_5^3 \left [ 
    \frac{1}{k_1} \left ( 1 - e^{-2 k_1 L_1}\right )  
    + \frac{1}{k_2} e^{2 (k_2 - k_1) L_1} 
    \left ( e^{-2 k_2 L_1} - e^{-2 k_2 (L_1 + \Delta L)}\right ) 
    \right ] \, .
\end{equation}
This relation reduces to the standard two-brane RS1 relation both for $\Delta L \to 0$ and $L_1 \to 0$. For large enough $k_1 L_1$ and $k_2 L_2$, the four dimensionful quantities $M_5, k_1, k_2$ and $M_{\rm P}$ can be taken of the same order. Consider $k_1 \sim k_2$ (in order not to introduce a new hierarchy) and $M_5 > k_2 > k_1$ 
(so that the fundamental Planck scale $M_5$ is the larger scale in the game).
It can be shown that any matter field located at the intermediate (IR) brane has a mass that is warped down as follows:
\begin{equation}
m_{\rm IR} = m_0 \, e^{- k_1 L_1} \, ,
\end{equation}
whereas fields located at the rightmost (DIR) brane have masses warped down as: 
\begin{equation}
m_{\rm DIR}  = m_0 \, e^{- k_1 L_1} \, e^{- k_2 \Delta L} .
\end{equation}
where $m_0$ is the mass at UV scale. This means that, if we place SM fields and DM fields on different branes, we have some flexibility to address at the same time 
the SM hierarchy problem and the DM hierarchy problem (namely, explaining why the experimentally observed Higgs mass is ${\cal O}(\Lambda_{\rm EW})$, 
and not as large as the would-be SM cut-off, and why the DM mass could be even lighter than that). 

It is useful to rephrase eq.~(\ref{eq:Mplanck3branes}) in a different way in order to better understand the relation between the Planck mass and the fundamental scale of gravity, $M_5$. This can be done under 
the assumption that the three-brane
setup can be seen in terms of a {\em nested two-brane set-up}. We first introduce the following warping factors: 
\begin{equation}
    \bar \xi = \exp(- k_2 \Delta L) \, ; \qquad \bar \omega = \exp (- k_1 L_1) \, ,
\end{equation}
where the latter is the same warping factor that we introduced in the two-branes setup. In terms of $\bar \xi$ and $\bar \omega$, Eq.~\eqref{eq:Mplanck3branes} becomes:
\begin{equation}
\label{eq:Mplanck3branes2}
    \bar M_{\rm P}^2 = \frac{M_5^3}{k_1} \, (1 - \bar \omega^2)
    + \frac{M_5^3}{k_2} \, \bar \omega^2 \, (1 - \bar \xi^2) \, .
\end{equation}
Let's now define an ``{\em effective reduced Planck mass}" $\bar M_{\rm eff}$ as the ``{\em Planck mass
as seen from the perspective of the rightmost brane}": 
\begin{equation}
\label{eq:meff}
   \bar M_{\rm eff}^2 =  \bar M_{\rm P}^2 - \frac{M_5^3}{k_1} \, \left (1 - \bar \omega^2 \right ) \sim {\cal O} \left ( \bar \omega^2 M_{\rm P}^2 \right ) \, ,
\end{equation}
where the last relation holds under the standard assumption $\bar M_{\rm P} \sim M_5 \sim k_1$. 
In terms of $\bar M_{\rm eff}$, we immediately have:
\begin{equation}
   \bar M_{\rm eff}^2 = \frac{M_5^3}{k_2} \, \bar \omega^2 \, \left ( 1 - \xi^2 \right ) \, .
\end{equation}
In this relation, we can redefine the dimensionful quantities as follows:
\begin{equation}
    \left \{
    \begin{array}{l}
    \tilde M_5 = \bar \omega \, M_5 \, ,\\
    \\
    \tilde k_2 = \bar \omega \, k_2 \,  , 
    \end{array}
    \right .
\end{equation}
such that:
\begin{equation}
\label{eq:nestedtwobranePlanckmass}
    \bar M_{\rm eff}^2 = \frac{\tilde M_5^3}{\tilde k_2}  \, \left (1 - \bar \xi^2 \right ) \, .
\end{equation}
We see that we have recovered the standard two-brane relation between the effective reduced Planck mass, $\bar M_{\rm eff}$, the ``{\em 
fundamental scale of gravity as seen from the rightmost brane}", $\tilde M_5$, and the warped curvature in the subregion $y \in [L_1,L_2]$, $\tilde k_2$.
In this relation, only the second warping factor, $\bar \xi$, is present, as the rightmost brane has no direct insight on the physics to the left of $L_1$.
It is also clear that, as in the original two-brane RS model, all dimensionful quantities in this relation are of the same order, $\bar M_{\rm eff} \sim \tilde M_5 \sim \tilde k_2$. 
This also proves the statement given above that, if  $\bar M_{\rm P} \sim M_5 \sim k_1$, then $\bar M_{\rm eff} \sim \bar\omega \bar M_{\rm P}$.

Eventually, it is immediate to define the universal inverse coupling between graviton KK modes and fields living on the DIR-brane, 
\begin{equation}\label{eq:DIRscale}
    \Lambda_{\rm DIR} = \bar \xi \, \bar M_{\rm eff} = 
    \bar \xi \, \bar \omega \, \bar M_{\rm P}\, ,
\end{equation}
that plays the same role of $\Lambda_{\rm IR, 2b}$ of the two-brane RS1 model. By replacing $\bar M_{\rm P}$ with $\bar M_{\rm eff}$ in Eq.~(\ref{eq:twobranesKKcouplingUV}), 
we can derive the functional dependence on the warping factors $\bar \xi$ and $\bar \omega$ of the inverse coupling of KK gravitons with fields located on the IR-brane. As it was the
case for the two-brane setup, we also get a non-universal coupling that depends on the KK number $n$. For the $n$-th KK graviton,
we get: 
\begin{equation}\label{eq:IRnscaleapprox}
    \Lambda^n_{\rm IR} \sim 8 \, \frac{J_2(x_n)}{x_n^2} \, \frac{\bar M_{\rm eff}}{\bar \xi^3} \sim 8 \, \frac{J_2(x_n)}{x_n^2} \, \frac{\bar \omega \bar M_{\rm P}}{\bar \xi^3} \, ,
\end{equation}
where in the last step we have defined
$\Lambda_{\rm IR} = \bar M_{\rm eff} = {\cal O}( \bar \omega \bar M_{\rm P})$. This is indeed what we will get computing directly the coupling between KK gravitons
and IR-fields (see below), with the only difference being the functional dependence on the ratio of the curvatures, $k_2/k_1$, something that cannot be derived from the point of view of
the rightmost brane, only. 

One last comment is in order, though: 
even if in the {\em nested two-brane setup} the relation 
in eq.~(\ref{eq:nestedtwobranePlanckmass}) holds, such that all 
KK gravitons interact with fields on the DIR or the IR branes
with couplings formally identical to those of a standard two-brane setup, but for the replacement of scales ${\cal O} (\bar M_{\rm P})$
with scales ${\cal O}(\bar M_{\rm eff})$, the KK zero mode still couples
with gravity with the standard $1/\bar M_{\rm P}$ coupling. The difference between the zero mode and the excited KK graviton couplings
with matter on the IR brane hints to the existence of a brane to the
left of what is seen as ``UV" as seen from the rightmost brane.

\subsection{KK graviton wave-functions in the three-brane setup}

The KK gravitons eigenfunctions in the two bulk subregions, $y \in [0, L_1]$ and $y \in [L_1, L_2]$, have the same functional form as in 
Eq.~\eqref{eq:eigenfunctions2branes}. However, in this case,  the coefficients of the two solutions and the mass spectrum $m_n$ of the KK modes 
must satisfy a different set of boundary conditions. Introducing the conformal coordinate $z$ such as:
\begin{equation}
    z = \left \{
    \begin{array}{lll}
    \frac{1}{k_1} \left ( e^{k_1 y} - 1 \right ) &\qquad {\rm for} & y \in [0, L_1] \, , \\
    && \\
    \frac{1}{k_2} e^{k_2 (y - L_1) + k_1 L_1} 
    + \frac{1}{k_1} \left ( e^{k_1 L_1} - 1 \right ) 
    - \frac{1}{k_2} e^{k_1 L_1}
    &\qquad {\rm for} & y \in [L_1, L_2] \, ,
    \end{array}
    \right .
\end{equation}
with the conformal weight:
\begin{equation}
    g(z) = \left \{
    \begin{array}{lll}
    k_1 z + 1 &\qquad {\rm for} & y \in [0, L_1] \, ,\\
    \\
    k_2 (z - z_1) + k_1 z_1 + 1 
    & \qquad {\rm for} & y \in [L_1, L_2] \, ,
    \end{array}
    \right .
\end{equation}
the $n$-th eigenfunction in the two bulk subregions can be written as \cite{Kogan:2000xc}: 
\begin{equation}
   \hat  \chi^{(n)} (z)
    =
    \left \{
    \begin{array}{c}
    \sqrt{\frac{g(z)}{k_1}} 
    \left [
    A_{1n} \, Y_2 \left ( \frac{m_n}{k_1} g(z) \right ) 
    + A_{2n} \, J_2 \left ( \frac{m_n}{k_1} g(z) \right ) 
    \right ]  \qquad {\rm for } \, y \in [0, L_1] \, ,\\
    \\
      \sqrt{\frac{g(z)}{k_2}} 
    \left [
    B_{1n} \, Y_2 \left ( \frac{m_n}{k_2} g(z) \right ) 
    + B_{2n} \, J_2
    \left ( \frac{m_n}{k_2} g(z) \right )
    \right ] \qquad {\rm for } \, y \in [L_1, L_2] \, ,
    \end{array}
    \right .
\end{equation}
whereas the zero mode eigenfunction is: 
\begin{equation}
    \hat \chi^{(0)}(z) = \frac{A_0}{\left [ g(z) \right ]^{3/2}}
\end{equation}
and, as for the two-brane case, $\chi^{(n)} (y)  = \hat \chi^{(n)}(z) / \sqrt{g(z)} $. 
The masses and the coefficients of the eigenfunctions can be obtained
asking for continuity of $\hat \chi^{(n)}(z(y))$ at $y = L_1$ and for the appropriate discontinuity of $d\hat \chi^{(n)}(z(y))/dz$ at $y = 0, L_1, L_2$. 
The boundary conditions, therefore, give a system of four equations that, in order to have a non-trivial solution, must satisfy the following condition: 
\begin{equation}
    \det \left (
    \begin{array}{cccc}
    Y_1 \left ( \frac{m_n}{k_1}\right ) &  
    J_1 \left ( \frac{m_n}{k_1}\right ) & 0 & 0  \\
    && \\
    Y_1 \left ( \frac{m_n}{k_1} g(L_1)\right ) &  
    J_1 \left ( \frac{m_n}{k_1} g(L_1)\right ) &
    - \sqrt{\frac{k_1}{k_2}} \, Y_1 \left ( \frac{m_n}{k_2} g(L_1)\right ) & - \sqrt{\frac{k_1}{k_2}} \, J_1 \left ( \frac{m_n}{k_2} g(L_1)\right ) \\
    && \\
     0 & 0 & Y_1 \left ( \frac{m_n}{k_2} g(L_2)\right ) &  
    J_1 \left ( \frac{m_n}{k_2} g(L_2)\right ) \\
    && \\
    Y_2 \left ( \frac{m_n}{k_1} g(L_1)\right ) &  
    J_2 \left ( \frac{m_n}{k_1} g(L_1)\right ) &
    - \sqrt{\frac{k_1}{k_2}} \, Y_2 \left ( \frac{m_n}{k_2} g(L_1)\right ) & - \sqrt{\frac{k_1}{k_2}} \, J_2 \left ( \frac{m_n}{k_2} g(L_1)\right )
    \end{array}
    \right ) = 0 \, , 
\end{equation}
where the first three rows arise from the boundary conditions on the derivative of the eigenfunctions at $0, L_1$ and $L_2$ and the fourth row from 
the continuity of the eigenfunctions at $L_1$, respectively.

Using the first three BC's we have: 
\begin{equation}
\label{eq:BCcoefficients1}
    \left \{
    \begin{array}{lll}
    A_{2n} & = & 
    - \frac{Y_1 \left ( \frac{m_n}{k_1} \right )}{
    J_1 \left ( \frac{m_n}{k_1} \right )} \, A_{1n} \, , \\
    &&\\
    B_{1n} & = &  - \sqrt{\frac{k_2}{k_1}} \, \frac{
    J_1 \left ( \frac{m_n}{k_2} g(L_2) \right )}{
    J_1 \left ( \frac{m_n}{k_1} \right )
    }
    \frac{
    \left [
   J_1 \left ( \frac{m_n}{k_1} \right )
   Y_1 \left ( \frac{m_n}{k_1} g(L_1) \right ) - 
   J_1 \left ( \frac{m_n}{k_1} g(L_1) \right )
   Y_1 \left ( \frac{m_n}{k_1} \right )
   \right ]
    }{
    \left [
   J_1 \left ( \frac{m_n}{k_2} g(L_1) \right )
   Y_1 \left ( \frac{m_n}{k_2} g(L_2) \right ) - 
   J_1 \left ( \frac{m_n}{k_2} g(L_2) \right )
   Y_1 \left ( \frac{m_n}{k_2} g(L_1) \right )
   \right ]
    } \, A_{1n} \, , \\
    &&\\
    B_{2n} & = & \sqrt{\frac{k_2}{k_1}} \, \frac{
    Y_1 \left ( \frac{m_n}{k_2} g(L_2) \right )}{
    J_1 \left ( \frac{m_n}{k_1} \right )
    }
    \frac{
    \left [
   J_1 \left ( \frac{m_n}{k_1} \right )
   Y_1 \left ( \frac{m_n}{k_1} g(L_1) \right ) - 
   J_1 \left ( \frac{m_n}{k_1} g(L_1) \right )
   Y_1 \left ( \frac{m_n}{k_1} \right )
   \right ]
    }{
    \left [
   J_1 \left ( \frac{m_n}{k_2} g(L_1) \right )
   Y_1 \left ( \frac{m_n}{k_2} g(L_2) \right ) - 
   J_1 \left ( \frac{m_n}{k_2} g(L_2) \right )
   Y_1 \left ( \frac{m_n}{k_2} g(L_1) \right )
   \right ]
    } \, A_{1n} \, ,
    \end{array}
    \right .
\end{equation}
from which we see that all coefficients are proportional to $A_{1n}$, 
that can be fixed imposing normalisation of the eigenfunctions,  eq.~(\ref{eq:twobranesnormalizationcond}). 
Being all coefficients proportional to $A_{1n}$, we can use the fourth row to derive the mass spectrum: 
\begin{equation}
\label{eq:BCmassspectrum}
    J_1 \left ( \frac{m_n}{k_2} g(L_2) \right ) = 
    Y_1 \left ( \frac{m_n}{k_2} g(L_2) \right ) \, 
    \frac{
    \left [
    J_2 \left ( \frac{m_n}{k_2} g(L_1) \right )
    + J_1 \left ( \frac{m_n}{k_2} g(L_1) \right ) \, 
     \frac{f_2 (m_n, k_1, L_1)}{f_1 (m_n, k_1, L_1)}
    \right ]
    }{
    \left [
    Y_2 \left ( \frac{m_n}{k_2} g(L_1) \right )
    - Y_1 \left ( \frac{m_n}{k_2} g(L_1) \right ) \,
    \frac{f_2 (m_n, k_1, L_1)}{f_1 (m_n, k_1, L_1)}
    \right ]
    } \, ,
\end{equation}
where: 
\begin{equation}
\left \{
\begin{array}{l}
f_2 (m_n, k_1, L_1) = 
Y_1 \left ( \frac{m_n}{k_1}\right ) \, 
J_2 \left ( \frac{m_n}{k_1} g(L_1)\right ) -
J_1 \left ( \frac{m_n}{k_1}\right ) \, 
Y_2 \left ( \frac{m_n}{k_1} g(L_1)\right ) \, , \\
\\
f_1 (m_n, k_1, L_1)= 
Y_1 \left ( \frac{m_n}{k_1}\right ) \, 
J_1 \left ( \frac{m_n}{k_1} g(L_1)\right ) -
J_1 \left ( \frac{m_n}{k_1}\right ) \, 
Y_1 \left ( \frac{m_n}{k_1} g(L_1)\right ) \, .
\end{array}
\right . 
\end{equation}

For $k_2 \Delta L > 1$, we can parametrise the mass of the $n$-th mode as in the two-brane case: 
\begin{equation}
    m_n = k_2 \, (x_n + \delta x_n)  
    \, e^{ - k_1 L_1} \, e^{-k_2 \Delta L}  = k_2 \, (x_n + \delta x_n) \,
    \bar \omega \, \bar \xi \, ,
\end{equation}
where $x_n$ is the $n$-th zero of the Bessel function $J_1(x)$ and $\delta x_n$ is a small shift. 

It can be shown that $\delta x_n$ is suppressed by powers of $\bar \xi$, $\bar \omega$ or both, depending on the position of the intermediate brane.  Therefore, 
the mass spectrum is given by: 
\begin{equation}
\label{eq:massspectrum3branes}
    m_n = k_2 x_n  \, \bar \omega \, \bar \xi  + \dots \, = k_2 \, x_n \, \left ( \frac{m_\text{DIR}}{m_0} \right ) \, ,
\end{equation}
as it was found in Ref.~\cite{Kogan:2000xc}. Notice that, in the limit $L_1 \to 0$ ($L_1 \to L_2$), the spectrum becomes that of a two-brane model on a segment with length $L_2$ 
($L_1$) and curvature $k_2$ ($k_1$), respectively. 

In order to derive explicit approximate formul\ae~for the mass spectrum $m_n$ and for the coefficients $A_{in}, B_{in}$ it is useful to expand in the two 
warping factors $\bar \xi$ and $\bar \omega$, if small. Two cases must be considered, though. If $L_1$ approaches the leftmost brane, $L_1 \to 0$, we have $\bar \xi \to \exp (- k_2 L_2)$ and $\bar \omega \to 1$. In this case, we should recover the two-brane model in the subregion $y \in [L_1,L_2]$ with curvature $k_2$. However, this scenario is not interesting
from a phenomenological point of view, since matter localized
on the intermediate brane will interact with an effective scale
that exponentially approaches $M_{\rm P}$. 
 On the other hand, for $L_1 \to L_2$ we have $\bar \xi \to 1$ and $\bar \omega \to \exp (- k_1 L_2)$. Again we shall recover a two-brane model, but in the subregion $y \in [0, L_1]$ with curvature $k_1$.  In order to derive an approximate expression for this case, we must expand first in $\bar \omega$ and then in $\bar \xi$, assuming that $\bar \xi \ll 1$. Formul\ae~for 
 the case $\bar \xi \to 1$ are cumbersome and not inspiring. 
 We will only consider the case $\bar \omega \ll \bar \xi \ll 1$, therefore. This is indeed the case considered in Ref.~\cite{Donini:2025cpl} (see, also, Refs.~\cite{Seung2,Lee}).
 
We obtain for the $n$-th shift in the mass spectrum: 
\begin{equation}
\label{eq:nthshiftcase2}
    \delta x_n = \frac{\pi}{32}  \, 
    \left ( 1 + \frac{k_2}{k_1} \right )  \, 
   \left [ x_n^4 \, \frac{Y_1 (x_n)}{J_2 (x_n)} \right ] \, \bar \xi^4 +
   {\cal O}(\bar \xi^5, \bar \omega^2) \, .
\end{equation}
At leading order in $\bar \omega \ll 1$, the coefficients of the wave-functions are: 
\begin{equation}
\label{eq:coeffwave-function}
    \left \{
    \begin{array}{l}
    A_{2n} = a_{2n} \, A_{1n} =
     - \frac{4}{\pi} \, \left ( \frac{k_1}{k_2}\right )^2 \, 
    \frac{1}{x_n^2} \, 
    \frac{1}{\bar \xi^2} \, \frac{1}{\bar \omega^2} 
    \, A_{1n} + \dots \\
    \\
    B_{1n} = b_{1n} \, A_{1n} = 
    \frac{1}{8} \, \left ( \frac{k_1}{k_2} \right )^{1/2} \, 
    \left ( 1 + \frac{k_2}{k_1} \right ) \, 
     \frac{x_n^2 \, J_1 \left (\frac{k_2}{k_1} \, x_n \, \bar \xi \right )}{J_1 \left (x_n \, \bar \xi \right )} \, 
    \frac{\bar \xi^2}{\bar \omega^2} \, A_{1n}  + \dots \\
    \\
    B_{2n} = b_{2n} \, A_{1n} =
    \frac{4}{\pi} \, \left ( \frac{k_1}{k_2} \right )^{1/2} 
    \,
    \frac{J_1 \left (\frac{k_2}{k_1} \, x_n \, \bar \xi \right )}{x_n^2 \, J_1 \left (x_n \, \bar \xi \right )} \,
    \, \frac{1}{\bar \xi^2 \, \bar \omega^2} \, A_{1n} + \dots
    \end{array}
    \right .
\end{equation}
It can be seen that all coefficients are proportional to $1/\bar \omega^2$.  On the other hand, $A_{2n}$ and $B_{2n}$ grow for small $\xi$,  whereas for $B_{1n}$ it is the opposite. 

We can now compute the normalisation factor $A_{1n}$ as in eq.~(\ref{eq:twobranenormalization}). For the $n$-th wave-function, 
we get:
\begin{equation}
    A_{1n} = \frac{\pi}{4} \, 
    k_2 \, \left ( \frac{k_2}{k_1} \right )^{1/2} \, \frac{x_n^2}{
    h_n (r_k, \bar \xi) \, J_2 (x_n)} \, \bar \xi^3 \, \bar \omega^3 \, ,
\end{equation}
where
\begin{eqnarray}
h^2_n (r_k, \bar \xi) &=& \frac{x_n^2 \bar \xi^2}{r_k} \, \left [ 
J_0^2(r_k \, x_n \, \bar \xi) + J_1^2(r_k \, x_n \, \bar \xi)\right ] 
- \frac{4}{r_k^3} \, J_1^2(r_k \, x_n \, \bar \xi) \\
&+& 2 \, \frac{J_1^2(r_k \, x_n \, \bar \xi)}{J_1^2(x_n \, \bar \xi)}
\, \left \{ \frac{x_n^2}{2} \, \left [ J_2^2(x_n) - \bar \xi^2 J_0^2(x_n \, \bar \xi) - \bar \xi^2 J_1^2(x_n \, \bar \xi)
\right ] + 2 J_1^2(x_n \, \bar \xi) \right \} \nonumber
\end{eqnarray}
and $r_k = k_2/k_1$ is the ratio between the curvature of the bulk to the right and to the left of the IR brane. In the limit $r_k \, x_n \, \bar \xi \ll 1$, the function $h_n (r_k, \bar \xi)$ reduces to: 
 \begin{equation}
  h_n (r_k, \bar \xi) \underset{r_k \, x_n \, \bar \xi \ll 1}{\simeq}
     \left ( \frac{k_2}{k_1}\right ) \, x_n  \, J_2 (x_n) + {\cal O}(\bar \xi^2)
\end{equation}
and the normalization coefficient becomes: 
\begin{equation}
A_{1n} \underset{r_k \, x_n \, \bar \xi \ll 1}{\simeq} 
\frac{\pi}{4} k_2 \, \left ( \frac{k_2}{k_1} \right )^{1/2} \, 
\frac{x_n^2}{J_2 (x_n)} \, \bar \xi^3 \, \bar \omega^3 + \dots \, .
\end{equation}
In this limit, the other three coefficients that define the wave-function on the whole  conformal segment $z \in [0,z_2]$ are:
\begin{equation}
\label{eq:finalcoefficientsomegallxil1}
    \left \{
    \begin{array}{l}
    A_{2n} \underset{r_k \, x_n \, \bar \xi \ll 1}{\simeq} - 
    k_2 \, \left (\frac{k_1}{k_2} \right )^{5/2} \, \frac{1}{J_2 (x_n)} \, \bar \xi \, \bar \omega + \dots \, , \\
    \\
    B_{1n} \underset{r_k \, x_n \, \bar \xi \ll 1}{\simeq} 
    \frac{\pi}{32} \, k_2 \, \left ( 1 + \frac{k_2}{k_1} \right ) \, \frac{x_n^4}{J_2(x_n)} \, \bar \xi^5 \,  
    \bar \omega + \dots \, , \\
    \\
    B_{2n} \underset{r_k \, x_n \, \bar \xi \ll 1}{\simeq}
    k_2 \frac{1}{J_2(x_n)} \, \bar \xi \, \bar \omega + \dots \, .
    \end{array}
    \right .
\end{equation}

The normalisation of the zero mode eigenfunction can be found
in the same way: 
\begin{equation}
    A_0 = \sqrt{k_1} \, \frac{1}{\sqrt{1 - \bar \omega^2}} \, 
    \left [1 + \frac{k_1}{k_2} \, \frac{\bar \omega^2 \, (1 -\bar \xi^2 )}{1 - \bar \omega^2 } \right ]^{-1/2}  = \frac{M_5^{3/2}}{\bar M_{\rm P}} \, .
\end{equation}

It is immediate to show that for $\bar \xi \to 1$ ({\em i.e.} $L_1 \to L_2$) the coefficients $A_{1n}$ and $A_{2n}$
(as long as $\bar \omega \ll \bar \xi \to 1$) give the two-brane wave-function obtained in Sect.~\ref{sec:twobranes} in the subregion 1, taking at the same time the limit $k_2 \to k_1 = k$. In this limit, we get the following mass spectrum: 
\begin{equation}
\lim_{\Delta L \to 0} m_n = k_1 \, x_n \, \exp{(- k_1 L_1)} \, ,
\end{equation}
{\em i.e.} the masses of a RS1 model with curvature $k_1$ and size $L_1$. 

\begin{figure}
	\centering
    \begin{tabular}{cc}
	\includegraphics[scale=0.55]{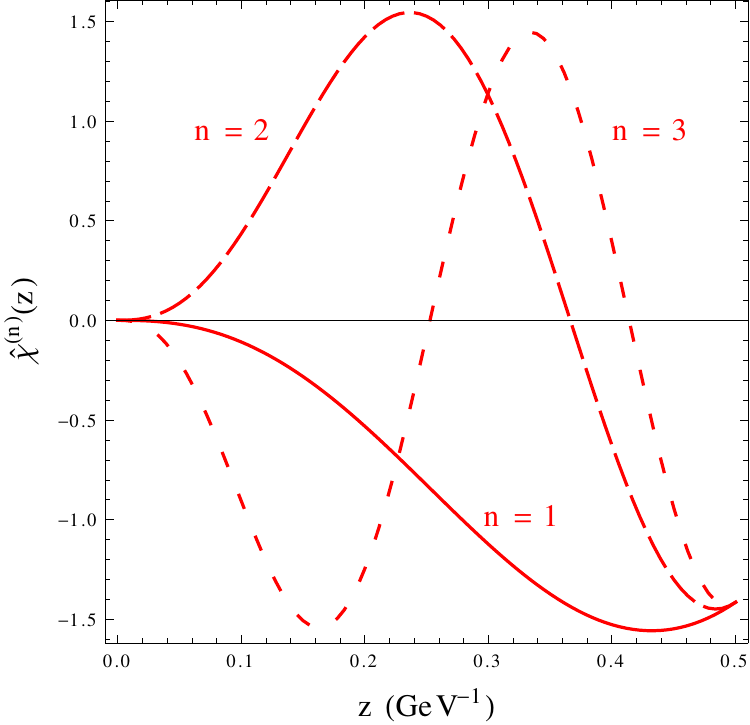} &
	\includegraphics[scale=0.58]{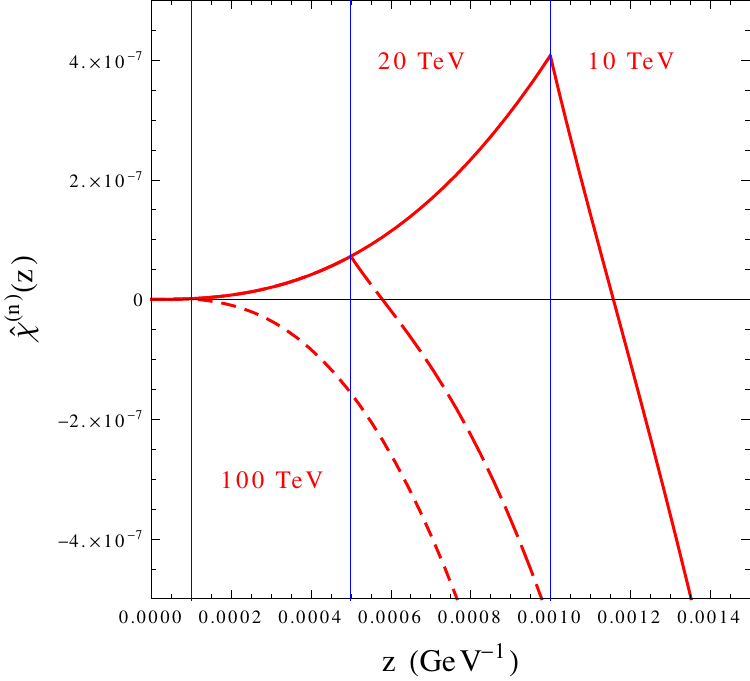}
	\end{tabular}
    \caption{Three-brane model KK graviton wave-functions as a function of the conformal coordinate $z$.
    Left panel: $\hat \chi^{(n)}(z)$, for KK number $n=1$ (solid), $n=2$ (dashed) and $n=3$ (dot-dashed) for $\bar M_{\rm eff}$ = 10 TeV.
    Right panel: 
    $\hat \chi^{(1)}(z)$ for three different values of the intermediate
    brane position corresponding to $\bar M_{\rm eff} = 10$ TeV (solid), $20$ TeV (dashed) and $100$ TeV (dot-dashed). The blue vertical lines represent $z(L_1)$ in the three cases. 
    }
    \label{fig:threebranesWF}
\end{figure}

In Fig.~\ref{fig:threebranesWF} we show the KK graviton wave-functions $\hat \chi^{(n)}(z)$ as a function of the conformal coordinate $z$ for $n=1, 2, 3$ and
$\bar M_{\rm eff} = 10$ TeV (left panel, solid, dashed and dot-dashed lines, respectively) and the first KK graviton wave-function $\hat \chi^{(1)}(z)$ as a function of the conformal coordinate $z$ (in a zoomed-frame) for three different values of $\bar M_{\rm eff} = 10, 20$ and 100 TeV (right panel, solid, dashed and dot-dashed lines, respectively).

\subsection{KK graviton couplings with matter on the IR and DIR branes}

Once we have the KK graviton wave-functions computed in the $\bar \omega \ll \bar \xi \ll 1$ scenario, we can derive the couplings of the graviton with fields located either on 
the DIR- or the IR-brane. In the former case, we have universal couplings: 
\begin{equation}
 \label{eq:Lambda_DIR}
a_n^{\rm DIR} = 
     \left ( \frac{g(z_2)}{M_5} \right )^{3/2} \, |\hat \chi^{(n)}(z_2)| \simeq 
     \left ( \frac{k_2}{k_1}\right )^{1/2} \, \frac{1}{\Lambda_{\rm DIR}} \qquad \forall \,  n \, .
\end{equation}
On the other hand, the couplings with fields located on the IR-brane (and on the UV-brane) are $n$-dependent, as it was the case in the two-brane setup, 
eq.~(\ref{eq:twobranesKKcouplingUV}). In this case, we have: 
 \begin{eqnarray}
 \label{eq:Lambda_IR}
     a_n^{\rm IR} &=& \frac{1}{\Lambda^n_{\rm IR}} = 
     \left ( \frac{g(z_1)}{M_5} \right )^{3/2} \, |\hat \chi^{(n)}(z_1)| \underset{r_k \, x_n \, \bar \xi \ll 1}{\simeq}  \left ( \frac{k_2}{k_1} \right )^{1/2} \, \frac{x_n^2}{8 \, J_2 (x_n)} \, 
     \frac{\bar \xi^3}{\bar \omega \bar M_{\rm P}} \nonumber \\
     &=& 
     \left ( \frac{k_2}{k_1} \right )^{1/2} \, \frac{x_n^2}{8 \, J_2 (x_n)} \, 
     \left ( \frac{\Lambda_{\rm DIR}}{\Lambda_{\rm IR}}\right )^3 \, \frac{1}{\Lambda_{\rm IR}}
 \end{eqnarray}
whereas the zero mode couples as usual, $a_0 = 1/ \bar M_{\rm P}$ (see Ref.~\cite{Kogan:2000cv}). Notice that we recover the result of eq.~(\ref{eq:IRnscaleapprox}), 
 albeit including the explicit dependence on the ratio $k_2/k_1$, that we could not obtain in the nested two-branes approach.
 In the last line we have introduced our IR scale definition: 
 \begin{equation}
 \label{eq:LambdaIRdef}
 \Lambda_{\rm IR} = \bar \omega \, \bar M_{\rm P} \, .
 \end{equation}
Notice that $\Lambda_{\rm IR}$ is, indeed, what we called 
$\bar M_{\rm eff}$ in eq.~(\ref{eq:meff}) and it has to be
understood as the scale you reach starting from $\bar M_{\rm P}$ 
after the first warping. In this sense, it differs from the
definition given in eq.~(2.40) of Ref.~\cite{Donini:2025cpl}, 
where it was defined as $\Lambda_{\rm IR}=(\bar \omega/ \bar \xi) \, \bar M_{\rm P}$, representing the coupling 
of the radion with matter on the IR brane. Although the latter 
definition is useful from a phenomenological point of view, the
definition given in this paper has a straightforward geometrical 
interpretation. 

  The corresponding Lagrangian is: 
 \begin{equation}
     {\cal L}_{\rm IR} = \frac{1}{\bar M_{\rm P}} \, h_{\mu\nu}^{(0)} (x) \, T^{\mu\nu} (x) + 
     \sum_{n \ge 1}^\infty \frac{1}{\Lambda^n_{\rm IR}} \, h_{\mu\nu}^{(n)} (x) \, T^{\mu\nu} (x) \, .
 \end{equation}

Eventually, we give for completeness the couplings of DIR and IR located particles with two KK gravitons (that arise expanding up to second order the metric around Minkowski): 
\begin{equation}
 \label{eq:Lambda_DIR_seagull}
a_{mn}^{\rm DIR} = 
     \left ( \frac{g(z_2)}{M_5} \right )^3 
     \, \hat \chi^{(m)}(z_2) 
     \, \hat \chi^{(n)}(z_2) \underset{B_{1n} \ll B_{2n}}{\simeq} 
     \left ( \frac{k_2}{k_1}\right ) \, \frac{1}{\Lambda^2_{\rm DIR}} \qquad \forall \, (m, n) \, ; 
\end{equation}
and
\begin{eqnarray}
\label{eq:Lambda_IR_seagull}
  a_{mn}^{\rm IR} &=& 
     \left ( \frac{g(z_1)}{M_5} \right )^3 
     \, \hat \chi^{(m)}(z_1) 
     \, \hat \chi^{(n)}(z_1) \nonumber \\
     &\underset{r_k \, x_n \, \bar \xi \ll 1}{\simeq} &
   \frac{1}{64}
    \left ( \frac{k_2}{k_1} \right ) \, \frac{x_m^2 \, x_n^2}{J_2(x_m) \, J_2 (x_n)} \, 
     \left ( \frac{\Lambda_{\rm DIR}}{\Lambda_{\rm IR}}\right )^6 \, \frac{1}{\Lambda^2_{\rm IR}}
     \qquad \forall \,  (m,n) \, . 
\end{eqnarray}

\section{Stabilization of the three-branes setup: excerpts}
\label{sect:threebranesradionsShort}

We present here a short reminder of how to stabilize the length
$L_2$ of the extra-dimension and the position of the intermediate
brane $L_1$ adapting the {\em Goldberger-Wise mechanism}
to a setup with three branes in an anti-de Sitter 5D metric. The interested reader is referred to the related App.~\ref{sect:threebranesradions}, where full details are given.

In an extra-dimensional setup with three branes (the UV , the IR and the DIR branes), the total length $L_2$ is obtained by summing two separate distances, $L_1$ and $\Delta L = L_2 - L_1$. Each of these two distances must be stabilized, in order to get a static background metric. It was shown in Ref.~\cite{Seung1} that, in a $N$-branes setup, a Goldberger-Wise stabilization mechanism requires $N-1$ radion fields, 
{\em i.e.} TWO radions are expected for a three-brane setup. A different conclusion was obtained in Ref.~\cite{Cai:2021mrw}, though (see, also, Ref.~\cite{Olechowski:2024wcf}). In that paper, it was claimed that only ONE radion field exists, independently from the number of branes present in the setup. We will come back to these two seemingly contradictory statements at the end of this Section and show how they indeed reconcile. 

We will now sketch how the two radions wave-functions are obtained and which are their masses and couplings with fields located on the DIR or the IR branes, following Ref.~\cite{Seung1}.

\subsection{Goldberger-Wise potentials and boundary conditions}

As we did in the two-brane system, a bulk scalar field $\varphi$ is added to the gravitational action. The bulk field has a simple bulk potential, as in the two-brane case: 
\begin{equation}
U(\varphi) =
\begin{cases}
\dfrac{m_1^2}{2}\,\varphi^2 
& \text{for } y \in (0, L_1), \\[6pt]
\dfrac{m_2^2}{2}\,\varphi^2 
& \text{for } y \in (L_1, L_2).
\end{cases}
\end{equation}
where $m_1,\,m_2$  are parameters with the dimension of a mass. 
Notice that, as was the case for the cosmological constant, there is no physical motivation for $m_1$ to be equal to $m_2$.
Three localized potentials $V_j(\varphi)$, with $j$ labeling the three branes, are added to the lagrangian. The complete action is, then:
\begin{eqnarray}
{\cal S} = {\cal S_{\rm grav}} &+& \int d^4 x \int_0^{L_2} dy \, \sqrt{\bar G^{(5)}} \, \left ( \frac{1}{2} \, \partial_M \varphi \partial^M \varphi - U(\varphi) \right ) \nonumber \\
&-& \sum_{j={\rm UV, IR, DIR}} 
\int d^4x \int_0^{L_2} dy \, \sqrt{- \bar g_j} \, \delta (y - y_j) \, V_j (\varphi) \, ,
\end{eqnarray}
where $y_0 = 0,\, y_1 = L_1$ and $y_2 = L_2$ and the brane terms $V_j$ are:
\begin{equation}
\left \{
\begin{array}{l}
V_{\rm UV} (\varphi) = \mu_{\rm UV} \left ( \varphi^2 - \frac{v_{\rm UV}^2}{r_{\rm UV}}\right )^2 \, , \\
\\
V_{\rm IR} (\varphi) = \mu_{\rm IR} \left ( \varphi^2 - \frac{v_{\rm IR}^2}{r_{\rm IR}}\right )^2 \, , \\
\\
V_{\rm DIR} (\varphi) = \mu_{\rm DIR} \left ( \varphi^2 - \frac{v_{\rm DIR}^2}{r_{\rm DIR}}\right )^2 \, .
\end{array}
\right .
\end{equation}
The parameters $\mu_j$ have the dimension of an inverse mass squared, whereas the field values $v_j$ have the dimension of a mass. The length scales $r_j $ have
been introduced so as to normalize properly the VEVs $v_i$, and they can be chosen convenientently. We will see that a convenient choice is: 
$r_{\rm UV} = r_{\rm IR} = L_1;\  r_{\rm DIR}= \Delta L$. 
Remind that  brane tension terms  $\sigma_j$ in eq.~(\ref{eq:threebranesaction}), needed to glue together the background metric in the two subregions and fulfill the orbifold symmetries, 
must be redefined accordingly, $\sigma_j = \mu_j v_j^4/r_j^2$. 

As it was understood in the case of the two-brane setup, also in 
this case we work in the {\em probe approximation}, {\em i.e.}
for $\epsilon_i \ll 1$ and $v_{\rm UV, IR, DIR} \ll M_5$. Under these
assumptions, we can safely neglect the back-reaction of the
bulk field over the metric. The ranges of parameters employed in our numerical scan in Sect.~\ref{sec:pheno} satisfy the assumptions
of the {\em probe approximation} and, therefore, neglecting the
back-reaction is fully justified for the purposes of the present work.

We decompose the bulk field as follows: 
\begin{equation}
    \varphi (x,y) = \varphi^{(0)} (x, y) + \delta \varphi  (x,y)\, ,
\end{equation}
where  $\varphi^{(0)} (x,y)$ is defined as:
\begin{equation}
    \varphi^{(0)} (x,y) = \left \{
    \begin{array}{l}
    \left [ \frac{1}{L_1} + \phi^{(0)}_1 (x) \right ] \, \varphi^{(0)}_1 (y) \, , \qquad y \in [0,L_1]  \\
    \\
        \left [ \frac{1}{\Delta L} + \phi^{(0)}_2 (x) \right ] \, \varphi^{(0)}_2 (y)  \, , \qquad y \in [L_1,L_2]
        \end{array}
        \right .
\end{equation}
being $\varphi^{(0)}_{1,2}(y)$ the zero mode wave-functions in the subregions 1 and 2, respectively. The 5D-field $\varphi^{(0)}(x,y)$ 
may develop  a $y$-dependent VEV: 
\begin{equation}
\left \{
\begin{array}{l}
\langle \varphi^{(0)}(x,0) \rangle =  \frac{1}{L_1} \, \langle \varphi^{(0)}_1(0) \rangle \, , \\
\\
\langle \varphi^{(0)}(x,L_1) \rangle =  \frac{1}{L_1} \, \langle \varphi^{(0)}_1(L_1) \rangle = \frac{1}{\Delta L} \, \langle \varphi^{(0)}_2(L_1) \rangle \, , \\
\\
\langle \varphi^{(0)}(x,L_2) \rangle =  \frac{1}{\Delta L} \, \langle \varphi^{(0)}_2(L_2) \rangle \, .
\end{array}
\right .
\end{equation}
The fluctuation $\delta \varphi (x,y)$, on the other hand, can be
expanded on 4D fields as usual: 
\begin{equation}
\delta \varphi (x,y) = \sum_{n= 1}^\infty \phi^{(n)} (x) \, \varphi^{(n)} (y) \, .
\end{equation}
Within the decompositions above, all KK modes wave-function have the same mass dimension, $[\varphi^{(0)}_i(y)] = [\varphi^{(n)}(y)] = 1/2$.
After solving the equations of motion for the bulk field KK modes, we
get two orthogonal KK towers with eigenfunctions in the extra-dimension given by: 
\begin{equation}
\label{eq:radion1nSHORT}
\hat \varphi^{(n)}_1(z) \simeq 
\left \{
\begin{array}{l}
\sqrt{6} \, k_1 \, \bar \omega^2 \, \frac{x^n_{\nu_1} }{f^n_{\nu_1} \,  J_{\nu_1+1} \left ( x^n_{\nu_1} \right )} \, \sqrt{\frac{g^3(z)}{k_1}} \, \left \{  J_{\nu_1} \left ( \frac{m_{n_1}}{k_1} \, g(z) \right ) + {\cal O} (\bar \omega^4) \right \} \, , \qquad( {\rm for} \; y \in [0,L_1]) \\
\\
0 \, , \qquad ({\rm for} \; y \in [L_1,L_2])
\end{array}
\right .
\end{equation}
and 
\begin{equation}
\label{eq:radion2nSHORT}
\hat \varphi^{(n)}_2(z) \simeq 
\left \{
\begin{array}{l}
0  \qquad ({\rm for} \; y \in [0,L_1]) \, ,  \\
\\
\sqrt{6} \, k_2 \, \left ( \bar \xi \, \bar \omega \right )^2  \frac{x^n_{\nu_2} }{f^n_{\nu_2}  \, J_{\nu_2+1} \left ( x^n_{\nu_2} \right )}\, \sqrt{\frac{g^3(z)}{k_2}} \, \left \{  J_{\nu_2} \left ( \frac{m_{n_2}}{k_2} \, g(z) \right ) + {\cal O} (\bar \xi^4) \right \}   \qquad ({\rm for} \; y \in [L_1,L_2])\, , \\
\end{array}
\right .
\end{equation}
where $\nu_i = 2 \sqrt{1 + m_i^2/4 k_i^2} = 2 \sqrt{1 + \epsilon_i}$ and
\begin{equation}
f^n_{\nu_i} = \left [ (x^n_{\nu_i})^2 + 2 \nu_i^2 - 2  \right ]^{1/2} \, .
\end{equation}
The masses of the KK modes of the bulk field are: 
\begin{equation}
\left \{
\begin{array}{ccc}
  m_{n_1} &=& k_1 \left ( x_{\nu_1}^n + \delta x^n_{\nu_1} \right ) \, ,  \\
  \\
  m_{n_2} &=& k_2 \left ( x_{\nu_2}^n + \delta x^n_{\nu_2} \right ) \, ,
\end{array}
\right .
\end{equation}
where $x_{\nu_i}^n = x_{2n} + {\cal O}(\epsilon_i)$ (being $x_{2n}$
the $n$-th zero of the Bessel function $J_2(x)$, as it was the case of
the two-brane setup) and $\delta x^n_{\nu_1}, \delta x^n_{\nu_2}$ are
${\cal O} (\bar \omega^{2 \nu_1})$ and ${\cal O}(\bar \xi^{2 \nu_2})$
corrections, respectively (see App.~\ref{sect:threebranesradions} for details).

Once we have derived the wave-functions of the $n$-th  $\varphi$ 
KK modes, we can compute their couplings with fields localized at the UV, IR or DIR branes. The couplings 
of two scalar KK modes $m$ and $n$ with a 4D field at $y = 0, L_1$ or $L_2$ are given by: 
\begin{eqnarray}
\lambda^j_{mn} &=& \int_0^{L_2} dy \, \sqrt{\bar G^{(5)} (y)} \, \delta (y - y_j) \, \varphi^{(m)}(y) \, \varphi^{(n)}(y) \nonumber \\
&=& 
\int_{z(0)}^{z(L_2)} dz \, \sqrt{\bar G^{(5)} (z)} \, \delta (z - z_j) \, \hat \varphi^{(m)}(z) \, \hat \varphi^{(n)}(z) \, .
\end{eqnarray}
Within the {\em probe approximation}, {\em i.e.} neglecting the
back-reaction of the brane energy density over the metric, it is immediate to see that $\varphi$ KK modes do not couple with field on any of the branes in the setup, as their coupling vanish at $y = 0, L_1$ and $L_2$ due to their boundary conditions (see, again, App.~\ref{sect:threebranesradions}). Only once
back-reaction effects are included does a non-vanishing
coupling of these modes to brane fields arise, suppressed by the back-reaction parameter $\ell^2$ (see, {\em e.g.}, Ref.~\cite{chivukula2024limitskaluzakleinportaldark}). 

When studying the zero mode wave-functions a different procedure must be considered. In the limit $\mu_i v_i^2 \gg 1$, the boundary conditions force the zero-mode wave-function to assume the values that we have fixed at the brane locations, eq.~(\ref{eq:zero modeVEV}). 
The wave-functions are, thus:
\begin{equation}
\label{eq:radion0solutionSHORT}
\left \{
\begin{array}{cc}
\hat \varphi^{(0)}_1 (z) =& \sqrt{\frac{g^3(z)}{k_1}} \, 
\left ( \frac{\tilde v_{\rm UV}}{1 - \omega^{2 \nu_1}} \right ) \, \left [
\omega^{2 \nu_1} \, \left (  \omega^{2-\nu_1} R_1 - 1 \right ) \,  \, g^{\nu_1}(z) + 
\left ( 1 - \omega^{2+\nu_1} R_1 \right ) \,  g^{- \nu_1}(z) \right ]  \, , \\
\\
 \hat \varphi^{(0)}_2 (z) =&  \sqrt{\frac{g^3(z)}{k_1}} \, 
 \left ( \frac{\Delta L}{L_1 }\right ) \, 
 \left ( \frac{\omega^2 \, \tilde v_{\rm IR}}{1 - \xi^{2 \nu_2}}\right ) \, 
 \left \{ \xi^{2 \nu_2} 
 \, \left (\xi^{2 - \nu_2} R_2 - 1 \right ) \,  \, 
 \left [\omega \, g(z) \right ]^{\nu_2}  \right .
  \\
  \\
&\qquad \qquad \qquad \qquad \qquad+ \left . 
 \, \left (1 - \xi^{2 + \nu_2} R_2 \right ) \, 
 \, \left [\omega \, g(z) \right ]^{- \nu_2} \right \}  \, ,
\end{array}
\right .
\end{equation}
where we have rescaled the brane VEV's of the bulk field $\varphi$ as follows: 
\begin{equation}
\left \{
\begin{array}{l}
\tilde v_{\rm UV} = \sqrt{k_1 L_1} \, v_{\rm UV} \, , \\
\\
\tilde v_{\rm IR} = \sqrt{k_1 L_1} \, v_{\rm IR} \, , \\
\\
\tilde v_{\rm DIR} = \sqrt{\frac{k_2}{k_1}} \, \sqrt{\frac{k_1 L_1}{k_2 \Delta L}} \, \sqrt{k_1 L_1} \, v_{\rm DIR} \, . \\
\end{array}
\right . 
\end{equation}
and $R_1, R_2$ are the ratios of the VEV's of the bulk field at the location of the three branes, $R_1 = \tilde v_{\rm IR}/\tilde v_{\rm UV}$, $R_2 = \tilde v_{\rm DIR}/\tilde v_{\rm IR}$. Notice that, as it was the case for the two-brane setup in Sect.~\ref{sec:GWtwobranes}, we are now considering $\omega$ and $\xi$ as free parameters to be fixed by a minimization procedure. 

After integration over the extra-dimension, we get: 
\begin{eqnarray}
{\cal S}_\varphi &=& \int d^4 x \int_0^{L_2} dy \, \sqrt{\bar G^{(5)}} \, \left \{ 
\frac{1}{2}  \bar G^{(5) \, MN} \partial_M \varphi \, \partial_N \varphi - \frac{m_i^2}{2} \varphi^2
\right \} \nonumber \\
&=& 
\frac{1}{2} \, \int d^4 x \left \{ K_1 (\omega) \partial_\mu \phi^{(0)}_1(x) \partial^\mu \phi^{(0)}_1 (x)  - V_1 (\omega) \, \left [ \frac{1}{L_1} + \phi^{(0)}_1(x) \right ] 
\left [ \frac{1}{L_1} + \phi^{(0)}_1 (x) \right ] \right \} \nonumber \\
&+&
\frac{1}{2} \, \int d^4 x \left \{ K_2 (\omega, \xi) \partial_\mu \phi^{(0)}_2(x) \partial^\mu \phi^{(0)}_2 (x)  - V_2 (\omega, \xi) \, 
\left [ \frac{1}{\Delta L} + \phi^{(0)}_2(x) \right ] \left [ \frac{1}{\Delta L} + \phi^{(0)}_2 (x) \right ] \right \} \, , \nonumber \\
\end{eqnarray}
where
\begin{equation}
\left \{
\begin{array}{l}
K_1 (\omega) = \int_0^{L_1} dy \sqrt{\bar G^{(5)}} \, \left ( \frac{1}{4} \, \eta_{\mu\nu} \, \bar G^{(5)\mu\nu} \right ) \varphi^{(0)}_1 (y) \varphi^{(0)}_1 (y) \, , \\
\\
V_1 (\omega) = \int_0^{L_1} dy \sqrt{\bar G^{(5)}} \, \left [ m_1^2  \varphi^{(0)}_1 (y) \varphi^{(0)}_1 (y) - \bar G^{(5)55} \, \partial_5  \varphi^{(0)}_1 (y)\, \partial_5 \varphi^{(0)}_1(y) \right ] \, ,
\end{array}
\right .
\end{equation}
and 
\begin{equation}
\left \{
\begin{array}{l}
K_2 (\omega,\xi) =  \int_{L_1}^{L_2} dy \sqrt{\bar G^{(5)}} \,  \left ( \frac{1}{4} \, \eta_{\mu\nu} \, \bar G^{(5)\mu\nu} \right )\varphi^{(0)}_2 (y) \varphi^{(0)}_2(y) \, , \\
\\
V_2(\omega,\xi) =  \int_{L_1}^{L_2} dy \sqrt{\bar G^{(5)}} \left [ m_2^2  \varphi^{(0)}_2 (y) \varphi^{(0)}_2 (y) - \bar G^{(5)55} \, \partial_5  \varphi^{(0)}_2 (y) \, \partial_5 \varphi^{(0)}_2 (y) \right ] \, .
\end{array}
\right .
\end{equation}
 
We must now identify minimize the effective potentials $V_1(\omega)$ and $V_2(\omega, \xi)$ with respect to the dynamical fields.
We introduce a ``{\em na\"ive}" metric ansatz for each
of the two regions of the orbifold:
\begin{equation}
\label{eq:GWmetric2branes}
G_{MN}  \, dx^M  \, dx^N = 
 \left \{
\begin{array}{l} 
e^{-2 k_1 \theta \, T_1(x) } \, g_{\mu\nu}(x) \, dx^\mu dx^\nu 
- T_1(x)^2 d \theta^2 \, , \qquad \qquad \qquad \qquad \; \theta \in [0, \theta_1]   \\
\\ 
e^{-2 \left [  k_2 (\theta - \theta_1) \, T_2(x) + k_1 \theta_1 \, T_1(x) \right ]  } \, g_{\mu\nu}(x) \, dx^\mu dx^\nu 
- T_2(x)^2 d \theta^2 \, , \qquad \theta \in [\theta_1,1] 
\end{array}
\right .
\end{equation}
where both $T_1(x)$ and $T_2(x)$ have the dimension of a coordinate and whose VEV's are 
$ \theta_1 \, \langle T_1(x) \rangle = L_1$ and 
$ (1-\theta_1) \, \langle T_2(x) \rangle = \Delta L$.
Remind that the two fields are not independent degrees of freedom: we have one massless graviscalar per segment of the orbifold, $\Sigma_1$ and $\Sigma_2$, intertwined by
the boundary conditions at $y = L_1$. 

After integration over the extra-dimension of the gravitational 5D action, we get (see App.~\ref{sect:threebranesradions} for details): 
\begin{eqnarray}
{\cal S}_T &=& {\cal S}_{4D}
+ \, 3 \, M_5^3 \, \left ( \frac{1}{k_1} - \frac{1}{k_2} \right ) \, \int d^4 x \, \sqrt{-g} \, \left [ \partial_\mu \left ( e^{-k_1 \, \theta_1 \, T_1} \right ) \right ]^2 \\
&& + \, 3 \, \left ( \frac{M_5^3}{ k_2} \right ) \, \int d^4 x \, \sqrt{-g} \, \left [ \partial_\mu \left ( e^{- k_2 \, (1-\theta_1) \, T_2 + k_1 \, \theta_1 \, T_1} \right ) \right ]^2
 + \dots \nonumber \\
 &=& {\cal S}_{4D} + \frac{1}{2} \int d^4 x \sqrt{-g} \, 
 \left [ 
 \left ( \partial_\mu \delta r_1 \right )^2 + 
 \left ( \partial_\mu \delta r_2 \right )^2  \right ] + \dots \, , 
\end{eqnarray}
where the two scalar degrees of freedom are defined as:
\begin{equation}
\label{eq:radionwavefunctions}
\left \{
\begin{array}{l}
r_1(x) = 
\sqrt{\frac{6 M_5^3}{k_2}} \,  
\sqrt{ \delta k } \, 
e^{- k_1 \, \theta_1 \,  T_1(x)} =
\bar r_1 \, e^{\delta r_1 (x)/\bar r_1}
= \bar r_1 + \delta r_1 + \dots  \, , \\
\\
r_2(x) = \sqrt{\frac{6 M_5^3 }{k_2} } \, e^{- \left [ k_2 \, (1-\theta_1) \,  T_2(x) + k_1 \, \theta_1 \, T_1(x) \right ] } 
=
\bar r_2 \, e^{\delta r_2 (x)/\bar r_2}
= \bar r_2 + \delta r_2 + \dots 
\, ,
\end{array}
\right .
\end{equation}
with $\delta k = (k_2-k_1)/k_1$ an adimensional parameter that
quantifies the departure from the evanescent limit.
The VEV's of the radion fields are: 
\begin{equation}
\left \{
\begin{array}{l}
\bar r_1 = \sqrt{\frac{6 M_5^3}{k_1}} \, \sqrt{\delta k} \, \bar \omega \, , \\
\\
\bar r_2 = \sqrt{\frac{6 M_5^3}{k_1}} \, \bar \omega \, \bar \xi \, . 
\end{array}
\right .
\end{equation}
with the following quantum fluctuations:
\begin{equation}
\label{eq:deltaradionwavefunctions}
\left \{
\begin{array}{l}
\delta r_1 = - \sqrt{\frac{6 M_5^3}{k_2}} \, \sqrt{\frac{k_2 - k_1}{k_1}} \, \bar \omega \, 
\left [ k_1 \, \theta_1 \, \delta T_1 (x) \right ]\rm\, , \\
\\
\delta r_2 = - \sqrt{\frac{6 M_5^3}{k_2}} \, \bar \omega 
\, \bar \xi \, 
\left [ k_2 \, (1 - \theta_1) \, \delta T_2 (x) 
        + k_1 \, \theta_1 \, \delta T_1 (x) \right ] \, .  
\end{array}
\right .
\end{equation}
In these expressions, it is understood that $k_2 \geq k_1$ in order not to have a tachyonic mode. Notice that, after integrating over the extra-dimension, the fields $\delta r_1$ and $\delta r_2$ appear formally in the lagrangian as independent fields, so that
two ``radions" are indeed present in the spectrum (as stressed in Ref.~\cite{Seung1}). 

Using the metric in eq.~(\ref{eq:GWmetric2branes}), 
we get for the effective potential:
\begin{equation}
V(\omega,\xi) = 
\left ( \frac{\tilde v_{\rm UV}^2}{L_1^2} \right )
\left \{ F_1(\omega) + r_k \, R_1^2 \, \omega^4 F_2(\xi)
\right \} \, , 
\end{equation}
where
\begin{equation}
F_i (x) = \frac{1}{\left ( 1 - x^{2 \nu_i} \right )} \, 
\left [ (\nu_i + 2) \, x^{2 \nu_i} \, 
\left ( 1 - R_i \, x^{2-\nu_i}\right )^2 + 
(\nu_i - 2) \, 
\left ( 1 - R_i \, x^{2+\nu_i}\right )^2
\right ] \, .
\end{equation} 

In order to get simple expressions, it is standard 
to consider both $\epsilon_i = m_i^2/(4k_i^2)$ to be ``small". 
One comment is in order, though:
whereas for the standard two-brane setup, the ```small" parameters are $\epsilon$ and $\bar \omega$ and, thus, 
expanding in $\bar \omega$ (first) and in $\epsilon$ (second) gives a straightforward recipe to find the minimum of the 
potential and, at the same time, accurate approximate analytic expressions for the radion mass, in the case of the three-brane setup at hand here the number of ``small" parameters is
significantly larger. In particular, we have $\bar \omega$
(that is very small, $\bar \omega \sim 10^{-14}$ to get
$\Lambda_{\rm IR} \sim 100$ TeV) followed by $\bar \xi$, 
$\epsilon_1$, $\epsilon_2$ and $\delta k$. In particular, the
value of $\bar \xi$ and one or both of the $\epsilon_i$'s
could switch, depending on the particular region of the parameter space we want to study. In order to have $\Lambda_{\rm DIR} \sim 1$ TeV, we need $\bar \xi \sim 10^{-2}$. This
does not leave much space for ```smallish" $\epsilon_i$'s. 
We have found, though, that certain features of the radion spectrum can be understood expanding in $\epsilon_2$ and
$\epsilon_1$. In particular, the ratio of the radion masses $m_{r_1}$ and $m_{r_2}$ as a function 
of $\bar \omega, \bar \xi, \epsilon_1, \epsilon_2$ and $\delta k$
can be understood by using an expansion in small $\epsilon_i$'s. This allow us to have an anlaytical approximate understanding of the ``decoupling" of the radion $r_1$ (whose mass is significantly larger than $\Lambda_{\rm DIR}$)
with respect to $r_2$ (whose mass is significantly smaller
than $\Lambda_{\rm DIR}$). This justifies the approach we
adopted in Ref.~\cite{Donini:2025cpl}, where only one light
radion was considered. 

\begin{figure}
	\centering
    \begin{tabular}{ccc}
         \includegraphics[scale=0.45]{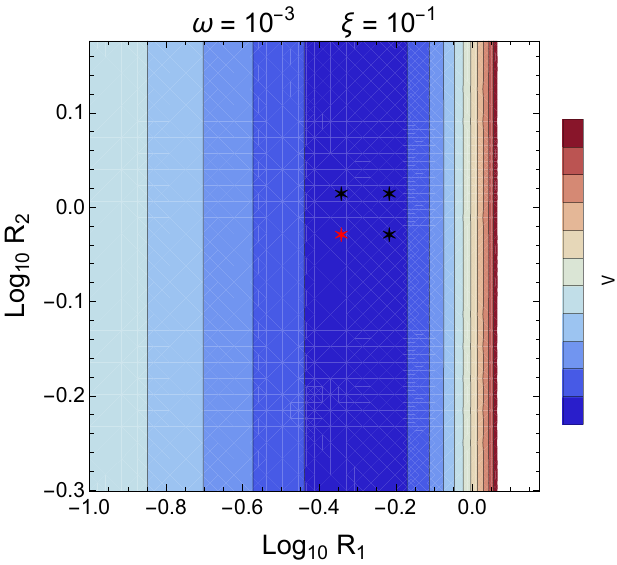} &
         \includegraphics[scale=0.45]{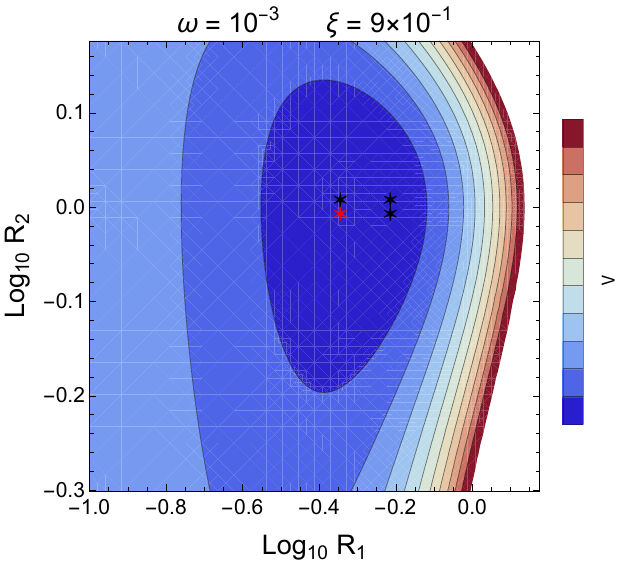} &
         \includegraphics[scale=0.45]{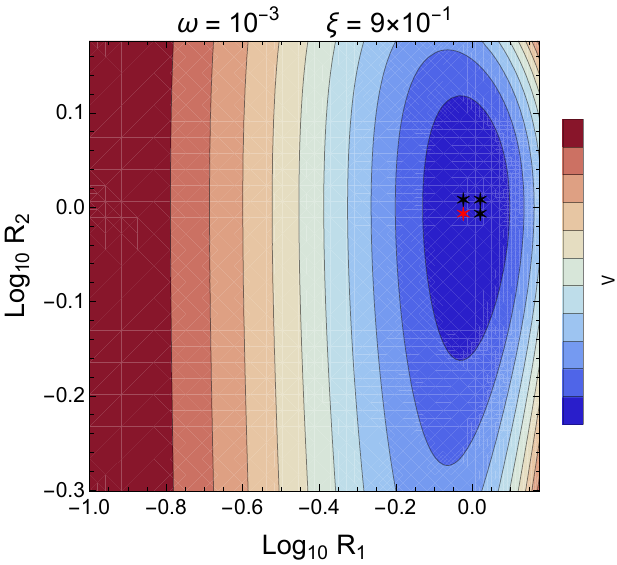}
    \end{tabular}
    \caption{ The potential $V(\omega, \xi)$ for several choices
    of ($\bar \omega, \bar \xi$) in the plane $(R_1, R_2)$. 
    Left panel: $\bar \omega = 1 \times 10^{-3}, \bar \xi = 1 \times 10^{-1}$; 
    Middle panel:  $\bar \omega = 1 \times 10^{-3}, \bar \xi =  9 \times 10^{-1}$; 
    Right panel:  $\bar \omega = 9 \times 10^{-1}, \bar \xi = 9 \times 10^{-1}$. 
    The red dot represents the numerical minimum of the potential, whereas the black dot is the analytical result. 
    In the white region, the potential is not real for the considered choice of $\bar \omega$ and $\bar \xi$.
    }
    \label{fig:potential}
\end{figure}

Derivatives of the full potential are, trivially, 
\begin{equation}
\left \{
\begin{array}{l}
\frac{\partial V}{\partial \omega} = 
\left ( \frac{\tilde v_{\rm UV}^2}{L_1^2} \right ) 
\left [\frac{d F_1(\omega)}{d \omega} + 4 r_k \, R_1^2 \, \omega^3 \, F_2(\xi) \right ] \, , \\
\\
\frac{d V}{d \xi} = \left ( \frac{\tilde v_{\rm UV}^2}{L_1^2} \right ) \, r_k \, R_1^2 \, \omega^4 \, \frac{d F_2(\xi)}{d \xi} \, .
\end{array}
\right .
\end{equation}
Clearly, the minimum of the potential in $\xi$ satisfies the
relation $d F_2 (\xi)/d \xi = 0$. Solving this equation
for fixed $\bar \xi$ in $R_2$ gives: 
\begin{equation}
\label{eq:R2barSHORT}
\left ( \bar R_2 \right )_\pm = 
\frac{\nu_2}{2} \left ( 1 \pm \Delta \right ) \, \bar \xi^{\nu_2 - 2} + {\cal O}(\bar \xi^5) \, ,
\end{equation}
where:
\begin{equation}
\Delta = \left ( 1 - \frac{4}{\nu_2 + 2} \right )^{1/2} \, .
\end{equation}
The two solutions in $\bar R_2$ (for any value of $\bar \xi$)
are always real, as $\nu_2 \ge 2$ for any value of $\epsilon_2$. 
Once we have found the minimum of the potential in $R_2$ for 
a given value of $\bar \xi$, we can minimize in $\omega$. 
The quadratic equation in $\omega$ now depends on the 
specific value of $\bar \xi$ through the second term, 
proportional to $F_2 (\bar \xi)$. Assuming that $\epsilon_2$ is
``small" (at least as small as $\bar \xi$) it can be shown
that $F_2 (\bar \xi) \sim \epsilon_2 + {\cal O}(\epsilon_2^2)$. 
By solving the first equation in $R_1$ for a fixed, ``small", 
$\bar \omega$, we find: 
\begin{equation}
\label{eq:R1barSHORT}
\left ( \bar R_1 \right )_\pm = \frac{\nu_1 (\nu_1 + 2)}{
2 \left [ \nu_1 + 2 + r_k F_2 (\bar \xi) \right ]} \, 
\left ( 1 \pm \Delta^\prime \right ) \, \bar \omega^{\nu_1 - 2} 
+ {\cal O} \left ( \bar \omega^5 \right ) \, ,
\end{equation}
where:
\begin{equation}
\Delta^\prime = \left ( 1 - \frac{4 
\left [ \nu_1 + 2 + r_k F_2 (\bar \xi) \right ]
}{(\nu_2 + 2)^2} \right )^{1/2} \, .
\end{equation}
In this case, depending on the value of $\bar \xi$, the solution
for $\bar R_1$ may be real or not. It is much better to 
draw a picture of the full potential in the plane $(R_1, R_2)$
than for $(\omega,\xi)$, due to the fact that the latter variables span several order of magnitudes and it is very difficult to give a pictorial understanding of the potential, 
whereas $R_1$ and $R_2$ are ${\cal O}(1)$ variables. 
In Fig.~\ref{fig:potential} we show $V(\bar \omega, \bar \xi)$
in the $(R_1,R_2)$ plane
for $\bar \omega = 1 \times 10^{-3}$ and $\bar \xi = 1 \times 10^{-1}$ (left panel),  $\bar \omega = 1 \times 10^{-3}$ and $\bar \xi = 9 \times 10^{-1}$ (middle panel) and $\bar \omega = 9 \times 10^{-1} = \bar \xi = 9 \times 10^{-1}$ (right pannel). 
These particular values of ($\bar \omega, \bar \xi$) have been
chosen in order to illustrate pictorially the behaviour of the 
potential. In each panel, the red star represents the numerical value of the minimum, and the black dots the four analytical results. It can be seen that only one of them corresponds
to the numerical result. In the other point, a computation of the Hessian shows that we are in presence of a saddle point. 
The solution in the ($R_1, R_2$) plane that corresponds to
the absolute minimum of the potential is 
$(\bar R_{1})_-, (\bar R_{2})_-$.
The color code implies a lower value
of the potential in the blue region, a large value in the red region. The white region depicts the ensemble of $(R_1,R_2)$
values for which the potential is complex. We can see that
the minimum of the potential, for a given pair ($\bar \omega, \bar \xi$), always corresponds to values of $R_1$ and $R_2$ that
are ${\cal O}(1)$.

At the minimum, the following relations can be derived:
\begin{equation}
\left \{
\begin{array}{l}
k_1 L_1 = - \frac{1}{\epsilon_1} \, \ln R_1 =
\frac{k_1^2}{m_1^2} \, \ln \left ( \frac{\tilde v_{\rm UV}}{\tilde v_{\rm IR}}\right ) + {\cal O} \left (\epsilon_1^2 \right ) + \dots \, , \\
\\
k_2   \Delta L = - \frac{1}{\epsilon_2} \, \ln R_2 =
\frac{k_2^2}{m_2^2} \, \ln \left ( \frac{\tilde v_{\rm IR}}{\tilde v_{\rm DIR}}\right ) + {\cal O} \left (\epsilon_2^2 \right ) + \dots \, , 
\end{array}
\right .
\end{equation}
where $\dots$ stand for terms of higher order in $\omega, \xi$ and the $\nu_i$'s have been expanded at second order
in the $\epsilon_i$'s. Note that these relations can be obtained
by simply inverting eqs.~(\ref{eq:R2barSHORT}) and (\ref{eq:R1barSHORT})
at the minimum: 
\begin{equation}
\left \{
\begin{array}{l}
\bar \omega = \left [ \frac{2 (\nu_1 + 2 + r_k \, F_2 (\bar \xi))}{\nu_1 (\nu_1 + 2)} \, \frac{1}{1 - \Delta^\prime} \, \bar R_{1,-} \right ]^{1/(\nu_1 - 2)} \, ,  \\
\\
\bar \xi = \left [ \frac{2}{\nu_2} \, \frac{1}{1 - \Delta} \, \bar R_{2,-} \right ]^{1/(\nu_2 - 2)} \, , 
\end{array}
\right .
\end{equation}
where these results are obtained up to ${\cal O}(\bar \omega^5)$
and ${\cal O}(\bar \xi^5)$, under the assumptions that both
$\bar \omega$ and $\bar \xi$ are ``small" (whereas no 
assumption on $\epsilon_1$ and $\epsilon_2$ is needed).

If we now expand the potential around the minimum 
at $(\bar r_1,\bar r_2)$ we have: 
\begin{eqnarray}
V(r_1,r_2) & = & V(\bar r_1, \bar r_2)  \nonumber \\
&+&\frac{1}{2} \, 
\left . \frac{\partial^2 V}{\partial r_1^2} \right |_{\bar r_1, \bar r_2} \, \delta r_1^2 +
\left . \frac{\partial^2 V}{\partial r_1 \, \partial r_2} \right |_{\bar r_1, \bar r_2} \, \delta r_1 \, \delta r_2 + 
\frac{1}{2} \left . \frac{\partial^2 V}{\partial r_2^2} \right |_{\bar r_1, \bar r_2} \, \delta r_2^2 + \dots \nonumber \\
& = & V(\bar r_1, \bar r_2) + \frac{1}{2} \, 
\left ( \delta r_1 \, \delta r_2 \right ) \, 
{\cal M}^2 (\bar r_1, \bar r_2) \, 
\left ( \begin{array}{c} \delta r_1 \\ \delta r_2 
\end{array} \right ) + \dots \, .
\end{eqnarray}
The potential computed at the minimum, $V(\bar r_1, \bar r_2)$, does not vanish, as it is the constant needed to cancel the corrections to the brane tension terms that are induced by the localized potential in the GW mechanism. This term is needed to have a stable background geometry once the bulk field $\Phi$
is added to the action.

The radion mass matrix ${\cal M}^2$ computed at the minimum
is, up to ${\cal O}(\bar \omega^5, \bar \xi^5)$: 
\begin{equation}
\label{eq:radionmassmatrix1}
{\cal M}^2 (\bar r_1, \bar r_2) =  
\frac{1}{\delta k} \, \left ( \frac{\tilde v_{\rm UV}}{L_1} \right )^2 \, 
\left ( \frac{k_2}{6 \, M_5^3}\right )
\times \left (
\begin{array}{cc}
A & \sqrt{\delta k} \, B \\
\\
\sqrt{\delta k} \, B & \delta k \, C
\end{array}
\right )  \, , \nonumber
\end{equation}
with
\begin{equation}
\left \{
\begin{array}{l}
A = \left . \frac{d^2 F_1(x)}{dx^2}\right |_{x = \bar \omega}
+ r_k \, (\bar R_1)_-^2 \, \bar \omega^2 \, 
\left (
12 \, F_2 (\bar \xi) 
+ \bar \xi^2 \, 
\left . \frac{d^2 F_2(x^\prime)}{dx^{\prime 2}}\right |_{x^\prime = \bar \xi} \right ) \, , \\
\\
B = - \frac{1}{2} \, r_k \, (\bar R_1)_-^2 \, \bar \omega^2 \, 
\bar \xi \, 
\left . \frac{d^2 F_2(x^\prime)}{dx^{\prime 2}}\right |_{x^\prime = \bar \xi} \, ,
\\
\\
C = r_k \, (\bar R_1)_-^2 \, \bar \omega^2 \,  
\left . \frac{d^2 F_2(x^\prime)}{dx^{\prime 2}}\right |_{x^\prime = \bar \xi} \, .
\end{array}
\right .
\end{equation}
It is clear from the structure of the matrix that, in the evanescent limit (see Ref.~\cite{Donini:2025cpl}), 
the radion $r_1$ decouples as its mass
is proportional to $1/\delta k$. On the other hand, we keep 
the $\delta k$ dependence explicit, as we want to study
the departure from this limit. In order to get some analytical 
insight in the radion mass spectrum it is now time to expand
in $\epsilon_1$ and $\epsilon_2$. We have found that, 
in order to get a minimum of the potential, the following
constraints must be fulfilled: 
\begin{equation}
\left \{
\begin{array}{l} 
\epsilon_1 \geq \epsilon_2 \, , \\
\\
r_k \, \left ( \frac{\epsilon_2}{\epsilon_1}\right ) \ll 1
\end{array}
\right .
\end{equation}
Under these assumptions, we have: 
\begin{eqnarray}
\label{eq:radionmassmatrix}
{\cal M}^2 (\bar r_1, \bar r_2) = &&  \\
&  
\frac{16}{\delta k} \, \left ( \frac{\tilde v_{\rm UV}}{L_1} \right )^2 \, 
\left ( \frac{k_2}{6 \, M_5^3}\right ) \, \bar \omega^{2 + 2 \epsilon_1} \, \epsilon_1^{3/2} \, 
\times \left (
\begin{array}{cc}
1 - \frac{1}{4} r_k  \, 
\left ( \frac{\epsilon_2}{\epsilon_1} \right ) 
& 
- \frac{\sqrt{\delta k}}{2} \, r_k \, \bar \xi^3 \, 
\left ( \frac{\epsilon_2}{\epsilon_1} \right )^{3/2}
\\
\\
- \frac{\sqrt{\delta k}}{2} \, r_k \,  \bar \xi^3 \, 
\left ( \frac{\epsilon_2}{\epsilon_1} \right )^{3/2} 
&
\delta k \, r_k \, \bar \xi^2 \, 
\left ( \frac{\epsilon_2}{\epsilon_1} \right )^{3/2}
\end{array}
\right ) + \dots \, , & \nonumber
\end{eqnarray}
where $\dots$ stand for terms ${\cal O} (\epsilon_i^2)$.
After diagonalization, we get: 
\begin{equation}
\left \{
\begin{array}{l}
m^2_{r_1} \simeq  \frac{16}{\delta k} \, \left ( \frac{\tilde v_{\rm UV}}{L_1} \right )^2 \, 
\left ( \frac{k_2}{6 \, M_5^3}\right ) \, \bar \omega^{2 + 2\epsilon_1} \, \epsilon_1^{3/2} \, 
\left [ 1 - \frac{1}{4} r_k  \, 
\left ( \frac{\epsilon_2}{\epsilon_1} \right ) \right ] \, , 
\\
\\
m^2_{r_2} \simeq 
16 \, \left ( \frac{\tilde v_{\rm UV}}{L_1} \right )^2 \, 
\left ( \frac{k_2}{6 \, M_5^3}\right ) \, \bar \omega^{2 + 2 \epsilon_1} \, r_k \, \bar \xi^2 \, \epsilon_2^{3/2} \, .
\end{array}
\right .
\end{equation}
It is easy to see that, even out of the evanescent limit, 
a significant hierarchy between the masses of the two scalar modes $r_1$ and $r_2$ exists: 
\begin{equation}
\label{eq:massratio}
\left ( \frac{m_{r_1}}{m_{r_2}}\right )^2 \sim 
\frac{1}{\delta k \, (1 + \delta k)} \, \frac{1}{\bar \xi^2} \, \left ( \frac{\epsilon_1}{\epsilon_2}
\right )^{3/2} \, \left [ 1 - \frac{1}{4} \left ( 1 + \delta k \right )  \, 
\left ( \frac{\epsilon_2}{\epsilon_1} \right ) \right ].
\end{equation}
At this stage, it is instructive to reassess the influence of backreaction on the spectrum. Using the results of 
Ref.~\cite{Seung1}, where the radion masses are computed
within the "classical" approach of Ref.~\cite{Csaki:2000zn} 
(under the assumption that the backreaction parameter $\ell$
is the same for both branes, $\ell_1 = \ell_2$, something that
should be tested as we are doing here for the curvatures $k_1$
and $k_2$), we find that our results are effectively unchanged 
as long as $\ell \leq 0.1$.

\begin{figure}
\centering
\begin{tabular}{cc}
\includegraphics[scale=0.45]{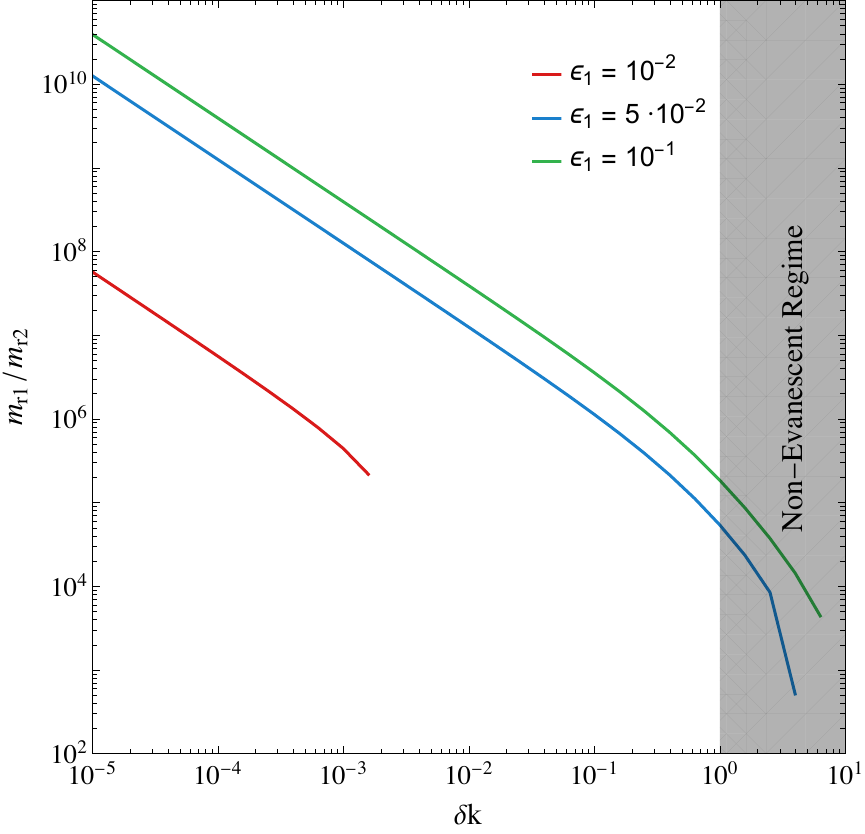} &
\includegraphics[scale=0.45]{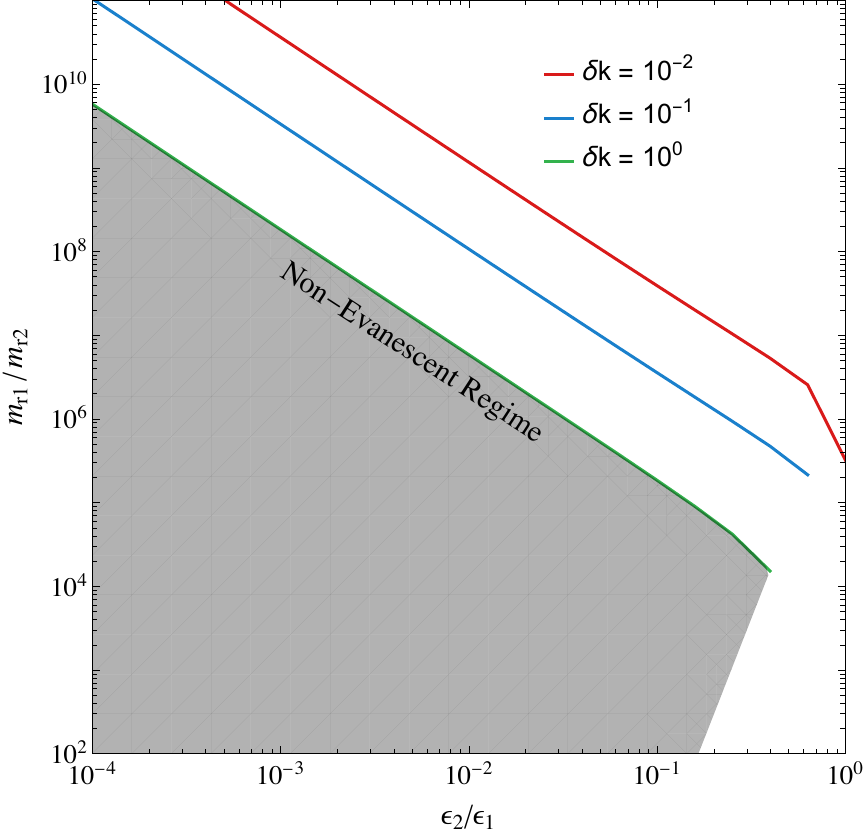}
\end{tabular}
\caption{The ratio of the masses of the scalar modes
$r_1$ and $r_2$. Left:
as a function of the curvature splitting
parameter $\delta k$, for several different values
of $\epsilon_1 = 0.1$ (green), $0.05$ (blue) and $0.01$ (red), with $\epsilon_2 = 0.01$ in all cases.
Right: 
as a function of $\epsilon_2/\epsilon_1$ for several different
values of $\delta k = 0.01$ (red), $0.1$ (blue) and $1$ (green), 
with $\epsilon_1 = 0.1$ in all cases. 
In both panels, the region where the difference between $k_1$ and
$k_2$ is ${\cal O}(1)$ ({\em i.e.}, the {\em non-evanescent regime}) is depicted as a grey area.
}
\label{fig:massratio}
\end{figure}

In Fig.~\ref{fig:massratio} we show the ratio of the masses
of the two ``radion modes" $r_1$ and $r_2$ as a function 
of the curvature splitting $\delta k$ for fixed $\epsilon_1$ and 
$\epsilon_2$ (left panel) or as a function of $\epsilon_2/\epsilon_1$
for fixed $\epsilon_1$ and $\delta k$ (right panel). In the left 
panel the parameter $\epsilon_2$ is kept fixed at $\epsilon_2 = 0.01$, whereas the parameter $\epsilon_1$ is varied keeping $\epsilon_1 \geq \epsilon_2$: 
$\epsilon_1 = 0.1$ (green), $0.05$ (blue) and $0.01$ (red). 
In the right panel, $\epsilon_1 = 0.1$, whereas $\delta k = 0.01$ (red), $0.1$ (blue) and $1$ (green). In both panels, we can see that
all lines stop for some value of $\delta k$, for which 
no minimum of the potential exists in the real plane. 
In all the considered parameter space we can see that
a huge hierarchy between $m_{r_1}$ and $m_{r_2}$ exists, 
with the former being always much larger than the latter. 
It can be seen that, even out of the evanescent limit
(that is, for $\delta k \geq 1$), the ratio $m_{r_1}/m_{r_2}$ can 
be as large as $10^3$ or $10^4$. 
This result is extremely important: 
in Ref.~\cite{Donini:2025cpl} (and in Refs.~\cite{Cacciapaglia,Koutroulis:2024wjl}) 
the Dark Matter relic abundance
was only computed assuming that $k_2 = k_1$, {\em i.e.}
when the curvatures in the two segments to the left
and to the right of the middle brane were identical. We have already stressed, though, that this limit is unphysical as, 
when the curvatures $k_1$ and $k_2$ are identical, 
the brane tension of the middle brane vanishes and the brane
{\em evaporates}. For this reason it is quite relevant
to see that one of the main simplification that we observed
in the evanescent limit, namely the decoupling of one of
the radion modes, is still valid {\em out of the evanescent limit}: it is still possible
to study a phenomenological setup in which only one radion 
is light and couples with SM matter on the middle brane. 

In order to diagonalize the radion mass matrix ${\cal M}^2$, 
we define the rotation matrix: 
\begin{equation}
\left ( 
\begin{array}{c}
\delta \tilde r_1 \\
\\
\delta \tilde r_2 
\end{array}
\right )
= 
\left ( 
\begin{array}{cc}
\cos \theta & \sin \theta \\
\\
- \sin \theta & \cos \theta
\end{array}
\right ) \, 
\left ( 
\begin{array}{c}
\delta r_1 \\
\\
\delta r_2 
\end{array}
\right ) \, , 
\end{equation}
where
\begin{equation}
\tan 2 \theta = 2 \frac{\sqrt{\delta k} \, B}{A - \delta k \, C} \simeq - \sqrt{\delta k} \, \bar \xi^3 \, 
\left ( \frac{\epsilon_2}{\epsilon_1} \right )^{3/2} + \dots \, ,
\end{equation}
that is a very small quantity also out of the evanescent limit, as it is suppressed by $\bar \xi^3$
and the ratio $(\epsilon_2/\epsilon_1)^{3/2}$. 
For this reason we can safely consider the approximation
$\delta \tilde r_1 \sim \delta r_1$ and $\delta \tilde r_2 \sim \delta r_2$ in the following and there is no need to modify
the radions wave-function when computing their couplings.

Eventually, the scalar action of the model is: 
\begin{eqnarray}
{\cal S} [r, \phi^{(0)}] &=& \frac{1}{2} \int d^4 x 
\sqrt{-g} \, 
\left \{ 
\partial_\mu \delta \tilde r_1 
\partial^\mu \delta \tilde r_1 + 
\partial_\mu \delta \tilde r_2 
\partial^\mu \delta \tilde r_2 
- m^2_{r_1} \, \delta \tilde r_1^2 - m^2_{r_2} \, \delta \tilde r_2^2
\right . \nonumber \\
&+& K_1(\bar r_1) \, \partial_\mu \phi_1^{(0)} \partial^\mu \phi_1^{(0)} 
+ K_2(\bar r_1,\bar r_2) \, \partial_\mu \phi_2^{(0)} \partial^\mu \phi_2^{(0)} \nonumber \\
&-& \left . V(\bar r_1, \bar r_2) 
\left [ L_1 \, \phi_1^{(0)} \right ]^2 
 - V(\bar r_1,\bar r_2) \left [ \Delta L \, \phi_2^{(0)} \right ]^2
\right \} \, .
\end{eqnarray}

After canonical normalization of the kinetic terms
of the fields $\phi_1^{(0)}$ and $\phi_2^{(0)}$, we have: 
\begin{equation}
\left \{
\begin{array}{l}
m_{\phi_1}^2 = 2 \frac{V (\bar r_1, \bar r_2 )}{K_1 (\bar r_1, \bar r_2)} = {\cal O } (m_1^2)
\\
\\
m_{\phi_2}^2 = 2 \frac{V (\bar r_1, \bar r_2)}{K_2 (\bar r_1, \bar r_2)} = {\cal O } (m_1^2/\bar \omega^{2 + 2 \epsilon_1}) 
\end{array}
\right .
\end{equation}
so that both decouple from the low-energy lagrangian, as it was the case in the two-brane setup. 

In summary, working out in detail the Goldberger-Wise stabilization method in the three-brane case, we find
that out of the four possible light scalar degrees of freedom
(two ``radions" and two bulk field zero modes), 
only one of them is light. The two bulk field zero-modes
get indeed a mass proportional to the bulk mass parameter
$m_1$ (or the same parameter {\em warped up} by some power
of $1/\bar \omega$), whereas the ``radion" mode $\tilde r_1$
has always a mass much larger than the light mode $\tilde r_2$, 
as long as $\epsilon_1 \geq \epsilon_2$ and $\delta k \epsilon_2 \ll 1$. This result reconciles the two different opinions present in the literature, namely that: 
\begin{itemize}
\item
in a Randall-Sundrum setup with $N$ branes there are $N -1$
radions, defined as {\em scalar modes related to the original 5D graviscalar zero mode} as in, {\em e.g.}, Refs.~\cite{Kogan:2000xc,Seung1}; this approach is stringy-inspired, in that the radions are essentially moduli fields, that set up the geometry of the compactified manifold;
\item
in a Randall-Sundrum setup with $N$ branes there
is only one radion, defined as a {\em light scalar mode related to the original 5D graviscalar zero mode}, as in, {\em e.g.}, 
Refs.~\cite{Cai:2021mrw,Olechowski:2024wcf}; this approach is phenomenologically-inspired, in that only light fields are relevant for low-energy physics.  
\end{itemize}
Both statements are therefore correct, the only difference is the definition of the word {\em radion}. Once we agree that 
there is only one {\em light radion}, we can safely consider its mass as a free parameter of the theory, $m_{r_2}$. 

\subsection{KK graviton and radion couplings with matter on the IR and DIR branes}
The spectrum and interactions of gravitons and radions between themselves and with SM fields located in the intermediate brane have been computed in Ref.~\cite{Seung1} in the evanescent limit. In Ref.~\cite{Donini:2025cpl}, we have also computed them 
using the wave-functions defined in eq.~(\ref{eq:deltaradionwavefunctions}). The interaction of the radion with matter living in a brane localized at $y=L$
is given by:
\begin{equation}
  S_{\phi,r} = \int d^4x \frac{\delta \tilde r_2(x) \, T(x)}{\sqrt{6} \, \Lambda_L}\,,
\end{equation}
where $T$ is the trace of the energy-momentum tensor and, out of the evanescent limit, we get for the interaction scale:
\begin{equation}
\Lambda_L =  \,  
\left \{ 
\begin{array}{l}
\left ( \frac{k_1}{k_2} \right )^{1/2} \, \frac{\bar \omega}{ \bar \xi} \, \bar M_{\rm P} + {\cal O}(\bar \omega^3) \sim 
\left ( \frac{k_1}{k_2} \right )^{1/2} \, \left ( \frac{\Lambda_{\rm IR}}{\Lambda_{\rm DIR}} \right ) 
\, \Lambda_{\rm IR}
\qquad (L = L_1) \, , \\
\\
\left ( \frac{k_1}{k_2} \right )^{1/2} \, \bar \omega \, \bar \xi \, \bar M_{\rm P} + {\cal O}(\omega^3) 
\sim \left ( \frac{k_1}{k_2} \right )^{1/2} \, \Lambda_{\rm DIR}  \qquad (L = L_2) \, . 
\end{array}
\right .
\end{equation}
The factor $(k_1/k_2)^{1/2}$ in front of the dimensionful scale
comes from the definition of $\bar M_{\rm P}$ as a function 
of $k_1, k_2, \bar \omega$ and $\bar \xi$, eq.~(\ref{eq:Mplanck3branes2}), 
once higher order terms in $\bar \omega$ and $\bar \xi$ have been neglected.
As we have just seen, there is no need to include the 
interaction of brane matter with the scalar mode 
$\delta \tilde r_1$, since this field decouples from the 
low-energy spectrum.

The couplings of the graviton and radion modes with DM (assumed
to be located on the DIR brane) and SM fields 
(on the IR brane) are
summarized in Tab.~\ref{tab:couplings}.
	
  \begin{table}[ht]
  	\centering
	\begin{tabular}{|c|c|c|}
	\hline
	Brane & $ h_{\mu\nu}^{n\geq1} $ & $r_2$  \\ \hline 
     & & \\
	DIR &$
	\left( \frac{k_1}{k_2} \right )^{1/2} \, 
    \Lambda_{\rm DIR} $ &  
	$\sqrt{6} \, \left ( \frac{k_1}{k_2} \right )^{1/2} \Lambda_{\rm DIR} $ \\
            & & \\
	IR  &  
$ \left \{
\begin{array}{c}
\frac{8 J_2(x_n) }{ x_n^2 } \, 
\left [ 
 \left( \frac{k_1}{k_2} \right )^{1/2} \, 
\left(\frac{\Lambda_{\rm IR}}{\Lambda_{\rm DIR}}\right)^3
\right ] \, \Lambda_{\rm IR} \qquad r_k \, x_n \, \bar \xi \ll 1 \\
\\
\left [ \frac{h_n (r_k, \bar \xi) \, J_1(x_n \, \bar \xi) }{ x_n J_1(r_k \, x_n \bar \xi) J_2 (x_n \, \bar \xi )} \right ]\, 
\left [ \left( \frac{k_1}{k_2} \right )^{1/2} \, 
\left(\frac{\Lambda_{\rm IR}}{\Lambda_{\rm DIR}}\right)
\right ] \, \Lambda_{\rm IR} \\
\\
\qquad \qquad \qquad \qquad \qquad \qquad \qquad 
r_k \, x_n \, \bar \xi \gg 1
\end{array}
\right .
$ &  
	$ \sqrt{6} \, \left [ \left( \frac{k_1}{k_2} \right )^{1/2}
    \left ( \frac{\Lambda_{\rm IR}}{\Lambda_{\rm DIR}}\right ) \right ] \, \Lambda_{\rm IR} $ \\
        & & \\
    \hline
		\end{tabular}
  \caption{Inverse couplings of the DM (SM) particles localized in the DIR (IR) brane with KK gravitons and the radion.
} 
  \label{tab:couplings}
\end{table}
Two remarks are in  order at this point: first, 
notice that, in the case of the KK graviton coupling with SM fields
on the IR brane, we show two relevant limits, corresponding to small and large $r_k \, x_n \, \bar \xi$. In the first case, the Bessel functions $J_2 (x_n \, \xi)$, $J_2 (r_k \, x_n \, \xi)$, $Y_2 (x_n \, \xi)$ and $Y_2 (r_k \, x_n \, \bar \xi)$ can be expanded at leading order in their argument (either $x_n \, \bar \xi$ or $r_k \, x_n \, \bar \xi$). In this limit, the dependence of the inverse coupling on the ratio of the DIR and IR
scales goes as $(\Lambda_{\rm DIR}/\Lambda_{\rm IR})^3$. On the other hand, for large $n$ (when the zeroes $x_n$ of the Bessel function $J_1 (x)$ are large enough for $x_n \, \bar \xi$ to be greater than 1) 
or when $r_k \gg 1$ we cannot expand in the argument of $(J_2,Y_2)$ and, therefore, the coupling is suppressed by $(\Lambda_{\rm DIR}/\Lambda_{\rm IR})$, only. For intermediate $n$ or moderate $\delta k$, both terms proportional 
to $Y_2$ and $J_2$ in $\hat \chi^{(n)}(z_1)$ plays a role and the dependence on $r_k$ is more complicated. The second relevant remark 
is that the couplings of both KK gravitons 
and radions with matter on the DIR brane are identical to those
in the evanescent limit (see Ref.~\cite{Donini:2025cpl}) once
$\Lambda_{\rm DIR}$ is replaced by $(k_1/k_2)^{1/2} \, \Lambda_{\rm DIR}$. This is not true for the coupling of KK gravitons with matter on the IR brane, as the coefficient in front of the combination 
$(k_1/k_2)^{1/2} \, \Lambda_{\rm IR}/\Lambda_{\rm DIR}$ depends
significantly on $r_k$.

We can also compute the couplings between KK gravitons and radions. 
They are defined in terms of the following parameters: 
\begin{eqnarray}
\chi_{nrr} &\equiv& 
- \frac{2}{J_0\left(x_n\right)}\int_{0}^{1}du\,\,  u^3J_2\left(x_n u\right) \, , \\
&& \nonumber \\
\tilde{\chi}_{mnr} &\equiv& 
\frac{2 \, x_m \, x_n}{J_0 \left(x_m \right) \, J_0 \left( x_n \right)} \, \int_{0}^{1}du\,\, u^3J_1\left(x_m u\right)J_1\left(x_n u\right) \, , \\
&& \nonumber \\
\chi_{lmn} &\equiv& 
- \frac{2}{J_0\left(x_l\right) \, J_0\left(x_m\right) \, J_0\left(x_n\right)} \, \int_{0}^{1}du\,\, u^3J_2\left(x_l u\right) \, J_2 \left(x_m u \right) \, J_2\left(x_n u\right) \, ,
\end{eqnarray}
as in Ref.~\cite{de_Giorgi_2021}. These couplings are summarized in Tab.~\ref{tab:couplings_bulk}. It can be seen that all couplings
are identical to those in the evanescent limit if we redefine
$\Lambda_{\rm DIR} \to (k_1/k_2)^{1/2} \, \Lambda_{\rm DIR}$.

  \begin{table}[ht]
  	\centering
	\begin{tabular}{|c|c|c|}
	\hline
	Interaction Type & Inverse coupling  \\ \hline 
     &  \\
	$r \, r \, r$ &
	$\sqrt{6} \, \left ( \frac{k_1}{k_2} \right )^{1/2} \Lambda_{\rm DIR} $ \\
            &  \\
	$G_n \, r \, r  $   &  
	$ \frac{1}{\chi_{rrn}}\, \left ( \frac{k_1}{k_2} \right )^{1/2} \Lambda_{\rm DIR}$ \\
        &  \\
        $G_m \, G_ n \, r $ &
	$\sqrt{\frac{2}{3}} \, \frac{1}{\tilde{\chi}_{nmr}} \, \left ( \frac{k_1}{k_2} \right )^{1/2} \, \frac{1}{k_2^2} \, \frac{1}{\left(\bar{\omega}\, \bar{\xi}\right)^2} \, \Lambda_{\rm DIR}$ \\
            &  \\
            $G_l \, G_ m \, G_ n$   &  
	$ \frac{1}{\chi_{nml}} \, \left ( \frac{k_1}{k_2} \right )^{1/2} \Lambda_{\rm DIR}$ \\
        &  \\
    \hline
		\end{tabular}
  \caption{Inverse couplings of KK gravitons and the radion $r_2$ between themselves.} 
\label{tab:couplings_bulk}
\end{table}

Eventually, we must compute the couplings that involve
two matter particles (either on the DIR brane or on IR brane) with two
radion modes or with one radion and one KK graviton. The corresponding inverse couplings, together with those in eqs.~(\ref{eq:Lambda_DIR_seagull}) and (\ref{eq:Lambda_IR_seagull})
in the two limits $r_k \, x_n \, \bar \xi \ll 1$ and $r_k \, x_n \, \bar \xi \gg 1$, are summarized in Tab.~\ref{tab:couplings_seagull}. 

 \begin{table}[ht]
  	\centering
	\begin{tabular}{|c|c|c|}
	\hline
    Interaction type & DIR & IR \\ \hline 
     &  & \\
     $ M M r_2 \, r_2$ &  
     $6 \,  \frac{k_1}{k_2}  \Lambda_{\rm DIR}^2 $ & $6\left [ \left ( \frac{k_1}{k_2} \right )^{1/2} \,
     \left ( \frac{\Lambda_{\rm IR}}{\Lambda_{\rm DIR}} \right )
     \right ]^2\Lambda_{\rm IR}^2$
     \\
     & & \\
     $M M G_n \, r_2$ & 
    
     $\sqrt{6} \, \frac{k_1}{k_2}  \Lambda_{\rm DIR}^2$ 
     &
      $ \left \{
     \begin{array}{c}     8\sqrt{6} \, \frac{ J_2 (x_n)}{ x_n^2} \, 
     \left [ \left ( \frac{k_1}{k_2} \right ) \,
     \left ( \frac{\Lambda_{\rm IR}}{\Lambda_{\rm DIR}} \right )^4
     \right ] \, \Lambda^2_{\rm IR} \qquad r_k \, x_n \, \bar \xi \ll 1 \\
     \\
     \sqrt{6} \, \left [ \frac{h_n (r_k, \bar \xi) \, J_1(x_n \, \bar \xi) }{ x_n J_1(r_k \, x_n \bar \xi) J_2 (x_n \, \bar \xi )} \right ] \, 
     \left [ \left ( \frac{k_1}{k_2} \right )^{1/2} \,
     \left ( \frac{\Lambda_{\rm IR}}{\Lambda_{\rm DIR}} \right )
     \right ]^2 \, \Lambda^2_{\rm IR} \\
     \\
     \qquad \qquad \qquad \qquad \qquad \qquad \qquad r_k \, x_n \, \bar \xi \gg 1
     \end{array}
     \right .
     $
     \\
     & & \\
     $M M G_m \, G_n$ &
     $ \frac{k_1}{k_2}  \Lambda_{\rm DIR}^2 $
     &
      $ 
      \left \{
      \begin{array}{c}
      64 \, \frac{J_2(x_m) \, J_2 (x_n)}{x_m^2 \, x_n^2} \, 
     \left [ \left ( \frac{k_1}{k_2} \right )^{1/2} \,
     \left ( \frac{\Lambda_{\rm IR}}{\Lambda_{\rm DIR}} \right )^3
     \right ]^2 \, \Lambda^2_{\rm IR} \qquad r_k \, x_n \, \bar \xi \ll 1 \\
     \\
     \left [ \frac{h_m (r_k, \bar \xi) \, J_1(x_m \, \bar \xi) }{ x_m J_1(r_k \, x_m \bar \xi) J_2 (x_m \, \bar \xi )} \right ] \, 
     \left [ \frac{h_n (r_k, \bar \xi) \, J_1(x_n \, \bar \xi) }{ x_n J_1(r_k \, x_n \bar \xi) J_2 (x_n \, \bar \xi )} \right ] \\
     \\
     \times
     \left [ \left ( \frac{k_1}{k_2} \right )^{1/2} \,
     \left ( \frac{\Lambda_{\rm IR}}{\Lambda_{\rm DIR}} \right )
     \right ]^2 \, \Lambda^2_{\rm IR} 
     \qquad \qquad \qquad r_k \, x_n \, \bar \xi \gg 1
     \end{array}
     \right .
     $ 
     \\
     & &  \\
    \hline
		\end{tabular}
  \caption{Inverse couplings between two matter particles (M), 
  either on the DIR (DM) or the IR (SM) brane, and two bulk particles (the radion $r_2$ or the KK gravitons $G_n$).} 
\label{tab:couplings_seagull}
\end{table}

Also in this case, it can be seen that couplings of KK gravitons 
and radions with fields on the DIR brane are identical to 
those in the evanescent limit, once $\Lambda_{\rm DIR}$ is replaced
by $(k_1/k_2)^{1/2} \, \Lambda_{\rm DIR}$. 

This means that the subtle cancellations in 
the amplitude of the processes ${ \rm DM} \, {\rm DM} \rightarrow G_m \, G_n$, $ { \rm DM} \, {\rm DM} \rightarrow G_m \, r_2$ and
${ \rm DM} \, {\rm DM} \rightarrow r_2 \, r_2$, that violate unitarity at ${\cal O}\left(s^3\right)$, ${\cal O}\left(s^2\right)$  and ${\cal O}\left(s^{3/2}\right)$ \cite{Chivukula_2020, 2206.10628,de_Giorgi_2021,deGiorgi:2021xvm}, are preserved
once the DIR scale $\Lambda_{\rm DIR}$ gets replaced by
$(k_1/k_2)^{1/2} \, \Lambda_{\rm DIR}$. The same statement is not 
as straightforward to proof for the processes ${ \rm SM} \, {\rm SM} \rightarrow G_m \, G_n$, $ { \rm SM} \, {\rm SM} \rightarrow G_m \, r_2$ and ${ \rm SM} \, {\rm SM} \rightarrow r_2 \, r_2$. Even though the radion couplings with IR matter becomes identical to those 
in the evanescent limit after replacing $\Lambda_{\rm DIR} \to (k_1/k_2)^{1/2} \, \Lambda_{\rm DIR}$ AND $\Lambda_{\rm IR} \to 
(k_1/k_2)^{1/2} \, \Lambda_{\rm IR}$, the same is not true
for the corresponding couplings of KK gravitons (see Tabs.~\ref{tab:couplings} and \ref{tab:couplings_seagull}).

\section{Phenomenology out of the {\em evanescent limit}}
\label{sec:pheno}
Having worked out analytically the departure from the evanescent limit that was adopted in Ref.~\cite{Donini:2025cpl}, it is now important to quantify (or, at least, have a qualitative understanding of) how much can we relax this limit whilst still retaining the phenomenological results obtained. The question is, therefore, which is the value of $\delta k$ for which the results of Ref.~\cite{Donini:2025cpl} still holds, or how they should be modified. 

First of all, we remind the important result of the previous Section, namely that {\bf only one light radion is present in the
low-energy spectrum of the model}, with a mass $m_{r_2}$
that can be treated as a free parameter. 
Notice that in eq.~(\ref{eq:massratio}), at fixed $\bar \xi$,
there are two parameters that contribute to the hierarchy between the masses of the two radions: $\delta k$
and $\epsilon_2/\epsilon_1$, under the constraint that
$ (1 + \delta k)  \, \left ( \epsilon_2/\epsilon_1 \right ) \ll 1$. As we have seen in the left panel of Fig.~\ref{fig:massratio} it can be seen how, even outside the evanescent limit, there is a huge hierarchy between $m_{r_1}$ and $m_{r_2}$ and, even if $\delta k \sim {\cal O}(10)$, we can still have $m_{r_1}/m_{r_2} \in [10^{3}, 10^4]$, 
depending on $\epsilon_2$. The dependence on the
$\epsilon_2/\epsilon_1$ ratio can be better seen in the right panel of Fig.~\ref{fig:massratio} that shows that, for the maximum value of $\epsilon_2/\epsilon_1$ for which positive real values for both masses can be obtained, we can still have
$m_{r_1}/m_{r_2} \sim {\cal O}(10^5)$. This means that the
the assumption of Ref.~\cite{Donini:2025cpl}, where only one light radion was considered, still holds for $\delta k$ as large as 10. We can safely ignore the effects of the second scalar mode $r_1$ on particles in the bulk, on the IR-brane or on 
the DIR-brane.

\begin{figure}
	\centering 
    \begin{tabular}{cc}
    \includegraphics[scale=0.3]{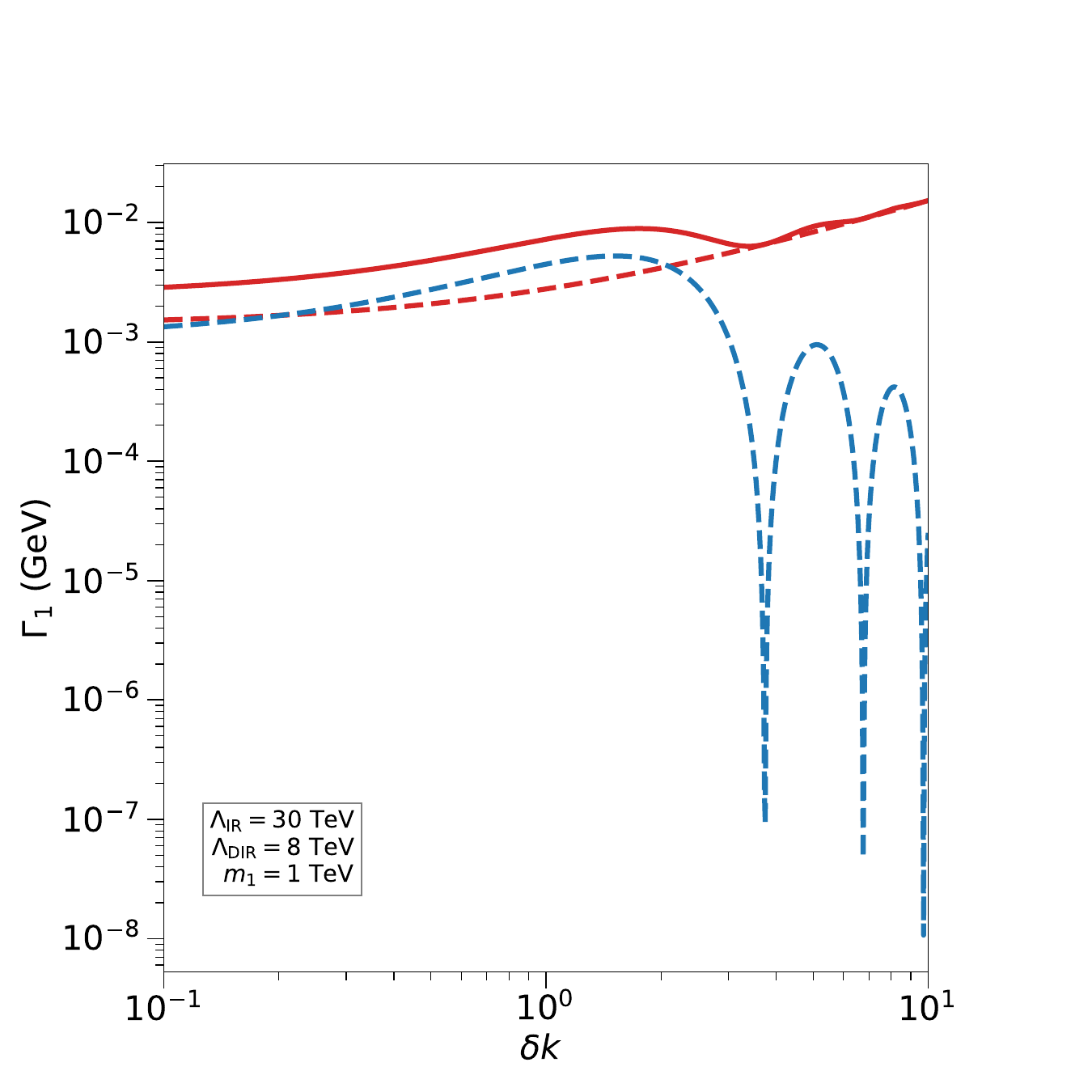} &
    \includegraphics[scale=0.3]{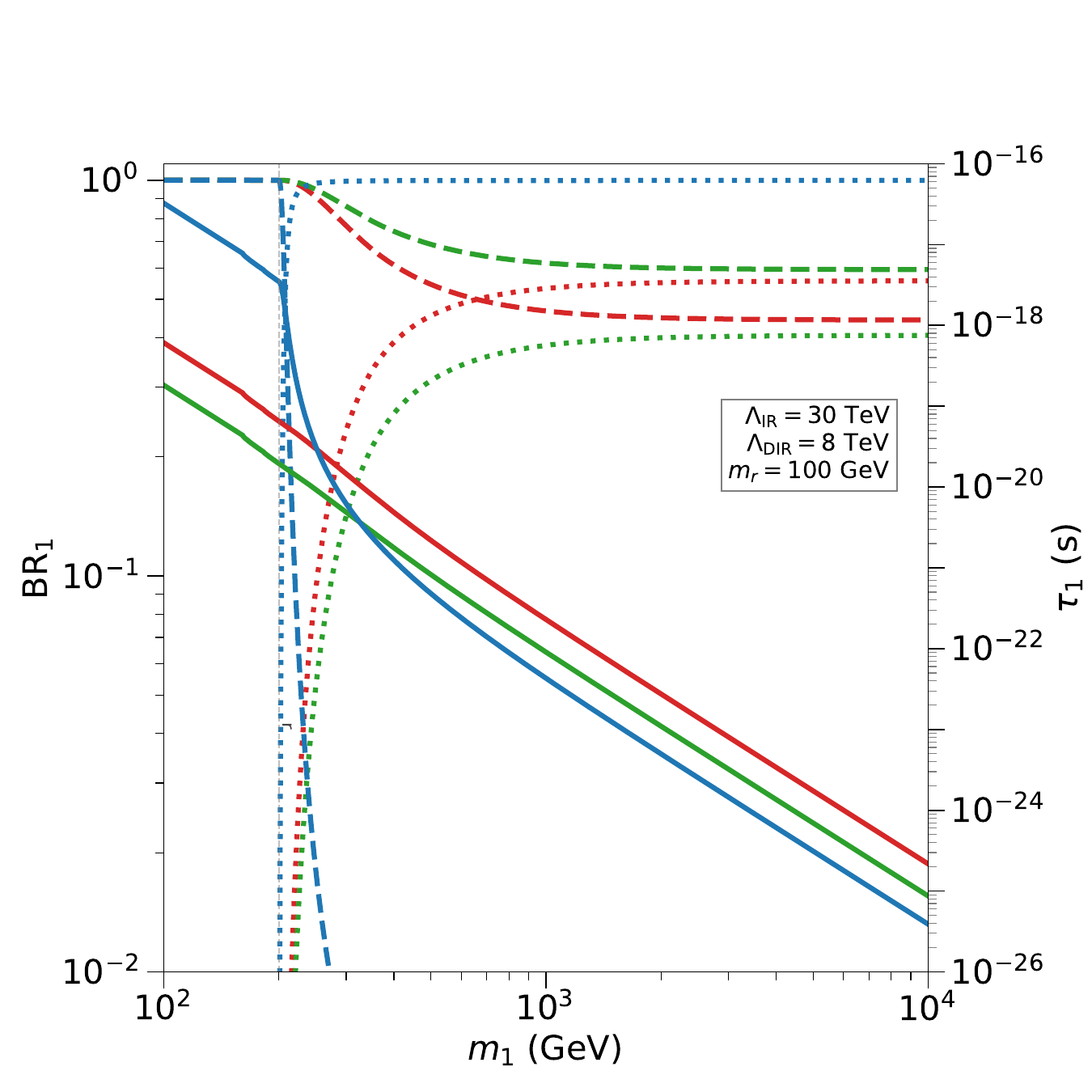}
    \end{tabular}
    \caption{
    Left panel: The total decay width of the first KK graviton $G_1$ as a function of $\delta k$. The DIR and IR scales are fixed to $\Lambda_{\rm DIR} = 8$ TeV and $\Lambda_{\rm IR} = 30$ TeV, respectively. The first KK graviton mass is $m_1 = 1$ TeV, whereas
    the radion mass is $m_{r_2} = 100$ GeV.
    The blue (dashed) line stands for $\Gamma \left ( G_1 \to {\rm SM \, SM}\right )$. The red (dashed) line stands for $\left ( G_1 \to r_2 \, r_2 \right )$. Eventually, the red (solid) line is the
    total decay width. 
    Right panel: 
    Branching ratios of the first KK graviton $G_1$ (scale on the left vertical axis) and its lifetime $\tau$ (scale on the right vertical axis), as a function of the KK graviton mass $m_1$. The solid lines correspond to $\tau$. Dashed and dotted lines represent the BR's
    ${\rm BR}(G_1 \to r_2 \, r_2)$ (dotted) and ${\rm BR} (G_1 \to {\rm SM \, SM})$ (dashed).
    The color code works as follows: red stands for $\delta k = 0.1$, green for $\delta k = 1$ and blue for $\delta k = 10$. 
    }
    \label{fig:decays}
\end{figure}

The next step is to analyse the effect of the couplings out of the evanescent limit shown in Tabs.~\ref{tab:couplings}, \ref{tab:couplings_bulk} and \ref{tab:couplings_seagull} on the decay rates of both the radion and the KK gravitons. Notice, first, that
the couplings of the radion with matter on the DIR- or the IR-branes
are only modified by a common factor $(k_2/k_1)^{1/2}$. This means
that the total decay width of $r_2$ is increased by a factor $(k_2/k_1)$, whereas the Branching Ratios (BR) into SM particles are unchanged with respect to the evanescent limit 
(see Fig.~3 of Ref.~\cite{Donini:2025cpl}). On the other hand, 
the coupling of KK gravitons between themselves, the radion and SM matter are not modified by a single multiplicative factor. 
This is because the dependence of the coupling between KK gravitons and SM matter on the IR-brane is more involved and, therefore, the BR's of the decay of a KK graviton $G_n$ into other (lighter) KK gravitons, the radion, DM and SM matter differ from those in the evanescent limit. For $\delta k$ small enough, an increase of the total width is to be expected
for the lowest-lying KK graviton modes, due to the factor $(k_2/k_1)^{1/2}$ of the coupling. This can be seen in the left panel of Fig.~\ref{fig:decays}, where the total decay width of the first
KK graviton $G_1$ is shown as a function of $\delta k$ for 
$m_{1} = 1$ TeV and $m_{r_2} = 100$ GeV. 
The DIR and IR scales are kept fixed at $\Lambda_{\rm DIR} = 8$ TeV
and $\Lambda_{\rm IR} = 30$ TeV, respectively. These values
correspond to $\bar \omega \sim 1.5 \times 10^{-14}$ and
$\bar \xi \sim 2.6 \times 10^{-1}$. The blue (dashed) line refers to $\Gamma \left( G_1 \to {\rm SM \, SM}\right)$, whereas the red (dashed) line stands for $\Gamma \left ( G_1 \to r_2 \, r_2 \right )$. 
We can see that for small $\delta k$, a $G_1$ with a mass around 1 TeV
decay similarly to SM particles or the radion. On the other hand, 
for an intermediate $\delta k$, the decay into SM particles increases slightly with respect to that into radions. However, for large $\delta k$, the involved $G_1 - {\rm SM} - {\rm SM}$ coupling dependence on $\delta k$ sets in and the width $\Gamma \left ( G_1 \to {\rm SM \, SM}\right )$ decreases with respect to $\Gamma \left ( G_1 \to r_2 \, r_2 \right )$. From that point on, the total width is saturated by decays into radions, that depends linearly on $\delta k$.

In the right panel of Fig.~\ref{fig:decays} we show the lifetime
of the first KK graviton $G_1$ (solid lines) and the Branching Ratios (BR) for the channels $G_1 \to {\rm SM \, SM}$ (dashed lines)
and $G_1 \to r_2 \, r_2$ (dotted lines), as a function of the
first KK graviton mass, $m_1$, for the same values of the DIR and IR scales. The radion mass is kept fixed at $m_{r_2} = 100$ GeV. 
The three colors represent different values of $\delta k$: red stands
for $\delta k = 0.1$ (within what we may call the evanescent limit); green for an intermediate value, $\delta k = 1$; and blue
for a hierarchical scenario, $\delta k = 10$.
Notice that for small values of $\delta k$, nearer to the evanescent limit ($\delta k = 0.1$), the results are identical to those in Fig. 4 (middle panel) of Ref.~\cite{Donini:2025cpl}. This certifies that the range of applicability of the evanescent approximation 
goes far beyond extremely small values of $\delta k$ (that is 
therefore proved to be more robust than at first sight). 
We can see that, when the decay channel into radions opens, 
the dependence on $\delta k$ is not trivial: if for small $\delta k$
at large values of $m_1$ $G_1 \to r_2 \, r_2$ dominates over
$G_1 \to {\rm SM \, SM}$ (as the partial width is proportional to $m_1^5$ in the former case, with respect to $m_1^3$ in the latter), 
the opposite is true for a moderate $\delta k$ (as we have seen
in the left panel for $m_1 = 1$ TeV). On the other hand, 
for very large $\delta k$ the BR into SM particles falls down, 
and the decay $G_1 \to r_2 \, r_2$ becomes dominant.

\begin{figure}
	\centering  \includegraphics[scale=0.55]{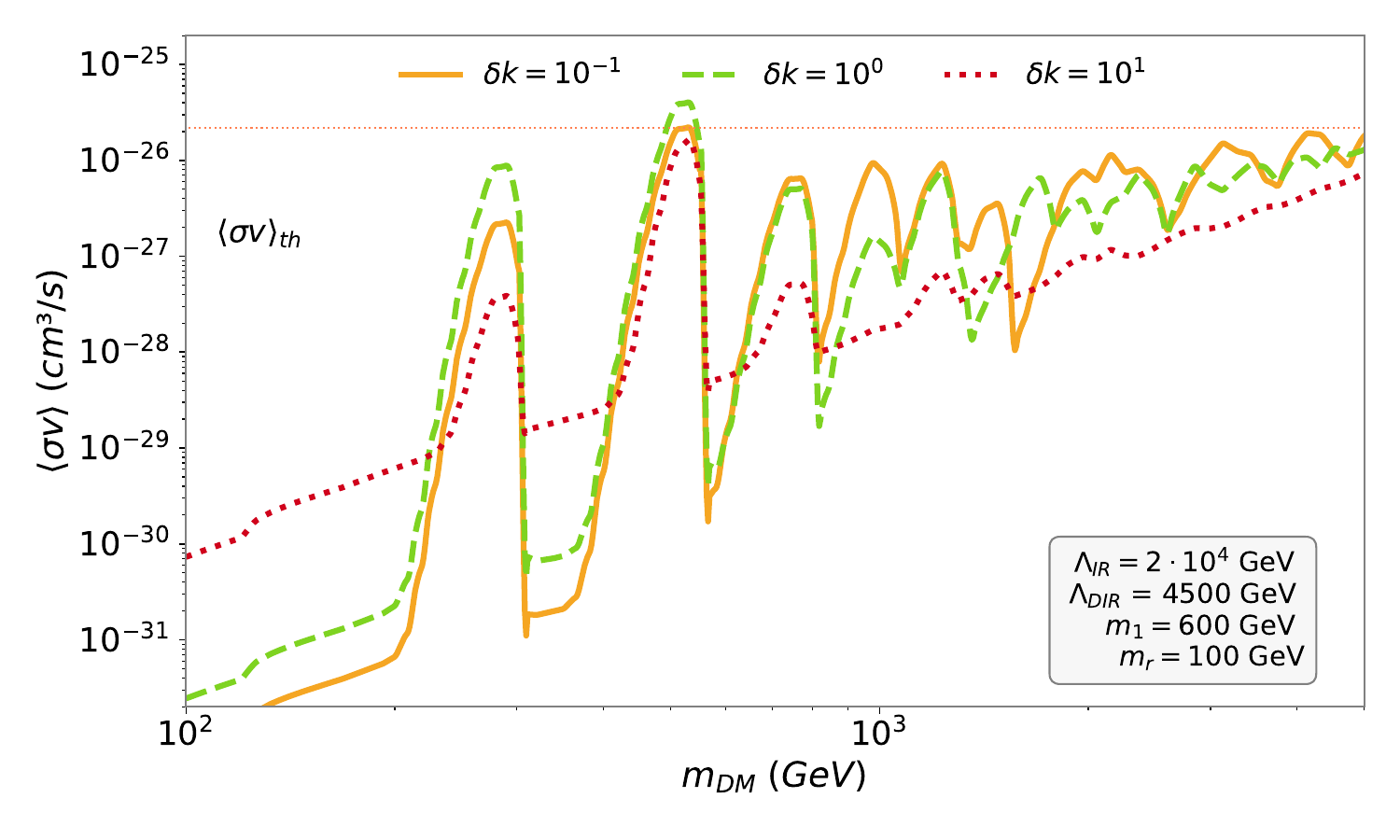} 
    \caption{
    Thermally-averaged annihilation cross-section 
    $\langle\sigma v\rangle_{{\rm DM}\rightarrow{\rm SM}}$
    as a function of the DM mass $m_{\rm DM}$ for different values of $\delta k$: $\delta k = 0.1$ (orange solid line), $\delta k = 1$ (green dashed line) and $\delta k = 10$ (red dotted line).
    The DIR and IR scales are $\Lambda_{\rm DIR} = 4.5$ TeV and
    $\Lambda_{\rm IR}$ = $20$ TeV, respectively, whereas the
    first KK graviton mass is $m_1 = 600$ GeV and the radion mass
    is $m_{r_2} = 100$ GeV. The value of the thermally-average
    cross-section that would reproduce the observed DM relic abundance,
    $\langle \sigma v\rangle_{\rm th}$, is shown as a red thin dotted 
    horizontal line.
    }
    \label{fig:Sigma_v_a_SM}
\end{figure}

After understanding how the decay rates of radions and KK gravitons
are modified when going beyond the evanescent limit, it is important
to study the behaviour of the different DM annihilation processes and the consequent production of a pair of SM or bulk particles. The two cases will be treated separately. Consider first DM annihilation 
into SM particles. Whereas the thermally-averaged cross-section 
$\langle\sigma v\rangle$ for the process ${\rm DM \, DM }\rightarrow r_2 \rightarrow {\rm SM \, SM}$ depends trivially on $\delta k$
through a common rescaling factor $(k_2/k_1)^{1/2}$ to the fourth power  (see the second column in Tab.~\ref{tab:couplings}), 
that same is not true for the process
${\rm DM \, DM }\rightarrow G_n \rightarrow {\rm SM \, SM}$. 
In this case, although the annihilation of two DM particles
into a KK gravitons is proportional to the same rescaling factor
$(k_2/k1)^{1/2}$, the coupling between the KK graviton and SM particles on the IR brane has a complicated dependence on $\delta k$ (see the first column of Tab.{\ref{tab:couplings}}). The total thermally-averaged cross-section $\langle\sigma v\rangle_{{\rm DM}\rightarrow{\rm SM}}$ is shown in Fig.~\ref{fig:Sigma_v_a_SM} as a function of the DM mass $m_{\rm DM}$ for different values of $\delta k$, 
$\delta k = 0.1$ (orange solid line), $\delta k = 1$ (green dashed line)
and $\delta k = 10$ (red dotted line). In this picture, 
$\Lambda_{\rm DIR} = 4.5$ TeV, $\Lambda_{\rm IR} = 20$ TeV, 
the first KK graviton mass is $m_1 = 600$ GeV and the radion mass
is $m_{r_2} = 100$ GeV. 

For a light DM mass ($m_{\rm DM} \leq 200$ GeV), we can distinctly
see an enhancement in the cross-section due to the multiplicative
factor $(k_2/k_1)^2$. A larger $\delta k$ corresponds to a larger
cross-section. This behaviour comes from the simple modification
of the amplitude with a virtual radion (as the KK gravitons are significantly off-shell). 
On the other hand, as soon as the DM mass increases (and processes
going through KK graviton resonances become relevant),
the involved $\delta k$ dependence in the amplitude
with virtual KK gravitons pours in and it is not easy to detect
any particular $\delta k$-dependent pattern. Indeed, for very large
DM mass, it seems that the opposite behaviour sets in: the larger
$\delta k$, the smaller the cross-section. 

\begin{figure}
	\centering  \includegraphics[scale=0.55]{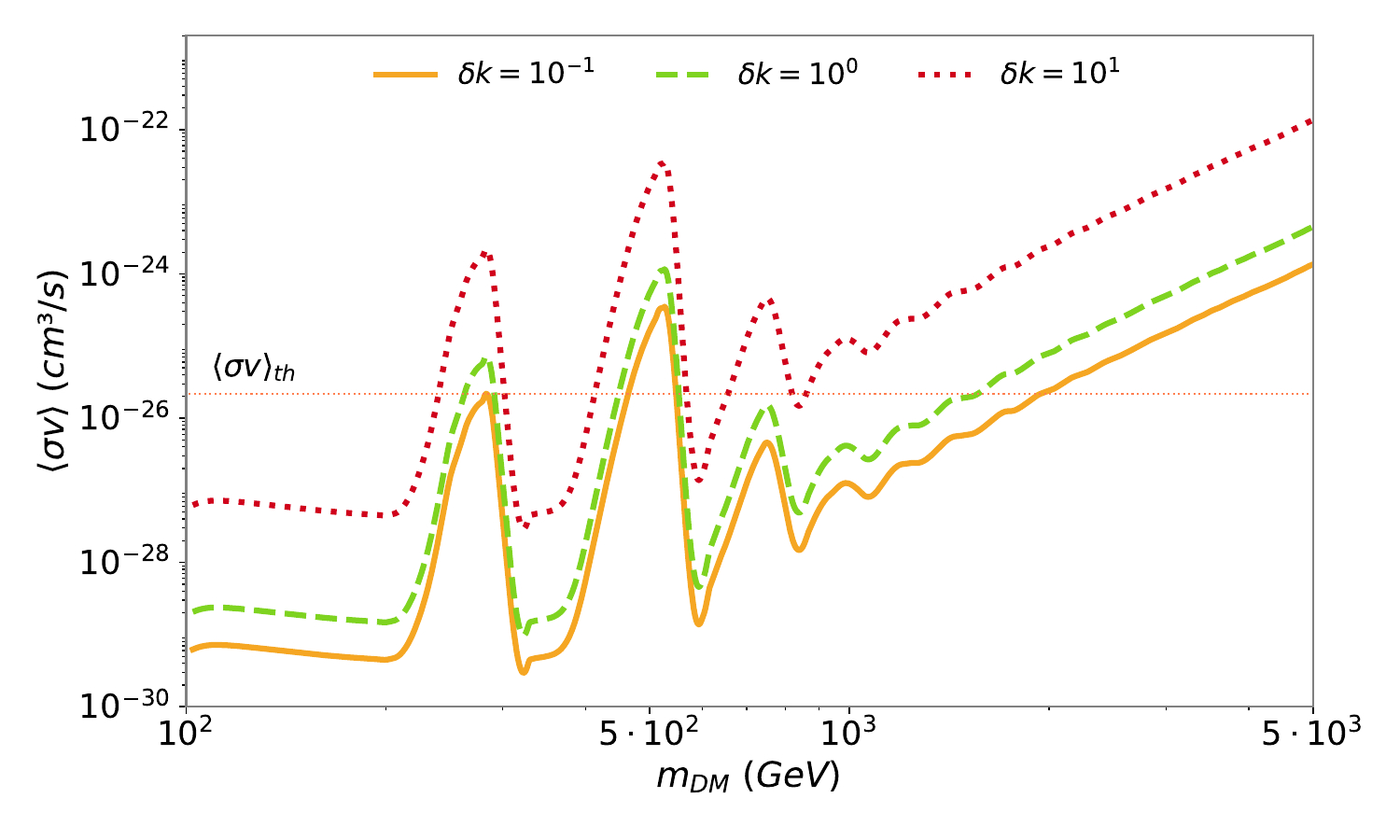} 
    \caption{
    Thermally-averaged annihilation cross-section 
    $\langle\sigma v\rangle_{{\rm DM} \rightarrow G_m \, G_n \, ; G_m \, r \, ;r \, r}$
    as a function of the DM mass $m_{\rm DM}$ for different values of $\delta k$: $\delta k = 0.1$ (orange solid line), $\delta k = 1$ (green dashed line) and $\delta k = 10$ (red dotted line).
    The DIR and IR scales are $\Lambda_{\rm DIR} = 4.5$ TeV and
    $\Lambda_{\rm IR}$ = $20$ TeV, respectively, whereas the
    first KK graviton mass is $m_1 = 600$ GeV and the radion mass
    is $m_{r_2} = 100$ GeV. The value of the thermally-average
    cross-section that would reproduce the observed DM relic abundance,
    $\langle \sigma v\rangle_{\rm th}$, is shown as a red thin dotted 
    horizontal line.    
    }
    \label{fig:Sigma_v_a_Bulk}
\end{figure}

It would then seem that a detailed study of the parameter space
of the three-brane model out of the evanescent limit
could be a difficult task, indeed. This is not the case, though.
In Fig.~~\ref{fig:Sigma_v_a_Bulk} we show the thermally-averaged
cross-section for the annihilation of DM particles into bulk 
particles, ${\rm DM \, DM} \rightarrow G_m \, G_n \, ; G_m \, r$ and $r \, r$. We can see that, for this channel, the dependence on $\delta k$
is very simple: since all vertices are multiplied by
a common factor $(k_2/k_1)^{1/2}$, the cross-section increases
as $(k_2/k_1)^2$. We can see this behaviour in the Figure, 
observing a uniform shift towards higher values of the cross-section 
for any value of the DM mass. 

\begin{figure}
	\centering  \includegraphics[scale=0.4]{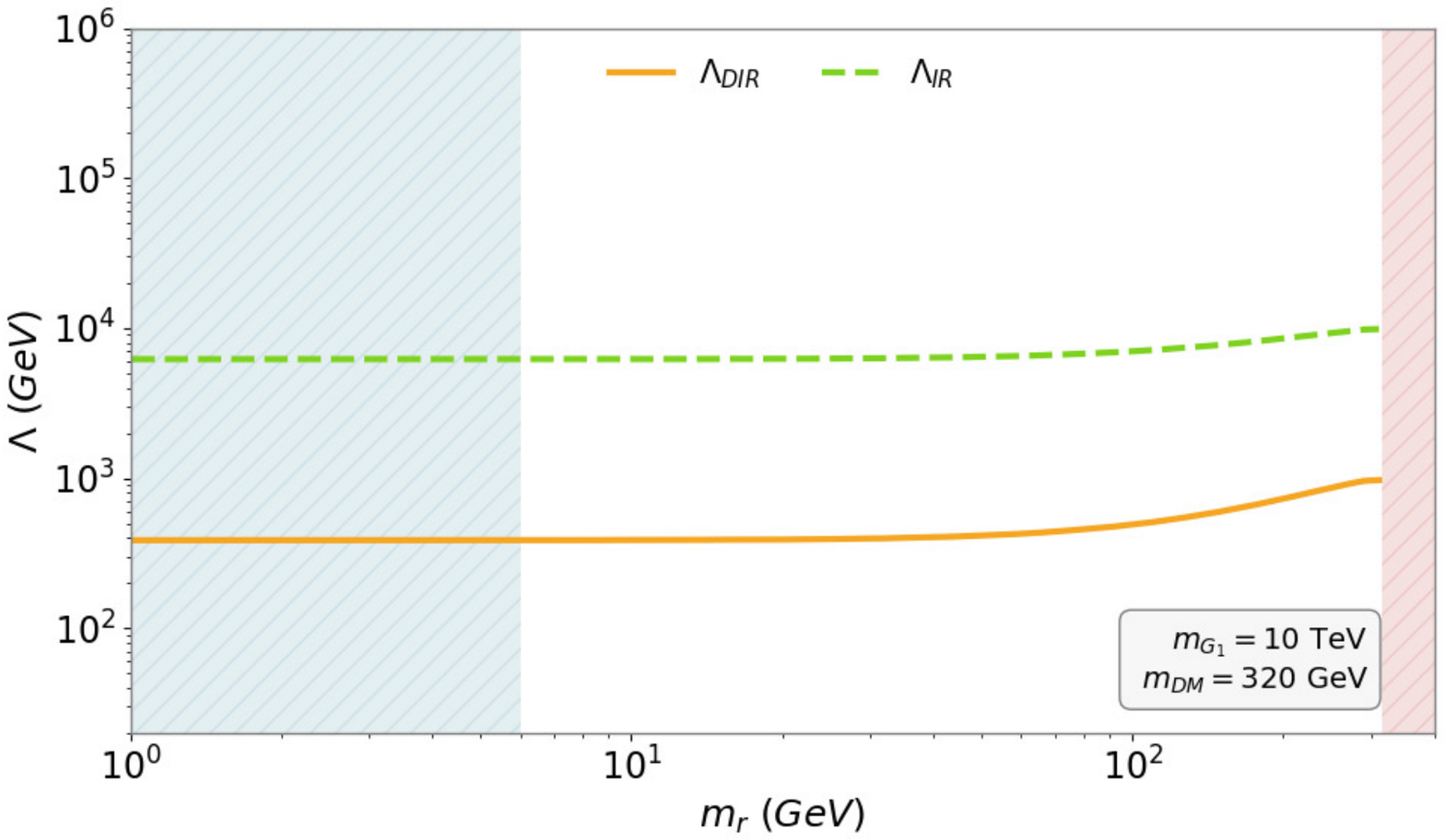} 
    \caption{
   The scales $\Lambda_{\rm DIR}$ (solid orange line) and $\Lambda_{\rm IR}$ (dashed green line) that reproduce the observed DM relic abundance $\Omega_{\rm DM}$ as a function of the radion mass $m_{r_2}$, for $m_{\rm DM} = 320$ GeV, $m_1 = 10$ TeV and $\delta k \ll 1$. 
   The results are taken from Fig.11 of Ref.~\cite{Donini:2025cpl}, 
   albeit casting $\Lambda_{\rm IR}$ in terms of the notation of this paper. The leftmost shaded region is excluded by experiments looking for Direct Detection of Dark Matter. The rightmost shaded region by consistency of the EFT. 
   }
    \label{fig:ourLambdaIR}
\end{figure}

This simple behavior allows us to draw an important conclusion: even out of the evanescent limit, annihilation into bulk particles
will dominate over that into SM particles in the total cross-section,
being the more so the larger $\delta k$. Therefore, the results
of Ref.~\cite{Donini:2025cpl} are still valid for most of the parameter space studied.  In particular, the left panels of Fig.~11 of Ref.~\cite{Donini:2025cpl}, that show the region of the parameter space
that reproduces the observed DM relic abundance while fulfilling experimental bounds, should not change siginificantly.
Notice, however that it is not possible to read the quantitative results out of those figures directly in terms of the parameters used in this paper. This is due to the different definition of $\Lambda_{\rm IR}$ between Ref.~\cite{Donini:2025cpl} and us. Results in Ref.~\cite{Donini:2025cpl} are given at fixed $m_1$ and $\Lambda_{\rm IR} = 1/\bar \xi \, \Lambda_{\rm IR}$ in the plane ($m_{\rm DM}, m_{r_2}$), from which the value 
of $\Lambda_{\rm DIR}$ that gives $\Omega_{\rm DM}$ is computed.
Since $\bar \xi$ depends in turn on $\Lambda_{\rm DIR}$, it is 
clear that an analogous scan using the notation of this paper will correspond to a different plane in the parameter space ($m_{\rm DM}, m_{r_2}, m_1, \Lambda_{\rm IR}$), resulting in a different $\Lambda_{\rm DIR}$. This can be seen in Fig.~\ref{fig:ourLambdaIR}, 
where we show the results of the middle left panel of Fig.~11 of
Ref.~\cite{Donini:2025cpl} for a fixed value of $m_{\rm DM} = 320$ GeV
and $m_1 = 10$ TeV, plotting the value of $\Lambda_{\rm DIR}$ for which
$\Omega_{\rm DM}$ is obtained (orange solid line), from which in 
turn we can compute the corresponding value of $\Lambda_{\rm IR}$
as defined in eq.~(\ref{eq:LambdaIRdef}) of this paper (green dashed line). We can see that, while $\Lambda_{\rm DIR}$ ranges from a few hundreds of GeV to 1 TeV (as in Fig.~11 of Ref.~\cite{Donini:2025cpl}), 
the corresponding $\Lambda_{\rm IR}$ goes from 6 to 10 TeV, approximately (whereas in the Fig.~11 of Ref.~\cite{Donini:2025cpl} is constant, $\Lambda_{\rm IR} = 100$ TeV). 
The leftmost and rightmost shaded regions are excluded by searches in Dark Matter Direct Detection experiments (left) and by consistency
of the Effective Field Theory (right). 

Finally, it is worth commenting on the observable consequences of departing from the evanescent limit. As the ratio $k_2/k_1$ increases within the natural range, the effective couplings of KK gravitons to fields on the IR brane can become moderately stronger, due to the corresponding decrease of the interaction scale. This implies slightly enhanced production rates of the lowest--lying KK modes at colliders, which would translate into mildly stronger bounds from resonance searches. Similarly, DM--nucleon scattering amplitudes mediated by the radion or by KK gravitons inherit the coherent rescaling $\Lambda_{\rm IR},\Lambda_{\rm DIR}\propto (k_2/k_1)^{-1/2}$, producing $\mathcal{O}(1)$ shifts in direct--detection cross sections. None of these effects, however, alter the qualitative structure of the phenomenology: the bulk--dominated annihilation pattern, the dominance of the light radion in direct detection, and the parametric behaviour presented in Ref.~\cite{Donini:2025cpl} remain essentially unchanged, with the results of Fig.~11 of that work simply undergoing a smooth coupling rescaling.

\section{Conclusions} 
\label{sec:conc}

The Randall-Sundrum model has been studied in depth and extensively in its original two-brane proposal. The phenomenological consequences of adding an extra brane, which allows SM particles and DM particles to inhabit different branes, have also been studied, albeit to a lesser extent. In particular, these latter works on a three-brane setup
consider the curvature to the left and to the right of the middle brane
($k_1$ and $k_2$, respectively, with $k_2 \geq k_1$ to avoid tachyonic modes) to be identical. In terms of our notation, this corresponds to $\delta k\ll1$, being $k_2 = k_1 (1 + \delta k)$. This limit has been called in Ref.~\cite{Donini:2025cpl}
the {\em evanescent limit} since, in order to have a stable
background metric for the three-brane setup, the middle brane is forced
to have a tension $f \propto (k_2-k1)$ that, necessarily, vanishes
for $\delta k \to 0$. The evanescent limit is very useful from a technical point of view, as the numerical and analytical computation to be carried on greatly simplify. However, there is no {\em physical}
motivation for $k_2$ to be identical to $k_1$ (although the requirement of not introducing new hierarchies between scales implies that they should be of the same order). 

It was, thus, the first aim of this paper to explore
how the phenomenological results of Ref.~\cite{Donini:2025cpl} are affected by a departure from the evanescent limit. 
To this scope, we have first computed analytically the wave-functions and spectrum of KK gravitons and radions, the latter introduced
following the Goldberger-Wise approach. The phenomenologically
interesting scenario assumes one UV-brane and two IR-branes
(called IR-brane and Deep IR-, or DIR-,brane), with SM matter living
on the IR brane and DM on the DIR one. The two IR-branes are
separated from the UV one by a huge warping factor, $\omega = \exp{(-k_1 L_1)}$, but are separated between themselves by a much smaller one, 
$\xi = \exp{(-k_2 \Delta L)}$. 
This means that $\omega$ is by far the smallest parameter in the setup. We have thus derived expansions in $\omega$
of the coefficients of the KK graviton wave-functions, that allow
for a better understanding of their behaviour as a function of the
other parameters of the model: $\xi$, the KK number $n$ and, most importantly, the ratio of the curvatures $k_2/k_1$, the parameter
we are interested in. Starting from the analytical wave-functions 
of KK gravitons, it is easy to derive then their couplings with matter
on the IR- or the DIR-brane. We have found that, whereas the coupling
of KK gravitons with matter on the DIR brane is simply rescaled
by a factor $(k_2/k_1)^{1/2}$, the coupling with matter on the IR brane
depends on $(k_2/k_1)$ in a involved way.

Once we turn to the radion sector of the model, we have
found that the presence of $N_b$ branes implies 
that $N_b-1$ 4D scalar modes ($r_1$ and $r_2$, for $N_b = 3$) related
to the original bulk graviscalar $h_{55}(x,y)$ are present in the spectrum of the theory (as it was stressed in a part of the literature on this subject). After a detailed analytical computation
of the two ``radions" effective potential and masses, 
we have found, however, that only one of them ($r_2$) is indeed light (as it was found in another part of the related literature). The heavy scalar $r_1$ decouples from the low-lying spectrum and does not interact
with matter neither on the IR- or the DIR-branes. 
We have analytically shown that the two would-be zero modes
of the bulk field $\varphi(x,y)$, added to the model in order to 
implement the Goldberger-Wise mechanism to stabilize the position 
of the branes in the extra-dimension, are heavy, too. They also decouple from the low-lying spectrum and do not play any role
in the phenomenology. The same happens to the corresponding
Kaluza-Klein modes ($\varphi^{(n)}$) that, in the limit of no
back-reaction, have vanishing couplings with matter on either brane. 
The general result that we have found is that, independently
on being in the evanescent limit or out of it, only one light radion
plays a significant role in the phenomenology of the model
(whose mass can be chosen as a free parameter, $m_{r_2}$). 
This was not an obvious result: in particular, since the mass
of the radion $r_1$ was found to be proportional to $1/(k_2 - k_1)$, 
it was not clear what would have happened to the radion spectrum 
for $k_2 \gg k_1$. What we have found in this paper is that the huge
hierarchy between the two radions masses depend also on other parameters of the model apart from $\delta k$, in particular
on the ratio $\epsilon_2/\epsilon_1$ between the bulk mass of the
field $\varphi$ in the two segments to the left and to the right
of the middle brane.

Another important result is the explicit computation of the 
dependence on the ratio $(k_2/k_1)$ of the couplings of the radion 
with matter on the IR or the DIR branes. We have found that, in both
cases, the couplings found in the evanescent limit are just rescaled
by a factor $(k_2/k_1)^{1/2}$. This means that all results
obtained for the radion in the evanescent limit can be understood
in the general (non-evanescent) model as a rescaling of the 
two scales $\Lambda_{\rm DIR, IR} (\delta k = 0) \to (k_2/k_1)^{1/2} \, \Lambda_{\rm DIR, IR}(\delta k \neq 0)$. 

We have also computed the dependence of the couplings between ``bulk"
fields (the radion and the KK gravitons) between themselves, finding
again that they are identical to those in the evanescent limit
but for a common rescaling of the $\Lambda_{\rm DIR}$ scale by
the same factor $(k_2/k_1)^{1/2}$. Together with the previous
results for the couplings of radion and KK graviton with DM on the
DIR-brane, it is therefore easy to proof that the subtle cancellations
between different amplitudes in the processes 
${\rm DM \, DM} \to G_m \, G_n \, ; G_m \, r \, ; r \, r$,
that reduce unitarity violation to ${\cal O}(s)$, still hold.

Using these results, both for the radion and the KK gravitons, we have first computed the decay
widths of a light radion into SM particles and of the first
KK graviton $G_1$. In both cases, we have found that the total 
width scales approximately with $(k_2/k_1)$ for $\delta k$ large enough.
We have then computed the annihilation cross-sections of DM into either SM particles or ``bulk" particles (the radion and the KK gravitons). We have found that the modifications of the coupling between KK gravitons and SM matter do not suffice to alter the main result of Ref.~\cite{Donini:2025cpl}, namely that the total cross-section 
is dominated by ${\rm DM \, DM} \to G_m \, G_n \, ; G_m \, r \, ; r \, r$. Since for these channels the corresponding couplings between the
radion and the KK gravitons and both DM and SM matter are obtained
by a rescaling of $\Lambda_{\rm DIR}$ and $\Lambda_{\rm IR}$, it is 
possible to re-interpret the phenomenological results of Ref.~\cite{Donini:2025cpl} out of the evanescent limit, modifying
accordingly the two scales.

A detailed phenomenological analysis of the model out of the evanescent limit is not the scope of this paper, but the results obtained here
will be useful in order to carry it on. What we can say for the moment is that most of the results obtained in the evanescent limit
should be still valid for $\delta k \leq 1$. For $k_2 \gg k_1$, on the
other hand, a specific computation should be performed. Remind, however, that in the spirit of the Randall-Sundrum approach to 
extra-dimensions, all dimensionful scales should be approximately of the same order, and hierarchies are introduced through
warpings ({\em i.e.} by the geometry of the space-time itself). 
Therefore, it is not `{\em natural}" for $k_2$ to be much larger than
$k_1$, without introducing a new, unjustified, hierarchy in the model. 
It remains, thus, to study the case $\delta k \in [1,10]$ and see
if the increase in $\delta k$ may alter the results of Ref.~\cite{Donini:2025cpl} in some region of the parameter space
where the KK gravitons play a significant role (as their coupling
with SM matter is the only one with a non-trivial dependence on $\delta k$). This will be the scope of further studies in this direction, in order to fully explore the parameter space of the three-brane Randall-Sundrum model.

\vspace{0.5 cm}

\section*{Acknowledgements}
We are grateful to Arturo de Giorgi, Juan Herrero Garc\'{\i}a, 
Giacomo Landini, Nuria Rius Dionis, Roberto Ruiz de Austri, Dipan Sengupta and Stefan Vogl for useful discussions. This work is partially supported by the Spanish \emph{Agencia Estatal de Investigación} through the MICINN/AEI (10.13039/501100011033) grant PID2023-148162NB-C21, the Generalitat Valenciana with
the grant PROMETEO/2022/069, and the \emph{Severo Ochoa} project MCIU/AEI CEX2023-001292-S. 
MGF is supported by MMT24-IFIC-01, that comes from the European Union's Recovery and Resilience Facility-Next Generation, in the framework of the General Invitation of the Spanish Government’s public business entity Red.es to participate in talent attraction and retention programmes within Investment 4 of Component 19 of the Recovery, Transformation and Resilience Plan. AMO acknowledges support from the Generalitat Valenciana programs Plan GenT Excellence Program CIDEGENT/2020/020, PROMETEO/2019/083 and CIACIF/2021/260.

\newpage

\appendix

\section{Stabilization of the two-brane setup: extended version}
\label{sec:GWtwobranes}

First of all, in order to stabilize the two-brane system, a bulk scalar field $\varphi$ is added to the gravitational action. The simplest bulk potential for $\varphi$ is $U(\varphi) = \frac{m^2}{2} \varphi^2$, where $m$  is a parameter with the dimension of a mass.  Two localized potentials $V_j(\varphi)$, with $j$ labelling the two branes, are added to the lagrangian. The full action is, then:
\begin{eqnarray}
{\cal S} = {\cal S_{\rm grav}} &+& \int d^4 x \int_0^{r_c} dy \, \sqrt{\bar G^{(5)}} \, \left ( \frac{1}{2} \, \partial_M \varphi \partial^M \varphi - U(\varphi) \right ) \nonumber \\
&-& \sum_{j={\rm UV, IR}} 
\int d^4x \int_0^{\pi r_c} dy \, \sqrt{- \bar g_j} \, \delta (y - y_j) \, V_j (\varphi) \, ,
\end{eqnarray}
where $y_{\rm UV} = 0$ and $y_{\rm IR} = \pi r_c$ and the brane terms $V_j$ are:
\begin{equation}
\left \{
\begin{array}{l}
V_{\rm UV} (\varphi) = \mu_{\rm UV} \left ( \varphi^2 - \frac{v_{\rm UV}^2}{\pi r_c} \right )^2 \, , \\
\\
V_{\rm IR} (\varphi) = \mu_{\rm IR} \left ( \varphi^2 - \frac{v_{\rm IR}^2}{\pi r_c} \right )^2 \, ,
\end{array}
\right .
\end{equation}
with $v_{\rm UV} > v_{\rm IR}$.
The parameters $\mu_j$ have the dimension of an inverse mass squared, whereas the VEV's $v_j$ have the dimension of a mass. The length $\pi r_c $ has
been introduced so as to normalize properly the VEV's $v_i$. The brane tension terms, $\sigma_j$ in eq.~(\ref{eq:RSbraneterms}), needed to glue together the background metric in the two subregions and fulfill the orbifold symmetries, can be 
written as $\sigma_j = \mu_j v_j^4/\pi^2 r_c^2$. 

We make the following ansatz: 
\begin{equation}
    \varphi (x,y) = \varphi^{(0)} (x, y) + \delta \varphi  (x,y)\, ,
\end{equation}
where:
\begin{equation}
\label{eq:zero modeRS2b}
    \varphi^{(0)} (x,y) =  \left [ \frac{1}{\pi r_c} + \phi^{(0)}_1 (x) \right ] \, \varphi^{(0)} (y) \, , 
\end{equation}
being $\varphi^{(0)}(y)$ the zero mode wave-function, and the 
KK modes can be expanded on 4D fields as usual: 
\begin{equation}
\delta \varphi (x,y) = \sum_{n= 1}^\infty \phi^{(n)} (x) \, \varphi^{(n)} (y) \, .
\end{equation}
Within the decompositions above, all KK mode wave-functions have the same mass dimension, $[\varphi^{(0)}(y)] = [\varphi^{(n)}(y)] = 1/2$.
Notice that the zero mode of the bulk field $\varphi$ is the only field that may develop a $y$-dependent VEV
({\em i.e.} a mode whose associated 4D plane-wave equation of motion has an eigenvalue $m_0 = 0$ with non-trivial eigenfunction in the extra-dimension): 
\begin{equation}
\label{eq:BC2bradion}
\left \{
\begin{array}{l}
\langle \varphi^{(0)}(x,0) \rangle =  \frac{1}{\pi r_c} \, \langle \varphi^{(0)} (0) \rangle \, , \\
\\
\langle \varphi^{(0)}(x,r_c) \rangle =  \frac{1}{\pi r_c} \, \langle \varphi^{(0)} (r_c) \rangle \, .
\end{array}
\right .
\end{equation}

The equations of motion for the bulk field zero and 
KK modes over the background metric $\bar G_{MN}$ 
in the extra-dimension are:
\begin{equation}
\left \{ 
\begin{array}{l}
\left [ \frac{d^2}{dy^2} - 4 k \, \frac{d}{dy} - m^2  \right ] \varphi^{(0)} (y) = 0 \, , \\
\\
\left [ \frac{d^2}{dy^2}  - 4 k \, \frac{d}{dy}  + \left ( m_n^2 \, e^{2 k y} - m^2 \right )  \right ] \varphi^{(n)} (y) = 0 \qquad  n \geq 1 \, ,  
\end{array}
\right .
\end{equation}
for which we find the following solutions in term of the conformal coordinate $z$ defined above: 
\begin{equation}
\label{eq:RS2bradionKKsolution1}
\left \{
\begin{array}{l}
\hat \varphi^{(0)} (z) = \sqrt{\frac{g^3(z)}{k}} \, \left [ C_{1} \, g^{\nu}(z) + C_{2} \, g^{- \nu}(z) \right ]  \, , \\
\\
\hat \varphi^{(n)} (z) = \sqrt{\frac{g^3(z)}{k}}\, \left [ D_{1n} \, Y_{\nu} \left ( \frac{m_n}{k} \, g(z) \right ) + D_{2n} \, J_{\nu} \left ( \frac{m_n}{k} \, g(z) \right ) \right ]  \, , 
\end{array}
\right .
\end{equation}
where $\nu = 2 \sqrt{1 + \frac{m^2}{4 k^2}}$, $m_n$ is the mass of the $n$-th KK mode and $g(z)$ has been defined when deriving the KK gravitons eigenfunctions. In order to get the eigenfunctions in the whole orbifold, we must glue together the solutions found so as to satisfy the boundary conditions 
at $y = 0$ and $y = \pi r_c$, as we have done for the KK gravitons tower. Eventually, we will enforce normalization of the eigenfuntions:
\begin{equation}
\left \{
\begin{array}{l}
\int_{z(0)}^{z(r_c)} dz \, \sqrt{\bar G^{(5)} (z)} \, | \hat \varphi^{(0)} (z) |^2 = 1 \, , \\
\\
\int_{z(0)}^{z(r_c)} dz \, \sqrt{\bar G^{(5)} (z)} \, | \hat \varphi^{(n)} (z) |^2 = 1 \, ,
\end{array}
\right .
\end{equation}
with $\hat \varphi^{(0,n)}(z) = \varphi^{(0,n)}[z(y)] / \sqrt{g(z)} $.

The brane-localized potential terms $V_j$ induce a discontinuity in the derivative of the wave-functions in the extra-dimension at the position of the branes, such that: 
\begin{equation}
    \left \{
    \begin{array}{l}
\left [
\frac{d \, \varphi^{(0)}}{dy}
\right ]_j = 
- \mu_j \left [  \left (  \frac{\langle \varphi^{(0)} (y_j) \rangle}{\pi r_c} \right )^2 -  \frac{v_j^2}{\pi r_c} \right ] 
\, \varphi^{(0)} (y_j) \, , \\
\\
\left [
\frac{d \, \varphi^{(n)}}{dy}
\right ]_j = 
- \mu_j \left [  \left ( \frac{\langle \varphi^{(0)} (y_j) \rangle}{\pi r_c} \right )^2 -  \frac{v_j^2}{\pi r_c} \right ] 
\, \varphi^{(n)} (y_j) \, , 
\end{array}
\right .
\qquad
\begin{array}{l}
\\
 j = {\rm UV, IR} \\
 \end{array}
\end{equation}
where the left hand side gives the discontinuity in the first derivative of the wave-functions in presence of a non-trivial boundary conditions:
\begin{equation}
\left [ \frac{d \, \varphi^{(0,n)}(y)}{dy} \right ]_j = \lim_{\epsilon \to 0} 
\left [ 
\left . \left ( \frac{d \, \varphi^{(0,n)}(y)}{dy} - 2 k \, \varphi^{(0,n)}(y) \right ) \right |_{y_j + \epsilon}  
- \left . \left ( \frac{d \, \varphi^{(0,n)}(y)}{dy} - 2 k \, \varphi^{(0,n)}(y) \right ) \right |_{y_j - \epsilon}
 \right ]  \, .
\end{equation}
In the limit of very large $\mu_j$ (stiff brane), the r.h.s. of the BC's dominates over the l.h.s. and, therefore: 
\begin{equation}
\label{eq:BC2branesfinal}
\left \{
\begin{array}{l}
\langle \varphi^{(0)}(x,0) \rangle = \sqrt{\pi r_c} \, v_{\rm UV} \, , \\
\\
\langle \varphi^{(0)}(x,\pi r_c) \rangle = \sqrt{\pi r_c} \, v_{\rm IR} \, , \\
\\
\langle \varphi^{(n)}(x,0) \rangle = \langle \varphi^{(0)}(x,\pi r_c) \rangle = 0  \, ,
\end{array}
\right .
\end{equation}
where the conditions for $\varphi^{(n)}$ come as a consequence of the first two lines (as the zero mode VEV saturates the bound both at $y = 0$ and $y = \pi r_c$). 

Let's study first the KK modes with $n \geq 1$. In this case, we have at $y = 0$ and $y = \pi r_c$:
\begin{equation}
\label{eq:BCradionKKsolution}
D_{2n} \, J_{\nu} \left ( \frac{m_n}{k} \, g(z) \right ) + D_{1n}  \, Y_{\nu } \left ( \frac{m_n}{k} \, g(z) \right ) = 0 \, , 
\end{equation}
where $m_n$ are the eigenvalues of the 4D equation of motion $(\Box - m_n^2) \phi^{(n)}(x) = 0$.
Using the boundary condition at $y= 0$ and $y = \pi r_c$ for  $\hat \varphi^{(n)} (z)$, we can relate $D_{1n}$ and $D_{2n}$. Under the assumption that the mass spectrum 
for this eigenfunction is: 
\begin{equation}
m_n = k \, (x_{\nu n} + \delta x_{\nu n} ) \, \bar \omega \, ,
\end{equation}
with $x_{\nu n}$ some ${\cal O}(1)$ coefficients to be determined, we can show that:
\begin{equation}
D_{1n} \propto \bar\omega^{2 \nu} \, D_{2n} \, . 
\end{equation}
and, eventually:
\begin{equation}
\left \{
\begin{array}{l}
J_\nu \left ( x_{\nu n} \right ) = 0 \, , \qquad \forall \, n \\
\\
\delta x_{\nu n} = - \frac{D_{1n}}{D_{2n}} \, \frac{Y_\nu \left (x_{\nu n} \right ) }{ J_{\nu + 1} \left (x_{\nu n} \right ) } 
\propto \bar \omega^{2 \nu} \,  \frac{Y_\nu \left (x_{\nu n} \right ) }{J_{\nu + 1} \left (x_{\nu n} \right ) }\, .
\end{array}
\right .
\end{equation}
The first relation states that $x_{\nu n}$ is the $n$-th zero of the Bessel function $J_{\nu} (x)$, whereas the second one
proves {\em a posteriori} that the shifts $\delta x_{\nu n}$ are, indeed, small corrections to the roots $x_{\nu n}$. If the ratio between the bulk mass parameter $m$ and the curvature $k$ is also small we can introduce yet another small parameter, traditionally called $\epsilon = m^2/4 k^2$ in the literature, to get:
\begin{equation}
x_{\nu n} = x_{2 n} + \epsilon \, \left . \frac{J^{(1,0)}(x)}{J_3(x)} \right |_{x = x_{2 n}} + {\cal O} \left ( \epsilon^2 \right ) \, ,
\end{equation}
where $x_{2 n}$ is the $n$-th zero of the Bessel function $J_2(x)$ and 
\begin{equation}
    J^{(1,0)}_m (x_n) = \left . \left ( \lim_{\gamma \to m} \frac{d J_\gamma (x)}{d \gamma} \right ) \right |_{x \to x_n} \, .
\end{equation}
The KK mode wave-functions are, therefore: 
\begin{equation}
\hat \varphi^{(n)} (z) = D_{2n} \, J_\nu \left [ x_{2 n} \, 
\bar \omega g(z) \, \right ] + {\cal O} \left ( \bar \omega^{2 \nu} \right ) \, .
\end{equation}
Normalizing the KK modes wave-function, we get at leading order in $\bar \omega$ and in $\epsilon$: 
\begin{equation}
D_{2n} = 2 \sqrt{2} \frac{k}{x_{2 n} \, J_0 (x_{2 n})} + {\cal O} \left ( \bar \omega^2, \epsilon \right ) \, ,
\end{equation}
so that, after normalization, we get for the action of the bulk scalar KK modes with $n \geq 1$: 
\begin{equation}
{\cal S} [ \phi^{(n)}] = \frac{1}{2} \, \sum_{n = 1}^\infty \, \int d^4 x \, \left ( \partial_\mu \phi^{(n)} \partial^\mu \phi^{(n)} - m_n^2 \phi^{(n)} \, \phi^{(n)} \right )  \, .
\end{equation}
We can see that the bulk scalar KK modes, in the absence of back-reaction, do not interact with brane fields, as their wave-function vanishes at $y = 0, \pi r_c$ \cite{Csaki:2000zn}. 

Let's focus now on the zero mode wave-function. Using the BC's in eq.~(\ref{eq:BC2branesfinal}) we get:
\begin{equation}
\label{eq:zero modecoefficients2b}
\left \{
\begin{array}{l}
C_1  = - \sqrt{\pi k r_c} \, v_{\rm UV} \, \omega^{2 \nu} \, \frac{\left ( 1 - R \, \omega^{2 - \nu} \right ) }{\left ( 1 - \omega^{2 \nu} \right )} \, ,\\
\\
C_2 = \sqrt{\pi k r_c} \, v_{\rm UV} \,  \frac{\left ( 1 - R \, \omega^{2 + \nu} \right ) }{\left ( 1 - \omega^{2 \nu} \right )} \, , 
\end{array}
\right .
\end{equation}
where $R$ is the ratio of the two VEV's, $R = v_{\rm IR}/v_{\rm UV}$. Notice that we do not expect a sizeable hierarchy between the two VEV's (that would be ``unnatural" within the RS1 model perspective, as all scales are supposed to be $\sim M_{\rm P}$).
The ratio $R$, therefore, is expected to be ${\cal O}(1)$.
We point out that in the definition of $C_1$ and $C_2$ we have introduced the warping factor $\omega$, that should not
be confused with $\bar \omega = \exp{ (- \pi k r_c)}$.
The reason will become clear in a few lines.

Plugging these coefficients into  $\varphi^{(0)}(y)$, 
we can integrate the action over the extra-dimension to get: 
\begin{eqnarray}
{\cal S} [ \phi^{(0)} ] = 
\frac{1}{2} \, \int d^4 x \left \{ K (\omega) \partial_\mu \phi^{(0)}(x) \partial^\mu \phi^{(0)} (x)  - V (\omega) \, \left [ \frac{1}{\pi r_c} + \phi^{(0)}(x) \right ] 
\left [ \frac{1}{\pi r_c} + \phi^{(0)} (x) \right ] \right \} \nonumber \\
\end{eqnarray}
where
\begin{equation}
\left \{
\begin{array}{l}
K (\omega) = \int_0^{\pi r_c} dy e^{- 2 k y} \varphi^{(0)} (y) \varphi^{(0)} (y) \, , \\
\\
V (\omega) = \int_0^{\pi r_c} dy e^{- 4 k y} \left [ m^2  \varphi^{(0)} (y) \varphi^{(0)} (y) + \partial_5  \varphi^{(0)} (y) \partial_5 \varphi^{(0)}(y) \right ] \, .
\end{array}
\right .
\end{equation}

The coefficient of the kinetic term is: 
\begin{eqnarray}
K(\omega) &=& \frac{1}{2 k^2 \, \omega^2} \, \left [ 
\frac{C_1^2}{1 + \nu} \, \frac{(1 - \omega^{2\nu+2})}{\omega^{2 \nu}} + \frac{C_2^2}{\nu - 1} (\omega^{2-2 \nu}-1) \, \omega^{2 \nu} + 2 C_1 C_2 (1 - \omega^2)
\right ] \nonumber \\
& = & \frac{(\pi k r_c) \, v_{\rm UV}^2}{2 k^2} \, \frac{\omega^{2 \nu - 2}}{\left ( 1 - \omega^{2 \nu} \right )^2} \, 
\left [ 
\frac{1}{\nu+1} \,  \left (1 - \omega^{2+2 \nu} \right ) \, \left (1 - \omega^{2 - \nu} \, R \right )^2 \right . \nonumber \\
&-& \left . \frac{1}{\nu - 1} \, 
\left (1 - \omega^{2-2 \nu} \right ) \, \left (1 - \omega^{2 + \nu} \, R \right )^2 
- 2  \left (1 -  \omega^2 \right ) \,  \left (1 - \omega^{2 - \nu} R\right ) \, \left (1 - \omega^{2 + \nu} \, R \right ) 
\right ]\, , \nonumber \\
&&
\end{eqnarray}
whereas the effective potential is:
\begin{equation}
V (\omega) = \left ( 1 - \omega^{2 \nu} \right ) \, \left [ (\nu + 2) \frac{C_1^2}{\omega^{2 \nu} } + (\nu - 2) C_2^2 \right ] 
= (\pi k r_c) \, v_{\rm UV}^2  \, F_\nu (\omega) \, ,
\end{equation}
with
\begin{equation}
\label{eq:Ffunction}
F_\nu (\omega) = \frac{1}{\left ( 1 - \omega^{2\nu} \right ) } \, 
 \left [ 
\left ( \nu + 2  \right ) \omega^{2\nu} \, \left ( 1 - \omega^{2 - \nu} \, R \right )^2 +
\left ( \nu - 2 \right )  \, \left ( 1 - \omega^{2 + \nu} \, R \right )^2
\right ] \, .
\end{equation}
 Minimizing  $V(\omega)$  with respect to  $\omega$, we get at leading order in $\epsilon$ the following relation: 
\begin{equation}
\label{eq:minimumRS1twobranes}
\pi k r_c = \frac{1}{\epsilon} \, \ln \left ( \frac{v_{\rm UV}}{v_{\rm IR}}\right ) + {\cal O} \left (\epsilon \right ) + \dots \, ,
\end{equation}
where $\dots$ stand for terms of higher order in $\bar \omega$ and we have expanded $\nu$ at first order in the squared bulk mass parameter $\epsilon$. We can see that the minimization of the potential fixes dynamically the value $\bar \omega$ that we have been using up to now.
This procedure, first outlined in Ref.~\cite{GW1}, relates the value of the warping factor between the UV and the IR brane to the ratio of the VEV's of the bulk field at the location 
of the two-branes, $v_{\rm UV}$ and $v_{\rm IR}$. 

As we explained at the beginning of the previous Section, in addition to the tower of KK gravitons (including the massless graviton zero mode and the massive KK excitations), 
the gravitational sector of the model contains a single scalar field, the zero mode of the graviscalar. Depending 
on the parametrization of the perturbed metric, this
field is called $h_{55}(x)$ or $F(x)$ (see eqs.~(\ref{eq:gravitonexp1}) and (\ref{eq:perturbedmetric}, respectively). Which is the relation between the graviscalar
(either $h_{55}$ or $F$) with the bulk field zero mode 
$\varphi^{(0)}(x)$? Adopting the {\em classical} approach
to GR, it was shown in Ref.~\cite{Csaki:2000zn} that a proper treatment of the perturbations of the metric require to solve the coupled scalar-gravitational equations, that take into account the back-reaction of the $\varphi$ bulk field on the metric (and viceversa). However, as it was shown in Ref.~\cite{GW2} (see also Ref.~\cite{Csaki:1999mp}), the relevant physics can be understood introducing the following ``{\em na\"ive ansatz}" for the metric\footnote{Notice that there is no simple transformation that relates this metric with the one proposed in eq.~(\ref{eq:perturbedmetric}). Although we can in principle
identify $\delta T(x)/r_c$ with $G$, we find that the relation $G = 2 F$ that arises from the Einstein's equation for $R_{\mu 5}$ cannot be fulfilled. 
For this reason, the results obtained using the two ans\"atze usually differ (see Ref.~\cite{Csaki:2000zn}). }:
\begin{eqnarray}
ds^2 &=& \tilde G_{MN} \, dx^M \, dx^N =  e^{-2 \, k \,  \theta \, T(x)} \, g(x) \, dx^2 - T^2(x) \, d \theta^2  \nonumber \\
&=& e^{-2 k r_c \left [1 + \frac{\delta T(x)}{r_c} \right ] \theta} \, g(x) \, dx^2 
- r_c^2  \left [ 1 + \frac{\delta T(x)}{r_c} \right ]^2 d \theta^2 \, ,
\end{eqnarray}
where $T(x)$ has the dimension of a coordinate (and whose VEV is $ \langle T(x) \rangle = r_c$), whilst $\theta \in [0,\pi]$. The metric $g_{\mu \nu} (x) = \eta_{\mu\nu} + \dots$ includes the standard 4D graviton. 

Computing the gravitational action with this metric, we get: 
\begin{equation}
\label{eq:gravactionT}
{\cal S} [T] = - M^3_5 \, \int d^4 x \int_0^{\pi} d \theta \, \sqrt{\tilde G} \, R^{(5)} + \dots \, , 
\end{equation}
where $R^{(5)}$ is the corresponding Ricci scalar and $\dots$ stand for the cosmological constant term and the brane actions whose background
contribution cancel after fine-tuning of the brane tensions (independently of the metric of choice). The determinant is in this case: 
\begin{equation}
\sqrt{\tilde G} = T \, \sqrt{-g} \, e^{-4 k \, T \, \theta} \, , 
\end{equation}
where we remind that $g_{\mu\nu}$ has signature $(+,-,-,-)$, and the Ricci scalar is: 
\begin{equation}
\label{eq:RicciT}
R^{(5)} = R^{(4) }\left(x,\theta\right)
+ 4\, \frac{k\, \theta}{T} \, \left ( 1 - \frac{3}{2} \, k \, \theta \, T \right ) \, e^{2 \, k \, \theta \, T}  \left ( \partial_\mu T \right )^2 
- \frac{2}{T} \left ( 1 - 3 \, k \, \theta \, T \right ) \, e^{2 \, k \, \theta \, T}  \, \Box T   \, , 
\end{equation}
where $R^{(4)}\left(x,\theta\right)$ is the Ricci scalar for the 4D metric $g_{\mu\nu}$. 
Inserting these two expressions into eq.~(\ref{eq:gravactionT}) and integrating over the extra-dimension we get:
\begin{eqnarray}
{\cal S} [T] &=& - \frac{M^3_5}{2k}\, \int d^4 x \sqrt{-g}\, e^{-2 \pi k T} \, \left ( 1 - e^{-2 \pi k T} \right ) \, R^{(4)}\left(x\right)  \nonumber \\
&-& \left ( \frac{M_5^3}{2k} \right ) \, \int d^4 x \sqrt{-g} \, \frac{1}{T^2} \, 
\left [- 1 + e^{-2 \pi k T} + 2 \pi k T \, e^{-2 \pi k T} +
6 \pi^2 k^2 T^2 \, e^{-2 \pi k T}  \right ] \, \left ( \partial_\mu T \right )^2 \nonumber \\
&-& \left ( \frac{M_5^3}{2k} \right ) \, \int d^4 x \sqrt{-g} \, \frac{1}{T} \, 
\left [ 1 - e^{-2 \pi k T} - 6 \pi k T \, e^{-2 \pi k T}
\right ] \, \Box T \, .
\end{eqnarray}
The first term represents the standard 4D gravity, where $g_{\mu\nu} = \eta_{\mu\nu} + h_{\mu\nu}/M_{\rm P}$.

On the other hand, integrating by parts the third term, we get eventually: 
\begin{equation}
{\cal S}_F = {\cal S}_{4D} + \frac{M_5^3}{2k}  \int d^4 x \, \frac{1}{T} \, \partial_\mu T \partial^\mu \left ( \sqrt{-g} \right ) \, \left (1 + \dots \right ) 
+ \, 3 \, \frac{M_5^3}{k} \, \int d^4 x \, \sqrt{-g} \, \left [ \partial_\mu \left ( e^{-\pi k T} \right ) \right ]^2 + \, {\rm S. \, T.} \, ,
\end{equation}
where the $\dots$ stand for the exponentially small correction to the 4D gravitational action due to the extra-dimension, the second term depends on the perturbation of the 4D metric 
$g_{\mu\nu}$, only, and S.T. stands for surface terms that vanish in the limit of an infinite 4D space-time. The third term, on the other hand, represents the kinetic term 
for a massless scalar field. If we normalize the field so as to get a canonical kinetic term: 
\begin{equation}
r(x) = \sqrt{\frac{6 M_5^3}{k} } \, e^{- \pi k T(x)}
= \bar r \, e^{\delta r (x)/ \bar r} \sim \bar r + \delta r (x) \, ,
\end{equation}
where $\delta r (x)$ is the quantum fluctuation of the radion field over the VEV $\bar r$. We get, eventually:
\begin{equation}
{\cal S}[r] = \frac{1}{2} \int d^4 x \sqrt{-g} \, 
\partial_\mu \delta r \, \partial^\mu \delta r \, .
\end{equation}
It is clear that the potential $V(\omega)$ is, indeed, an effective potential for the scalar $r(x)$ that, in turn, is related to the original scalar perturbation of the metric 
($T$, under the metric ansatz used above). Minimizing $V(\omega)$ corresponds to give a VEV to $T$, such that $\langle T \rangle = r_c$. The complete zero mode term of our 4D action is, therefore:
\begin{equation}
{\cal S} [r, \phi^{(0)}] = \frac{1}{2} \int d^4 x \sqrt{-g} \, \left \{ \partial_\mu \delta r \partial^\mu \delta r + K(r) \, \partial_\mu \phi^{(0)} \partial^\mu \phi^{(0)} - V(r) \left [\frac{1}{\pi r_c} + \phi^{(0)} \right ]^2 
\right \} \, .
\end{equation}

If we now expand the potential over the minimum, and we keep
terms of the lagrangian up to second order in the quantum fluctuations, we may recognize a mass term for the radion field: 
\begin{equation}
{\cal S} [r, \phi^{(0)}] = \frac{1}{2} \int d^4 x \sqrt{-g} \, \left \{ \partial_\mu \delta r \partial^\mu \delta r + K(\bar r) \, \partial_\mu \phi^{(0)} \partial^\mu \phi^{(0)} 
- m_r^2 \, \delta r^2 - V(\bar r) \, \phi^{(0)} \phi^{(0)}
\right \} \, ,
\end{equation}
where the radion mass is: 
\begin{eqnarray}
\label{eq:radionmass2branes}
m_r^2 & \sim & \frac{1}{(\pi r_c)^2} \, \left ( \frac{\pi r_c k^2}{M_5^3} \right ) \, \left ( \frac{m^2}{k^2} v_{\rm IR}^2 \right ) \, e^{-2 \pi k r_c} + \dots
= \frac{m^2}{k^2} \, \left ( \frac{k}{\pi r_c} \, \frac{v_{\rm IR}^2 e^{-2 \pi k r_c}}{M_{\rm P}^2} \right ) + \dots \nonumber \\
& \sim & \epsilon^2 \, 
\left ( \frac{k^2}{M_{\rm P}^2 }\right ) \, 
\left ( \frac{v_{\rm IR}^2}{M^2_{\rm P} }\right ) \, 
\left [ \frac{\Lambda^2}{\ln \left ( v_{\rm UV} / v_{\rm IR}\right ) }\right ] + \dots \, ,
\end{eqnarray}
where in the last line we have used eq.~(\ref{eq:minimumRS1twobranes}).
Under the standard assumption that scales in the RS1 setup are ${\cal O} (M_{\rm P})$, we see that the radion mass is warped down by the exponential factor and, therefore, 
it should be ${\cal O} (\Lambda)$. However, due to the coefficient $\epsilon^2$, it can actually be much lighter than 
$\Lambda$ and, therefore, give a phenomenology different
from that of the KK gravitons.

As we have expanded around the minimum of $V(r)$, the would-be mixing between $r$ and the bulk field zero mode $\phi^{(0)}$ necessarily vanishes.
On the other hand, after rescaling the field $\phi^{(0)}$ in order to have a canonically normalized kinetic term, we have for its mass term:
\begin{equation}
m_0^2 =\frac{V(\bar r)}{K(\bar r)} = 2 k^2 \, \left ( \frac{m^2}{k^2} \right ) + \dots \, ,
\end{equation}
where $\dots$ stand for terms of higher-order in $\bar \omega$ and $m^2/k^2$. We can see that, contrary to the case of the radion mass, the leading-order mass of the bulk field zero mode $\phi^{(0)}$ is  ${\cal O}(m)$. Since, in the RS paradigm, all scales are assumed to be ${\cal O} (M_{\rm P})$, the bulk field zero mode is much heavier than the KK modes and than the radion, and it decouples from the low-energy spectrum of the model. 

\section{Stabilization of the three-branes setup: extended version}
\label{sect:threebranesradions}

We will now derive with full details how the two radions wave-functions are obtained and which are their masses and couplings with fields located on the DIR or the IR branes, following Ref.~\cite{Seung1}.

\subsection{Goldberger-Wise potentials and boundary conditions}

First of all, in order to stabilize the three-brane system, a bulk scalar field $\varphi$ is added to the gravitational action. The bulk field has a simple bulk potential, as
in the two-brane case: $U(\varphi) = \frac{m_i^2}{2} \varphi^2$, where $m_1,\,m_2$  are parameters with the dimension of a mass 
(the index $i = 1,2$ refers to the subregions $y \in [0, L_1]$ and $y \in [L_1,L_2]$, respectively). Notice that, as it was
the case for the cosmological constant in each of the two subregions, there is no physical motivation for $m_1$ to be
equal to $m_2$.
Three localized potentials $V_j(\varphi)$, with $j$ labelling the three-branes, are added to the lagrangian. The full action is, then:
\begin{eqnarray}
{\cal S} = {\cal S_{\rm grav}} &+& \int d^4 x \int_0^{L_2} dy \, \sqrt{\bar G^{(5)}} \, \left ( \frac{1}{2} \, \partial_M \varphi \partial^M \varphi - U(\varphi) \right ) \nonumber \\
&-& \sum_{j={\rm UV, IR, DIR}} 
\int d^4x \int_0^{L_2} dy \, \sqrt{- \bar g_j} \, \delta (y - y_j) \, V_j (\varphi) \, ,
\end{eqnarray}
where $y_0 = 0,\, y_1 = L_1$ and $y_2 = L_2$ and the brane terms $V_j$ are:
\begin{equation}
\left \{
\begin{array}{l}
V_{\rm UV} (\varphi) = \mu_{\rm UV} \left ( \varphi^2 - \frac{v_{\rm UV}^2}{r_{\rm UV}}\right )^2 \, , \\
\\
V_{\rm IR} (\varphi) = \mu_{\rm IR} \left ( \varphi^2 - \frac{v_{\rm IR}^2}{r_{\rm IR}}\right )^2 \, , \\
\\
V_{\rm DIR} (\varphi) = \mu_{\rm DIR} \left ( \varphi^2 - \frac{v_{\rm DIR}^2}{r_{\rm DIR}}\right )^2 \, .
\end{array}
\right .
\end{equation}
The parameters $\mu_j$ have the dimension of an inverse mass squared, whereas the field values $v_j$ have the dimension of a mass. The lengths $r_j $ have
been introduced so as to normalize properly the VEVs $v_i$, and they can be chosen convenientently. We will see that a convenient choice is: 
$r_{\rm UV} = r_{\rm IR} = L_1;\  r_{\rm DIR}= \Delta L$. 
Remind that  brane tension terms  $\sigma_j$ in eq.~(\ref{eq:threebranesaction}), needed to glue together the background metric in the two subregions and fulfill the orbifold symmetries, 
must be redefined accordingly, $\sigma_j^4 = \mu_j v_j^4/r_j^2$. 

We make the following ansatz: 
\begin{equation}
    \varphi (x,y) = \varphi^{(0)} (x, y) + \delta \varphi  (x,y)\, ,
\end{equation}
where  $\varphi^{(0)} (x,y)$ may be decomposed as follows:
\begin{equation}
\label{eq:zero modedecomp}
    \varphi^{(0)} (x,y) = \left \{
    \begin{array}{l}
    \left [ \frac{1}{L_1} + \phi^{(0)}_1 (x) \right ] \, \varphi^{(0)}_1 (y) \, , \qquad y \in [0,L_1]  \\
    \\
        \left [ \frac{1}{L_2-L_1} + \phi^{(0)}_2 (x) \right ] \, \varphi^{(0)}_2 (y)  \, , \qquad y \in [L_1,L_2]
        \end{array}
        \right .
\end{equation}
being $\varphi^{(0)}_{1,2}(y)$ the zero mode wave-functions in the subregions 1 and 2, respectively. The 5D-field $\varphi^{(0)}(x,y)$ 
may develop  a $y$-dependent VEV: 
\begin{equation}
\left \{
\begin{array}{l}
\langle \varphi^{(0)}(x,0) \rangle =  \frac{1}{L_1} \, \langle \varphi^{(0)}_1(0) \rangle \, , \\
\\
\langle \varphi^{(0)}(x,L_1) \rangle =  \frac{1}{L_1} \, \langle \varphi^{(0)}_1(L_1) \rangle = \frac{1}{\Delta L} \, \langle \varphi^{(0)}_2(L_1) \rangle \, , \\
\\
\langle \varphi^{(0)}(x,L_2) \rangle =  \frac{1}{\Delta L} \, \langle \varphi^{(0)}_2(L_2) \rangle \, .
\end{array}
\right .
\end{equation}
Notice that, as it was the case for the two-brane setup, this is the only field that may develop a $y$-dependent VEV.
The corresponding 4D-fluctuations, $\phi^{(0)}_{1,2}(x)$, could in principle mix with the projection of the massless graviscalar $h^{(0)}_{55}(x)$ onto each of the two segments and form the two ``radion" fields. For this reason, they should be treated separately from the fluctuation $\delta \varphi (x,y)$, that can be
expanded on 4D fields as usual: 
\begin{equation}
\delta \varphi (x,y) = \sum_{n= 1}^\infty \phi^{(n)} (x) \, \varphi^{(n)} (y) \, .
\end{equation}
Within the decompositions above, all KK modes wave-function have the same mass dimension, $[\varphi^{(0)}_i(y)] = [\varphi^{(n)}(y)] = 1/2$. The equations of motion in the 5th coordinate for its zero mode and the 
KK modes to be solved in the two subregions, $y \in [0, L_1]$ and $y \in [L_1,L_2]$, are: 
\begin{equation}
\left \{ 
\begin{array}{l}
\left [ \frac{d^2}{dy^2} - 4 k_i \, \frac{d}{dy} - m_i^2  \right ] \varphi^{(0)}_i (y) = 0 \, , \\
\\
\left [ \frac{d^2}{dy^2}  - 4 k_i \, \frac{d}{dy}  + \left ( m_n^2 \, e^{2 k_i y} - m_i^2 \right )  \right ] \varphi^{(n)} (y) = 0 \qquad  n \geq 1 \, ,  
\end{array}
\right .
\end{equation}
for which we find the following solutions in term of the conformal coordinate $z$ defined in Sect.~\ref{sect:threebranes}: 
\begin{equation}
\label{eq:radionKKsolution1}
\left \{
\begin{array}{l}
\hat \varphi^{(0)}_1 (z) = \sqrt{\frac{g^3(z)}{k_1}} \, \left [ C_{11} \, g^{\nu_1}(z) + C_{21} \, g^{- \nu_1}(z) \right ]  \, , \\
\\
\hat \varphi^{(n)}_1 (z) = \sqrt{\frac{g^3(z)}{k_1}}\, \left [ D_{11n} \, Y_{\nu_1} \left ( \frac{m_{n_1}}{k_1} \, g(z) \right ) + D_{21n} \, J_{\nu_1} \left ( \frac{m_{n_1}}{k_1} \, g(z) \right ) \right ]  \, , 
\end{array}
\right .
\end{equation}
and
\begin{equation}
\label{eq:radionKKsolution2}
\left \{
\begin{array}{l}
\hat \varphi^{(0)}_2 (z) =  \sqrt{\frac{g^3(z)}{k_2}} \, \left [ C_{12} \, g^{\nu_2}(z) + C_{22} \, g^{- \nu_2}(z) \right ]  \, , \\
\\
\hat \varphi^{(n)}_2 (z) = \sqrt{\frac{g^3(z)}{k_2}} \, \left [ D_{12n} \, Y_{\nu_2} \left ( \frac{m_{n_2}}{k_2} \, g(z) \right ) 
+ D_{22n} \, J_{\nu_2} \left ( \frac{m_{n_2}}{k_2} \, g(z) \right ) \right ]  \, , 
\end{array}
\right .
\end{equation}
where $\nu_i = 2 \sqrt{1 + \frac{m_i^2}{4 k_i^2}}$, $m_{n_i}$ is the mass of the $n$-th KK mode in the subregion $i=1,2$ and $g(z)$ has been defined when deriving the KK gravitons eigenfunctions. In order to get the eigenfunctions for the zero modes and their corresponding KK towers in the whole orbifold, we must glue together the solutions found in each subregion so as to satisfy the boundary conditions at $y = 0$, $y = L_1$ and $y = L_2$, as we have done for the KK gravitons tower. Eventually, we will enforce normalization of the eigenfuntions:
\begin{equation}
\left \{
\begin{array}{l}
\int_{z(0)}^{z(L_2)} dz \, \sqrt{\bar G^{(5)} (z)} \, | \hat \varphi^{(0)} (z) |^2 = 1 \, , \\
\\
\int_{z(0)}^{z(L_2)} dz \, \sqrt{\bar G^{(5)} (z)} \, | \hat \varphi^{(n)} (z) |^2 = 1 \, ,
\end{array}
\right .
\end{equation}
with $\hat \varphi^{(0,n)}(z) = \varphi^{(0,n)}[z(y)] / \sqrt{g(z)} $.

The brane-localized potential terms $V_j$ induce a discontinuity in the derivative of the wave-functions in the extra-dimension at the position of the branes, such that: 
\begin{equation}
\label{eq:kink}
    \left \{
    \begin{array}{l}
\left [
\frac{d \, \varphi^{(0)}}{dy}
\right ]_j = 
- \mu_j \left [  \left (  \frac{\langle \varphi^{(0)}_i (y_j) \rangle}{l_i} \right )^2 -  \frac{v_j^2}{r_j} \right ] 
\, \varphi^{(0)} (y_j) \, , \\
\left [
\frac{d \, \varphi^{(n)}}{dy}
\right ]_j = 
- \mu_j \left [  \left ( \frac{\langle \varphi^{(0)} (y_j) \rangle}{l_i} \right )^2 -  \frac{v_j^2}{r_j} \right ] 
\, \varphi^{(n)} (y_j) \, , 
\end{array}
\right .
\qquad
\begin{array}{l}
 j = {\rm UV, IR, DIR} \\
 \\
 l_1 = L_1; \, l_2 = \Delta L 
 \end{array}
\end{equation}
where the left hand side gives the discontinuity in the first derivative of the wave-functions in presence of a non-trivial boundary conditions:
\begin{equation}
\left [ \frac{d \, \varphi_i^{(0,n)}(y)}{dy} \right ]_j = \lim_{\epsilon \to 0} 
\left [ 
\left . \left ( \frac{d \, \varphi_i^{(0,n)}(y)}{dy} - 2 k_i \, \varphi_i^{(0,n)}(y) \right ) \right |_{y_j + \epsilon}  
- \left . \left ( \frac{d \, \varphi_{i^\prime}^{(0,n)}(y)}{dy} - 2 k_{i^\prime} \, \varphi_{i^\prime}^{(0,n)}(y) \right ) \right |_{y_j - \epsilon}
 \right ]  \, ,
\end{equation}
with $i, i^\prime$ labelling the subregions to the right and to the left of each brane location, respectively.  Notice that, 
taking into account periodicity of the orbifold under $y \to y + 2 L_2$,  the two subregions bordering the brane at $y = 0$ and $y = L_2$  are the same ({\em i.e.} $i = i^\prime$, 
whereas for $y = L_1$ we have $i = 2$ and $i^\prime = 1$). The normalizations of the VEV of the field $\varphi^{(0)}(x,y)$ are $L_1$ and $\Delta L$, respectively, as in 
eq.~(\ref{eq:zero modedecomp}). 

On the r.h.s. of the boundary conditions on the derivative of the wave-function only the VEV of the zero mode, $ \langle \varphi^{(0)}(y_j) \rangle$, is present. 
This is a consequence of the
fact that, after compactification, this is the only field that can acquire a VEV. If we take the limit in which
the dimensionless quantities $\mu_j v_j^2 \gg 1$, then the r.h.s.'s of the boundary conditions in eq.~(\ref{eq:kink}) dominates over the corresponding l.h.s.'s.
Therefore, the VEV of the zero mode solution $\varphi^{(0)}_i$ at the brane location $y_j$ is driven to the parameter that we have included in the brane-localized potentials $V_i$, 
\begin{equation}
\label{eq:zero modeVEV}
\left \{
\begin{array}{l}
\langle \varphi^{(0)}_1 (0) \rangle =  \sqrt{g (0)  } \, 
\langle\hat \varphi^{((0)}_1 (0) \rangle =
\sqrt{\frac{L^2_1}{r_{\rm UV}}} \, v_{\rm UV} = 
\sqrt{L_1} \, v_{\rm UV}\, ; \\
\\
\langle \varphi^{(0)}_1 (L_1) \rangle =  
\sqrt{g \left ( L_1 \right ) } \, 
\langle \hat \varphi^{((0)}_1 (L_1) \rangle= 
\sqrt{\frac{L^2_1}{r_{\rm IR}}} \, v_{\rm IR} = 
\sqrt{L_1} \, v_{\rm IR} \, ; \\
\\
\langle \varphi^{(0)}_2 (L_1) \rangle =  
\sqrt{g \left ( L_1  \right ) } \, 
\langle \hat \varphi^{((0)}_2 (L_1) \rangle =  
\sqrt{\frac{\Delta L^2}{r_{\rm IR}}} \, v_{\rm IR} = 
\sqrt{\frac{\Delta L^2}{L_1}} \, v_{\rm IR} \, ; \\
\\
\langle \varphi^{(0)}_2 (L_2) \rangle =  
\sqrt{g \left (  L_2 \right ) } \, 
\langle \hat \varphi^{((0)}_2 (L_2) \rangle =  
\sqrt{\frac{\Delta L^2}{r_{\rm DIR}}} \, v_{\rm DIR} = 
\sqrt{\Delta L} \, v_{\rm DIR} \, .
\end{array}
\right .
\end{equation}
Notice that the values of the zero mode wave-functions at $y = L_1$ may differ, due to their different normalization. In the same limit, 
the right-hand side of the boundary conditions in eq.~(\ref{eq:kink}) forces the wave-function of the KK modes of the scalar field $\varphi$
to be: 
\begin{equation}
\label{eq:KKmodeVEV}
\left \{
\begin{array}{l}
\varphi^{(n)}_1 (0)  = \varphi^{(n)}_1 (L_1) = 0 \, , \\
\\
\varphi^{(n)}_2 (L_1)  = \varphi^{(n)}_2 (L_2) = 0 \, . 
\end{array}
\right . \qquad \forall \, n \geq 1
\end{equation}

\subsection{Wave-functions of the KK modes of the bulk scalar}

We first compute the wave-functions and the masses of the KK modes
of the bulk field $\varphi$. Using the solutions in the two subregions that we have derived in the limit of no back-reaction, eq.~(\ref{eq:radionKKsolution1}), we get for the BC's: 
\begin{equation}
\label{eq:BCradionKKsolution}
D_{2in} \, J_{\nu_i } \left ( \frac{m_{n_i}}{k_i} \, g(z_j) \right ) + D_{1in}  \, Y_{\nu_i } \left ( \frac{m_{n_i}}{k_i} \, g(z_j) \right ) = 0 \, , 
\end{equation}
where $i= 1,2$  and  $z_j = z(0), z(L_1)$ or $z(L_2)$, depending if we are in subregion 1 or 2.
Using the boundary condition at $y= 0$ and $y = L_1$ for  $\hat \varphi^{(n)}_1 (z)$, we can relate the $D_{11n}$ and $D_{21n}$. Under the assumption that the mass spectrum 
for this eigenfunction is: 
\begin{equation}
m_{n_1} = k_1 \, (x^n_{\nu_1} + \delta x_{\nu_1} ) \, \bar \omega \, ,
\end{equation}
with $x^n_{\nu_1}$ to be ${\cal O}(1)$ coefficients to be determined, we can show that
\begin{equation}
D_{11n} \propto \bar \omega^{2 \nu_1} D_{21n} \, . 
\end{equation}
From this, we can immediately show (as for the two-brane setup) that: 
\begin{equation}
\left \{
\begin{array}{l}
J_{\nu_1} \left ( x^n_{\nu_1} \right ) = 0 \, , \qquad \forall \, n \\
\\
\delta x_{\nu_1} = - \frac{D_{11n}}{D_{21n}} \, \frac{Y_{\nu_1} \left (x^n_{\nu_1} \right ) }{ J_{\nu_1 + 1} \left (x^n_{\nu_1} \right ) } 
\propto \bar \omega^{2 \nu_1} \,  \frac{Y_{\nu_1} \left (x^n_{\nu_1} \right ) }{J_{\nu_1 + 1} \left (x^n_{\nu_1} \right ) }\, ,
\end{array}
\right .
\end{equation}
{\em i.e.} $x^n_{\nu_1}$ is the $n$-th zero of the Bessel function $J_{\nu_1} (x)$. If the ratio between the bulk mass parameter $m_1$ and the curvature $k_1$ is also small, 
we get:
\begin{equation}
x^n_{\nu_1} = x_{2n} + \epsilon_1 \, \left . \frac{J^{(1,0)}(x)}{J_3(x)} \right |_{x = x_{2n}} + {\cal O} \left ( \epsilon_1^2 \right ) \, ,
\end{equation}
where $x_{2n}$ is the $n$-th zero of the Bessel function $J_2(x)$, $\epsilon_1 = m_1^2/4 k_1^2$ and 
\begin{equation}
    J^{(1,0)}_m (x_n) = \left . \left ( \lim_{\gamma \to m} \frac{d J_\gamma (x)}{d \gamma} \right ) \right |_{x \to x_n} \, .
\end{equation}
In order to compute the wave-function $\varphi^{(n)}_2 (z)$ we use the boundary condition at $y= L_1$ and $y = L_2$. Under the assumption that the mass spectrum 
for this eigenfunction is: 
\begin{equation}
m_{n_2} = k_2 \, \left ( x^n_{\nu_2} + \delta x_{\nu_2} \right )  \, \bar \omega \, \bar \xi \, ,
\end{equation}
we get
\begin{equation}
D_{12n} \propto  \bar \xi^{2 \nu_2} D_{22n}
\end{equation}
and
\begin{equation}
\left \{
\begin{array}{l}
J_{\nu_2} \left ( x^n_{\nu_2} \right ) = 0 \, , \qquad \forall n \\
\\
\delta x_{\nu_2} = - \frac{D_{12n}}{D_{22n}} \, \frac{Y_{\nu_2} \left (x^n_{\nu_2} \right ) }{ J_{\nu_2 + 1} \left (x^n_{\nu_2} \right ) } 
\propto \bar \xi^{2 \nu_2} \,  \frac{Y_{\nu_2} \left (x^n_{\nu_2} \right ) }{J_{\nu_2 + 1} \left (x^n_{\nu_2} \right ) }\, ,
\end{array}
\right .
\end{equation}
{\em i.e.} $x^n_{\nu_2}$ are the zeroes of the Bessel function $J_{\nu_2} (x)$. If the ratio between the bulk mass parameter $m_2$ and the curvature $k_2$ is also small, 
we get:
\begin{equation}
x^n_{\nu_2} = x_{2n} + \epsilon_2 \, \left . \frac{J^{(1,0)}(x)}{J_3(x)} \right |_{x = x_{2n}} + {\cal O} \left ( \epsilon_2^2 \right ) \, ,
\end{equation}
where $\epsilon_2 = m_2^2/4 k_2^2$.

Remind that a trivial solution $D_{i1n} = D_{i2n} = 0$ is also consistent with the BC's. It is possible to show \cite{Seung1} that two independent solutions 
on the whole segment $y \in [0,L_2]$ for the scalar KK modes can be built by taking:
\begin{equation}
\label{eq:radion1n}
\hat \varphi^{(n)}_1(z) = 
\left \{
\begin{array}{l}
D_{21n} \, \sqrt{\frac{g^3(z)}{k_1}} \, \left \{  J_{\nu_1} \left ( \frac{m_{n_1}}{k_1} \, g(z) \right ) + {\cal O} (\bar \omega^4) \right \} \, , \qquad( {\rm for} \; y \in [0,L_1]) \\
\\
0 \, , \qquad ({\rm for} \; y \in [L_1,L_2])
\end{array}
\right .
\end{equation}
and 
\begin{equation}
\label{eq:radion2n}
\hat \varphi^{(n)}_2(z) = 
\left \{
\begin{array}{l}
0 \, , \qquad ({\rm for} \; y \in [0,L_1])  \\
\\
D_{22n} \, \sqrt{\frac{g^3(z)}{k_2}} \, \left \{  J_{\nu_2} \left ( \frac{m_{n_2}}{k_2} \, g(z) \right ) + {\cal O} (\bar \xi^4) \right \} \, .  \qquad ({\rm for} \; y \in [L_1,L_2]) \\
\end{array}
\right .
\end{equation}
Being the complete solution for $\hat \varphi^{(n)}_1(z)$ vanishing in subregion 2 (and viceversa), the normalization can be computed integrating 
over $y \in [0, L_1]$ ($y \in [L_1,L2]$), only. We find: 
\begin{equation}
\left \{
\begin{array}{l}
D_{21n} \simeq \sqrt{6} \, k_1 \, \bar \omega^2 \, \frac{x^n_{\nu_1} }{f^n_{\nu_1} \,  J_{\nu_1+1} \left ( x^n_{\nu_1} \right )} \, , \\
\\
D_{22n} \simeq \sqrt{6} \, k_2 \, \left ( \bar \xi \, \bar \omega \right )^2  \frac{x^n_{\nu_2} }{f^n_{\nu_2}  \, J_{\nu_2+1} \left ( x^n_{\nu_2} \right )} \, ,
\end{array}
\right .
\end{equation}
where
\begin{equation}
f^n_{\nu_i} = \left [ (x^n_{\nu_i})^2 + 2 \nu_i^2 - 2  \right ]^{1/2} \, .
\end{equation}

Notice that the two sets of KK towers are orthogonal between them, as they are only non-vanishing in subregion 1 or 2 and, thus: 
\begin{equation}
\int_{z(0)}^{z(L_2)} dz \, \sqrt{\bar G^{(5)} (z)} \, \hat \varphi_1^{(n)\star}(z) \, \hat \varphi_2^{(m)}(z) = 0 \, .
\end{equation}

Once we have derived the wave-functions of the $n$-th  $\varphi$ 
KK modes, we can compute their couplings with fields localized at the UV, IR or DIR branes. The couplings 
of two scalar KK modes $m$ and $n$ with a 4D field at $y = 0, L_1$ or $L_2$ are given by: 
\begin{eqnarray}
\lambda^j_{mn} &=& \int_0^{L_2} dy \, \sqrt{\bar G^{(5)} (y)} \, \delta (y - y_j) \, \varphi^{(m)}(y) \, \varphi^{(n)}(y) \nonumber \\
&=& 
\int_{z(0)}^{z(L_2)} dz \, \sqrt{\bar G^{(5)} (z)} \, \delta (z - z_j) \, \hat \varphi^{(m)}(z) \, \hat \varphi^{(n)}(z) \, .
\end{eqnarray}
Within the {\em probe approximation}, it can be seen 
that $\varphi$ KK modes do not couple to fields on any of the branes
in the setup, as their wave-functions vanish at $y = 0, L_1$ and $L_2$
due to the boundary conditions in eq.~(\ref{eq:KKmodeVEV}). Only once
back-reaction effects are included does a non-vanishing coupling of these modes to brane fields arise, suppressed by the back-reaction parameter $\ell^2$ (see, {\em e.g.}, Ref.~\cite{chivukula2024limitskaluzakleinportaldark}).

\subsection{The GW effective potential and the two ``radion" fields wave-functions}

When studying the zero mode wave-functions a different procedure must be considered. In the absence of the GW localized boundary terms, the only solution compatible
with $m_0 = 0$ is the trivial solution $C_{1i} = C_{2i} = 0$ in both subregions.  However,  in the limit $\mu_i v_i^2 \gg 1$, the r.h.s. of the BC's 
force the zero mode wave-function to assume the values that we have fixed at the brane locations, eq.~(\ref{eq:zero modeVEV}). We get for the coefficients of the wave-function: 
\begin{equation}
\left \{
\begin{array}{l}
C_{11} = \sqrt{k_1 L_1} \, \omega^{2+\nu_1} \, \frac{\left [ 1 - \omega^{\nu_1 - 2} (v_{\rm UV}/v_{\rm IR}) \right ]}{1 - \omega^{2 \nu_1}} \, v_{\rm IR} \, , \\
\\
C_{21} = \sqrt{k_1 L_1} \, \frac{\left [ 1 - \omega^{\nu_1+2} (v_{\rm IR}/v_{\rm UV}) \right ]}{1 - \omega^{2 \nu_1}}  \,  v_{\rm UV}
\end{array}
\right .
\end{equation}
in the subregion $y \in [0, L_1]$, and: 
\begin{equation}
\left \{
\begin{array}{l}
C_{12} = \sqrt{k_2 \Delta L}  \, ( \xi \omega)^{2 + \nu_2} \, \frac{\left [ 1 - \sqrt{\frac{k_1}{k_2}} \, \sqrt{\frac{k_2 \Delta L }{k_1 L_1}} \, 
\xi^{\nu_2 - 2} (v_{\rm IR}/v_{\rm DIR}) \right ]}{1 - \xi^{2 \nu_2}} \, v_{\rm DIR} \, , \\
\\
C_{22} = \sqrt{\frac{k^2_2 \Delta L^2}{k_1 L_1}} \, \sqrt{\frac{k_1}{k_2}} \, \omega^{2 - \nu_2} \, \frac{\left [ 1 - \sqrt{\frac{k_2}{k_1}} \, \sqrt{\frac{k_1 L_1 }{k_2 \Delta L}} \, 
\xi^{\nu_2 + 2} (v_{\rm DIR}/v_{\rm IR}) \right ]}{1 - \xi^{2 \nu_2}} \, v_{\rm IR} 
\end{array}
\right .
\end{equation}
in the subregion $y \in [L_1, L_2]$. It is useful to rescale the brane VEV's of the bulk field $\varphi$ as follows: 
\begin{equation}
\left \{
\begin{array}{l}
\tilde v_{\rm UV} = \sqrt{k_1 L_1} \, v_{\rm UV} \, , \\
\\
\tilde v_{\rm IR} = \sqrt{k_1 L_1} \, v_{\rm IR} \, , \\
\\
\tilde v_{\rm DIR} = \sqrt{\frac{k_2}{k_1}} \, \sqrt{\frac{k_1 L_1}{k_2 \Delta L}} \, \sqrt{k_1 L_1} \, v_{\rm DIR} \, . \\
\end{array}
\right . 
\end{equation}
Eventually, 
\begin{equation}
\left \{
\begin{array}{l}
C_{11} = \frac{\omega^{2 \nu_1}}{1 - \omega^{2 \nu_1}} \, \left [  \omega^{2-\nu_1} R_1 - 1 \right ] \, \tilde v_{\rm UV} \, , \\
\\
C_{21} = \frac{1}{1 - \omega^{2 \nu_1}} \, \left [ 1 - \omega^{2+\nu_1} R_1 \right ] \,  \tilde v_{\rm UV} \, 
\end{array}
\right .
\end{equation}
and: 
\begin{equation}
\label{eq:zero modecoefficients}
\left \{
\begin{array}{l}
C_{12} = \sqrt{\frac{k_1}{k_2}} \, \left ( \frac{k_2 \Delta L}{k_1 L_1 }\right ) \, \frac{\omega^{2+\nu_2} \xi^{2 \nu_2} }{1 - \xi^{2 \nu_2}}
 \, \left [\xi^{2 - \nu_2} R_2 - 1 \right ] \, \tilde v_{\rm IR} \, , \\
\\
C_{22} = \sqrt{\frac{k_1}{k_2}} \, \left ( \frac{k_2 \Delta L}{k_1 L_1 }\right ) \, \frac{\omega^{2-\nu_2}}{1 - \xi^{2 \nu_2}}
 \, \left [1 - \xi^{2 + \nu_2} R_2 \right ] \, \tilde v_{\rm IR} \, ,
\end{array}
\right .
\end{equation}
where $R_1 = \tilde v_{\rm IR}/\tilde v_{\rm UV}$ and $R_2 = \tilde v_{\rm DIR}/\tilde v_{\rm IR}$. 

Notice that, as it was the case for the two-brane setup 
in Sect.~\ref{sec:GWtwobranes}, we are now considering
$\omega$ and $\xi$ as free parameters to be fixed by a minimization procedure. In order to stabilize the distances between the three branes, we need first to compute
the effective potential for the zero modes of the bulk field, integrating over the extra-dimension the action:
\begin{eqnarray}
{\cal S}_\varphi &=& \int d^4 x \int_0^{L_2} dy \, \sqrt{\bar G^{(5)}} \, \left \{ 
\frac{1}{2}  \bar G^{(5) \, MN} \partial_M \varphi \, \partial_N \varphi - \frac{m_i^2}{2} \varphi^2
\right \} \nonumber \\
&=& 
\frac{1}{2} \, \int d^4 x \left \{ K_1 (\omega) \partial_\mu \phi^{(0)}_1(x) \partial^\mu \phi^{(0)}_1 (x)  - V_1 (\omega) \, \left [ \frac{1}{L_1} + \phi^{(0)}_1(x) \right ] 
\left [ \frac{1}{L_1} + \phi^{(0)}_1 (x) \right ] \right \} \nonumber \\
&+&
\frac{1}{2} \, \int d^4 x \left \{ K_2 (\omega, \xi) \partial_\mu \phi^{(0)}_2(x) \partial^\mu \phi^{(0)}_2 (x)  - V_2 (\omega, \xi) \, 
\left [ \frac{1}{\Delta L} + \phi^{(0)}_2(x) \right ] \left [ \frac{1}{\Delta L} + \phi^{(0)}_2 (x) \right ] \right \} \, , \nonumber \\
\end{eqnarray}
where
\begin{equation}
\left \{
\begin{array}{l}
K_1 (\omega) = \int_0^{L_1} dy \sqrt{\bar G^{(5)}} \, \left ( \frac{1}{4} \, \eta_{\mu\nu} \, \bar G^{(5)\mu\nu} \right ) \varphi^{(0)}_1 (y) \varphi^{(0)}_1 (y) \, , \\
\\
V_1 (\omega) = \int_0^{L_1} dy \sqrt{\bar G^{(5)}} \, \left [ m_1^2  \varphi^{(0)}_1 (y) \varphi^{(0)}_1 (y) - \bar G^{(5)55} \, \partial_5  \varphi^{(0)}_1 (y)\, \partial_5 \varphi^{(0)}_1(y) \right ] \, ,
\end{array}
\right .
\end{equation}
and 
\begin{equation}
\left \{
\begin{array}{l}
K_2 (\omega,\xi) =  \int_{L_1}^{L_2} dy \sqrt{\bar G^{(5)}} \,  \left ( \frac{1}{4} \, \eta_{\mu\nu} \, \bar G^{(5)\mu\nu} \right )\varphi^{(0)}_2 (y) \varphi^{(0)}_2(y) \, , \\
\\
V_2(\omega,\xi) =  \int_{L_1}^{L_2} dy \sqrt{\bar G^{(5)}} \left [ m_2^2  \varphi^{(0)}_2 (y) \varphi^{(0)}_2 (y) - \bar G^{(5)55} \, \partial_5  \varphi^{(0)}_2 (y) \, \partial_5 \varphi^{(0)}_2 (y) \right ] \, .
\end{array}
\right .
\end{equation}
 
We must now identify the ``radion fields" and minimize the effective potentials $V_1(\omega)$ and $V_2(\omega, \xi)$ with respect to the dynamical fields.
As we have done for the two-brane case (see Sect.~\ref{sec:GWtwobranes}), we introduce a ``{\em na\"ive}" metric ansatz for each
of the two regions of the orbifold:
\begin{equation}
\label{eq:GWmetric2branes_App}
G_{MN}  \, dx^M  \, dx^N = 
 \left \{
\begin{array}{l} 
e^{-2 k_1 \theta \, T_1(x) } \, g_{\mu\nu}(x) \, dx^\mu dx^\nu 
- T_1(x)^2 d \theta^2 \, , \qquad \qquad \qquad \qquad \; \theta \in [0, \theta_1]   \\
\\ 
e^{-2 \left [  k_2 (\theta - \theta_1) \, T_2(x) + k_1 \theta_1 \, T_1(x) \right ]  } \, g_{\mu\nu}(x) \, dx^\mu dx^\nu 
- T_2(x)^2 d \theta^2 \, , \qquad \theta \in [\theta_1,1] 
\end{array}
\right .
\end{equation}
where both $T_1(x)$ and $T_2(x)$ have the dimension of a coordinate and whose VEV's are 
$ \theta_1 \, \langle T_1(x) \rangle = L_1$ and 
$ (1-\theta_1) \, \langle T_2(x) \rangle = \Delta L$.
Remind that the two fields are not independent degrees of freedom: we have one massless graviscalar per segment of the orbifold, $\Sigma_1$ and $\Sigma_2$, intertwined by
the boundary conditions at $y = L_1$. 
As for the two-brane case, standard 4D gravity is implied by the induced metric  $g_{\mu \nu} (x) = \eta_{\mu\nu} + \dots$. 
The determinant of the metric is: 
\begin{equation}
\sqrt{\tilde G} = 
\left \{
\begin{array}{ll}
T_1 (x) \, \sqrt{-g} \, e^{-4 k_1 \, \theta \, T_1(x) } \, ,& \qquad  \theta \in [0, \theta_1]   \\
& \\
T_2 (x) \, \sqrt{-g} \, e^{-4 \left [ k_2 \, (\theta - \theta_1) \, T_2(x) + k_1 \, \theta_1 \, T_1(x) \right ] } \, , & \qquad \theta \in [\theta_1, 1]
\end{array}
\right .
\end{equation}
where, again,  $g_{\mu\nu}$ has signature $(+,-,-,-)$. The Ricci scalar in the two segments is: 
\begin{equation}
\label{eq:RicciT1}
\left . R^{(5)} \right |_{\Sigma_1} = R^{(4)}\left(x,\theta\right) 
+ 4\, \frac{k_1\, \theta}{T_1} \, \left ( 1 - \frac{3}{2} \, k_1 \, \theta \, T_1 \right ) \, e^{2 \, k_1 \, \theta \, T_1}  \left ( \partial_\mu T_1 \right )^2 
- \frac{2}{T_1} \left ( 1 - 3 \, k_1 \, \theta \, T_1 \right ) \, e^{2 \, k_1 \, \theta \, T_1}  \, \Box T_1   \, , 
\end{equation}
and
\begin{eqnarray}
\label{eq:RicciT2}
\left . R^{(5)} \right |_{\Sigma_2} = R^{(4) }\left(x,\theta\right)
&+& 4\, \frac{k_1 \, \theta_1}{T_2} \, \left ( 1 - 3 \, k_2 \, (\theta-\theta_1) \, T_2 \right ) \, e^{2 \, \rho (\theta,x)}  \left ( \partial_\mu T_1 \right )  \, \left ( \partial^\mu T_2 \right )
\\
&+& 4\, \frac{k_2 \, (\theta - \theta_1)}{T_2} \, \left ( 1 - \frac{3}{2} \, k_2 \, (\theta-\theta_1) \, T_2 \right ) \, e^{2 \, \rho (\theta, x)}  \left ( \partial_\mu T_2 \right )^2 \nonumber \\
&-& 6 \, k_1^2 \, \theta_1^2 \, e^{2 \, \rho (\theta, x)}  \left ( \partial_\mu T_1 \right )^2 + 6 \, k_1 \, \theta_1 \,  e^{2 \, \rho(\theta, x)}  \, \Box T_1 \nonumber \\
&-& \frac{2}{T_2} \left ( 1 - 3 \, k_2 \, (\theta-\theta_1) \, T_2 \right ) \, e^{2 \, \rho (\theta, x)}  \, \Box T_2  \, ,  \nonumber
\end{eqnarray}
where $R^{(4)}\left(x,\theta\right)$ is the Ricci scalar for the 4D metric $g_{\mu\nu}$ in the two segments and we have introduced the function $\rho (\theta,x)$ as a short-hand for
$\rho (\theta, x) = k_2 \, (\theta - \theta_1) \, T_2(x) + k_1 \, \theta_1 \, T_1(x)$. 

Integrating over the extra-dimension we get for the first segment:
\begin{eqnarray}
{\cal S}_1 [T_1] &=& - \, \frac{M^3_5}{2k_1} \, \int d^4 x \sqrt{-g} \,e^{-2 \, k_1 \, \theta_1 \,  T_1}\, \left ( 1 - e^{-2 \, k_1 \, \theta_1 \,  T_1} \right ) \, R^{(4)}\left(x\right) \nonumber \\
&-& \left ( \frac{M_5^3}{2k_1} \right ) \, \int d^4 x \sqrt{-g} \, \frac{1}{T_1^2} \, 
\left [- 1 + e^{-2 k_1 \theta_1 T_1} + 2 \, k_1 \, \theta_1 \, T_1 \, e^{-2 \, k_1 \theta_1 T_1} +
6 \, k_1^2 \, \theta_1^2 \, T_1^2 \, e^{-2 k_1 \theta_1 T_1}  \right ] \, \left ( \partial_\mu T_1 \right )^2 \nonumber \\
&-& \left ( \frac{M_5^3}{2k_1} \right ) \, \int d^4 x \sqrt{-g} \, \frac{1}{T_1} \, 
\left [ 1 - e^{-2 \, k_1 \theta_1 \, T_1} - 6  \, k_1 \theta_1 \, T_1 \, e^{2 \, k_1 \, \theta_1 \, T_1}
\right ] \, \Box T_1 \, , 
\end{eqnarray}
and, after integrating by parts the last term, we get: 
\begin{equation}
{\cal S}_1 [T_1] = - \, \frac{M^3_5}{2k_1} \, \int d^4 x \sqrt{-g} \, e^{-2 \, k_1 \, \theta_1 \,  T_1}\, \left ( 1 - e^{-2 \, k_1 \, \theta_1 \,  T_1} \right ) \, R^{(4)}\left(x\right) +
3 \, \frac{M_5^3}{k_1} \int d^4 x \sqrt{-g} \, \left [ \partial_\mu \left ( e^{- k_1 \, \theta_1 \, T_1 }\right ) \right ]^2 + \dots \, , 
\end{equation}
where $\dots$ stand for surface terms that vanish for infinite 4D space-time and terms proportional to $\partial_\mu \sqrt{-g}$ (that depict interactions between $T_1$ 
and 4D gravitons). On the other hand, when we integrate over the extra-dimension in the second segment, we get: 
\begin{eqnarray}
{\cal S}_2 [T_1, T_2] &=&  - \, 
\left ( \frac{M^3_5}{2k_2} \right ) \, \int d^4 x \sqrt{-g} \,e^{-2 \, \rho\left(1,x\right)}\, \left ( e^{-2 \, k_1 \, \theta_1 \,  T_1} - e^{-2 \, \rho\left(1,x\right)} \right ) \, R^{(4)}\left(x\right)  \nonumber \\
& - & \left ( \frac{M_5^3}{2k_2} \right ) \, \int d^4 x \sqrt{-g} \, \frac{e^{-2 \, \rho (1,x)}}{T_2^2} \, 
\left [1 - e^{2 \, k_2 (1-\theta_1) T_2} + 2 \, k_2 \, (1 - \theta_1) \, T_2  - 6 \, k_2^2 \, (1 - \theta_1)^2 \, T_2^2  \right ] \, \left ( \partial_\mu T_2 \right )^2  \nonumber \\
& - & \left ( \frac{M_5^3}{2k_2} \right ) \, \int d^4 x \sqrt{-g} \, \frac{2 k_1 \, \theta_1 \, e^{-2 \, \rho (1,x)} }{T_2} \, 
\left [1- e^{2 \, k_2 (1- \theta_1) T_2} +
6 \, k_2 \, (1- \theta_1) \, T_2  \right ] \, \left ( \partial_\mu T_1 \right ) \, \left ( \partial^\mu T_2 \right )  \nonumber \\
&-&\left ( \frac{M_5^3}{2k_2} \right ) \, \int d^4 x \sqrt{-g} \, 6 k_1^2 \, \theta_1^2 \, e^{-2 \, \rho (1,x)}  \, \left [ 1 - e^{2 \, k_2 (1-\theta_1) \, T_2} \right ] \, 
\left ( \partial_\mu T_1 \right )^2 \nonumber \\
&+&\left ( \frac{M_5^3}{2k_2} \right ) \, \int d^4 x \sqrt{-g} \, 6 k_1 \, \theta_1 \, e^{-2 \, \rho (1,x)}  \, \left [ 1 - e^{2 \, k_2 (1-\theta_1) \, T_2} \right ] \, 
\Box T_1 \nonumber \\
&+& \left ( \frac{M_5^3}{2k_2} \right ) \, \int d^4 x \sqrt{-g} \, \frac{e^{-2 \, \rho(1,x)}}{T_2} \, 
\left [ 1 - e^{2 \, k_2 (1-\theta_1) \, T_2} + 6  \, k_2 (1-\theta_1) \, T_2 \right ] \, \Box T_2 \, .
\end{eqnarray}
Once we integrate by parts the last two terms, we get:
\begin{eqnarray}
{\cal S}_2 [T_1,T_2] &=& - \, \frac{M^3_5}{2k_2} \, \int d^4 x \sqrt{-g} \, e^{-2 \, \rho\left(1,x\right)}\, \left ( e^{-2 \, k_1 \, \theta_1 \,  T_1} - e^{-2 \, \rho\left(1,x\right)} \right ) \, R^{(4)} \left(x\right)\nonumber \\
 &+&
3 \, \frac{M_5^3}{k_2} \int d^4 x \sqrt{-g} \, \left [ \partial_\mu \left ( e^{- \rho(1,x) }\right ) \right ]^2 \nonumber \\
&-& 3 \, \frac{M_5^3}{k_2} \int d^4 x \sqrt{-g} \, \left [ \partial_\mu \left ( e^{- k_1 \, \theta_1 \, T_1 }\right ) \right ]^2 
+ \dots \, , 
\end{eqnarray}
where, again, $\dots$ stand for surface terms that vanish for infinite 4D space-time and  interactions with 4D gravitons.

Summing the actions for the two segments, we get:  
\begin{eqnarray}
{\cal S}_T &=& {\cal S}_{4D}
+ \, 3 \, M_5^3 \, \left ( \frac{1}{k_1} - \frac{1}{k_2} \right ) \, \int d^4 x \, \sqrt{-g} \, \left [ \partial_\mu \left ( e^{-k_1 \, \theta_1 \, T_1} \right ) \right ]^2 \\
&& + \, 3 \, \left ( \frac{M_5^3}{ k_2} \right ) \, \int d^4 x \, \sqrt{-g} \, \left [ \partial_\mu \left ( e^{- k_2 \, (1-\theta_1) \, T_2 + k_1 \, \theta_1 \, T_1} \right ) \right ]^2
 + \dots \nonumber \\
 &=& {\cal S}_{4D} + \frac{1}{2} \int d^4 x \sqrt{-g} \, 
 \left [ 
 \left ( \partial_\mu \delta r_1 \right )^2 + 
 \left ( \partial_\mu \delta r_2 \right )^2  \right ] + \dots \, , 
\end{eqnarray}
where the two scalar fields are defined as:
\begin{equation}
\label{eq:radionwavefunctions}
\left \{
\begin{array}{l}
r_1(x) = 
\sqrt{\frac{6 M_5^3}{k_2}} \,  
\sqrt{ \delta k } \, 
e^{- k_1 \, \theta_1 \,  T_1(x)} =
\bar r_1 \, e^{\delta r_1 (x)/\bar r_1}
= \bar r_1 + \delta r_1 + \dots  \, , \\
\\
r_2(x) = \sqrt{\frac{6 M_5^3 }{k_2} } \, e^{- \left [ k_2 \, (1-\theta_1) \,  T_2(x) + k_1 \, \theta_1 \, T_1(x) \right ] } 
=
\bar r_2 \, e^{\delta r_2 (x)/\bar r_2}
= \bar r_2 + \delta r_2 + \dots 
\, ,
\end{array}
\right .
\end{equation}
where $\delta k = (k_2-k_1)/k_1$ is a parameter that
states how near or far we are from the evanescent limit.
The VEV's of the radion fields are: 
\begin{equation}
\left \{
\begin{array}{l}
\bar r_1 = \sqrt{\frac{6 M_5^3}{k_1}} \, \sqrt{\delta k} \, \bar \omega \, , \\
\\
\bar r_2 = \sqrt{\frac{6 M_5^3}{k_1}} \, \bar \omega \, \bar \xi \, . 
\end{array}
\right .
\end{equation}
whereas the quantum fluctuations over the VEV's are:
\begin{equation}
\left \{
\begin{array}{l}
\delta r_1 = - \sqrt{\frac{6 M_5^3}{k_2}} \, \sqrt{\frac{k_2 - k_1}{k_1}} \, \bar \omega \, 
\left [ k_1 \, \theta_1 \, \delta T_1 (x) \right ]\rm\, , \\
\\
\delta r_2 = - \sqrt{\frac{6 M_5^3}{k_2}} \, \bar \omega 
\, \bar \xi \, 
\left [ k_2 \, (1 - \theta_1) \, \delta T_2 (x) 
        + k_1 \, \theta_1 \, \delta T_1 (x) \right ] \, .  
\end{array}
\right .
\end{equation}
In these expressions, it is understood that $k_2 \geq k_1$ in order not to have a tachyonic mode. Notice that, after integrating over the extra-dimension, the fields $\delta r_1$ and $\delta r_2$ appear formally in the lagrangian as independent fields, so that
two ``radions" are indeed present in the spectrum (see Ref.~\cite{Seung1}). 

Using the metric choice in eq.~(\ref{eq:GWmetric2branes_App}), 
we get for the coefficients of the kinetic terms:
\begin{equation}
\left \{
\begin{array}{l}
K_1 (\omega) =\frac{1}{2 k_1^2} \frac{1}{\omega^2}\, 
\left [ 
\frac{C_{11}^2}{\nu_1+1} \, \frac{(1 - \omega^{2\nu_1+2})}{\omega^{2 \nu_1}} 
+ \frac{C_{21}^2}{\nu_1 - 1} \, \omega^2 \, 
(1 - \omega^{2 \nu_1-2}) 
+ 2 \, C_{11} C_{21} \, \left ( 1 - \omega^2 \right )
\right ] \, , \\
\\
K_2 (\omega, \xi) =  \frac{1}{2 k_2^2 } \, \frac{1}{\left ( \omega \xi \right )^2}\, 
\left [ 
\frac{C_{12}^2}{\nu_2+1} \,
\frac{\left (1 - \xi^{2\nu_2+2} \right )}{(\omega \xi )^{2 \nu_2}}
+ \frac{C_{22}^2}{\nu_2 - 1} \, \omega^{2 \nu_2} \, \xi^2
\, \left ( 1- \xi^{2 \nu_2 - 2}\right )
+ 2 \, C_{12} C_{22} \, (1 - \xi^2)
\right ] \, ,
\end{array}
\right .
\end{equation}
and for the effective potentials: 
\begin{equation}
\left \{
\begin{array}{l}
V_1 (\omega) =  \left ( 1 - \omega^{2 \nu_1} \right ) \, 
\left [ (\nu_1 + 2) \, 
\frac{1}{\omega^{2 \nu_1}} \, C_{11}^2 + 
(\nu_1 - 2) \, C_{21}^2 \right ] \, , \\
\\
V_2 (\omega, \xi) =  \omega^{2 \nu_2} \, \left (1 -  \xi^{2 \nu_2} \right ) \, 
\left [ (\nu_2 + 2) \, 
\frac{1}{\omega^{4 \nu_2}} \, 
\frac{1}{\xi^{2 \nu_2} } \, C_{12}^2 + 
(\nu_2 - 2) \, C_{22}^2 \right ] \, .
\end{array}
\right .
\end{equation}
Replacing the coefficients $C_{ij}$ with the results of eq.~(\ref{eq:zero modecoefficients}) we get for the kinetic terms:
\begin{equation}
\left \{
\begin{array}{lll}
K_1 (\omega) &=& \frac{\tilde v_{\rm UV}^2}{2 k_1^2} \, 
\frac{\omega^{2 \nu_1 - 2}}{
\left ( 1 - \omega^{2 \nu_1} \right )^2} \, \left \{
\frac{\left ( 1 - \omega^{2 \nu_1 + 2} \right )}{(\nu_1 +1)}
\, \left ( 1 - \omega^{2 - \nu_1} \, R_1 \right )^2 
\right . \\
&-& \left . \frac{\left ( 1 - \omega^{-2 \nu_1 + 2} \right )}{(\nu_1 - 1)}
\, \left ( 1 - \omega^{2 + \nu_1} \, R_1 \right )^2 
- 2 (1 - \omega^2) \, 
\left ( 1 - \omega^{2 - \nu_1} \, R_1 \right ) \, 
\left ( 1 - \omega^{2 + \nu_1} \, R_1 \right )
\right \}\, , \\
\\
K_2 (\omega, \xi) &=& 
\frac{\tilde v_{\rm UV}^2 }{2 k_1 k_2} \, 
\left ( \frac{\Delta L}{L_1}\right )^2 \, R_1^2 \, 
\frac{\omega^2 \, \xi^{2 \nu_2 - 2}}{
\left ( 1 - \xi^{2 \nu_2} \right )^2} \, \left \{
\frac{\left ( 1 - \xi^{2 \nu_2 + 2} \right )}{(\nu_2 +1)}
\, \left ( 1 - \xi^{2 - \nu_2} \, R_2 \right )^2 
\right . \\
&-& \left . \frac{\left ( 1 - \xi^{-2 \nu_2 + 2} \right )}{(\nu_2 - 1)}
\, \left ( 1 - \xi^{2 + \nu_2} \, R_2 \right )^2 
- 2 (1 - \xi^2) \, 
\left ( 1 - \xi^{2 - \nu_2} \, R_2 \right ) \, 
\left ( 1 - \xi^{2 + \nu_2} \, R_2 \right )
\right \}
\, .
\end{array}
\right .
\end{equation}
and for the effective potentials: 
\begin{equation}
\left \{
\begin{array}{l}
V_1 (\omega) =  \tilde v_{\rm UV}^2 \frac{1}{\left ( 1 - \omega^{2 \nu_1} \right )} \, \left [  (\nu_1 + 2) \omega^{2 \nu_1} 
\left (1 - \omega^{2-\nu_1} \, R_1 \right )^2
 + (\nu_1 - 2)\left (1 - \omega^{2+\nu_1} \, R_1 \right )^2 \right ] \, , \\
\\
V_2 (\omega, \xi) =   \tilde v_{\rm UV}^2 \, R_1^2 \, \left ( \frac{k_2}{k_1}\right ) \,  \left ( \frac{\Delta L}{L_1}\right )^2
\frac{\omega^4}{\left (1 -  \xi^{2 \nu_2} \right )} \, 
\left [ (\nu_2 + 2) \, \xi^{2\nu_2} \,  
\left (1 - \xi^{2-\nu_2} \, R_2 \right )^2
 + (\nu_2 - 2)\left (1 - \xi^{2+\nu_2} \, R_2 \right )^2\right ] \, ,
\end{array}
\right .
\end{equation}

The full effective potential:
\begin{equation}
V(\omega,\xi) = \frac{1}{L_1^2} \, V_1(\omega) + \frac{1}{\Delta L^2} \, V_2 (\omega, \xi)
\end{equation}
should be now minimized with respect to $r_1$ and $r_2$
(that are functions of $\omega$ and $\xi$). Let's simplify the full potential introducing
the three-brane analogue of eq.~(\ref{eq:Ffunction}): 
\begin{equation}
\label{eq:Ffunction3branes}
F_i (x) = \frac{1}{\left ( 1 - x^{2 \nu_i} \right )} \, 
\left [ (\nu_i + 2) \, x^{2 \nu_i} \, 
\left ( 1 - R_i \, x^{2-\nu_i}\right )^2 + 
(\nu_i - 2) \, 
\left ( 1 - R_i \, x^{2+\nu_i}\right )^2
\right ] \, ,
\end{equation}
which allow us to write: 
\begin{equation}
V(\omega,\xi) = 
\left ( \frac{\tilde v_{\rm UV}^2}{L_1^2} \right )
\left \{ F_1(\omega) + r_k \, R_1^2 \, \omega^4 F_2(\xi)
\right \} \, .
\end{equation}
Derivatives of the full potential are, trivially, 
\begin{equation}
\left \{
\begin{array}{l}
\frac{\partial V}{\partial \omega} = 
\left ( \frac{\tilde v_{\rm UV}^2}{L_1^2} \right ) 
\left [\frac{d F_1(\omega)}{d \omega} + 4 r_k \, R_1^2 \, \omega^3 \, F_2(\xi) \right ] \, , \\
\\
\frac{d V}{d \xi} = \left ( \frac{\tilde v_{\rm UV}^2}{L_1^2} \right ) \, r_k \, R_1^2 \, \omega^4 \, \frac{d F_2(\xi)}{d \xi} \, .
\end{array}
\right .
\end{equation}
Clearly, the minimum of the potential in $\xi$ satisfies the
relation $d F_2 (\xi)/d \xi = 0$. Solving this equation
for fixed $\bar \xi$ in $R_2$ gives: 
\begin{equation}
\label{eq:R2bar}
\left ( \bar R_2 \right )_\pm = 
\frac{\nu_2}{2} \left ( 1 \pm \Delta \right ) \, \bar \xi^{\nu_2 - 2} + {\cal O}(\bar \xi^5) \, ,
\end{equation}
where:
\begin{equation}
\Delta = \left ( 1 - \frac{4}{\nu_2 + 2} \right )^{1/2} \, .
\end{equation}
The two solutions in $\bar R_2$ (for any value of $\bar \xi$)
are always real, as $\nu_2 \ge 2$ for any value of $\epsilon_2$. 
Once we have found the minimum of the potential in $R_2$ for 
a given value of $\bar \xi$, we can minimize in $\omega$. 
The quadratic equation in $\omega$ now depends on the 
specific value of $\bar \xi$ through the second term, 
proportional to $F_2 (\bar \xi)$. Assuming that $\epsilon_2$ is
``small" (at least as small as $\bar \xi$) it can be shown
that $F_2 (\bar \xi) \sim \epsilon_2 + {\cal O}(\epsilon_2^2)$. 
By solving the first equation in $R_1$ for a fixed, ``small", 
$\bar \omega$, we find: 
\begin{equation}
\label{eq:R1bar}
\left ( \bar R_1 \right )_\pm = \frac{\nu_1 (\nu_1 + 2)}{
2 \left [ \nu_1 + 2 + r_k F_2 (\bar \xi) \right ]} \, 
\left ( 1 \pm \Delta^\prime \right ) \, \bar \omega^{\nu_1 - 2} \, ,
\end{equation}
where:
\begin{equation}
\Delta^\prime = \left ( 1 - \frac{4 
\left [ \nu_1 + 2 + r_k F_2 (\bar \xi) \right ]
}{(\nu_2 + 2)^2} \right )^{1/2} \, .
\end{equation}
Inverting eqs.~(\ref{eq:R2bar}) and (\ref{eq:R1bar}) we get: 
\begin{equation}
\left \{
\begin{array}{l}
\bar \omega = \left [ \frac{2 (\nu_1 + 2 + r_k \, F_2 (\bar \xi))}{\nu_1 (\nu_1 + 2)} \, \frac{1}{1 - \Delta^\prime} \, \bar R_{1,-} \right ]^{1/(\nu_1 - 2)} \, ,  \\
\\
\bar \xi = \left [ \frac{2}{\nu_2} \, \frac{1}{1 - \Delta} \, \bar R_{2,-} \right ]^{1/(\nu_2 - 2)} \, , 
\end{array}
\right .
\end{equation}
where these results are obtained up to ${\cal O}(\bar \omega^5)$
and ${\cal O}(\bar \xi^5)$, under the assumptions that both
$\bar \omega$ and $\bar \xi$ are ``small". Now, expanding
in $\epsilon_1$ and $\epsilon_2$, we get: 
\begin{equation}
\left \{
\begin{array}{l}
k_1 L_1 = - \frac{1}{\epsilon_1} \, \ln R_1 =
\frac{k_1^2}{m_1^2} \, \ln \left ( \frac{\tilde v_{\rm UV}}{\tilde v_{\rm IR}}\right ) + {\cal O} \left (\epsilon_1^2 \right ) + \dots \, , \\
\\
k_2   \Delta L = - \frac{1}{\epsilon_2} \, \ln R_2 =
\frac{k_2^2}{m_2^2} \, \ln \left ( \frac{\tilde v_{\rm IR}}{\tilde v_{\rm DIR}}\right ) + {\cal O} \left (\epsilon_2^2 \right ) + \dots \, , 
\end{array}
\right .
\end{equation}
where $\dots$ stand for terms of higher order in $\omega, \xi$ and the $\nu_i$'s have been expanded at second order
in the $\epsilon_i$'s. 

\bibliographystyle{JHEP}
\bibliography{paper}
	
\end{document}